\documentclass[12pt]{article}
\usepackage{cite}
\usepackage{amsmath, amsthm, amssymb,slashed,url}
\usepackage{ifpdf}
\ifpdf
  \usepackage[pdftex]{graphicx}
  \usepackage{epstopdf}
\else
  \usepackage[dvips]{graphicx}
\fi
\parskip=\baselineskip
\def\Bbb{\mathbb}
\def\Tr{{\rm Tr}}
\def\M{{\mathcal M}}
\def\16{{\bf 16}}
\def\1{{\bf 1}}
\def\2{{\bf 2}}
\def\3{{\bf 3}}
\def\4{{\bf 4}}
\def\Bbb{\mathbb}
\def\Tr{{\mathrm {Tr}}}
\def\16{{\bf 16}}
\def\1{{\bf 1}}
\def\2{{\bf 2}}
\def\3{{\bf 3}}
\def\4{{\bf 4}}

\def\bar{\overline}
\def\tilde{\widetilde}
\def\R{{\Bbb{R}}}\def\Z{{\Bbb{Z}}}
\def\N{{\mathcal N}}
\def\hat{\widehat}
\font\teneurm=eurm10 \font\seveneurm=eurm7 \font\fiveeurm=eurm5
\newfam\eurmfam
\textfont\eurmfam=\teneurm \scriptfont\eurmfam=\seveneurm
\scriptscriptfont\eurmfam=\fiveeurm

\font\teneusm=eusm10 \font\seveneusm=eusm7 \font\fiveeusm=eusm5
\newfam\eusmfam
\textfont\eusmfam=\teneusm \scriptfont\eusmfam=\seveneusm
\scriptscriptfont\eusmfam=\fiveeusm
\def\eusm#1{{\fam\eusmfam\relax#1}}
\font\tencmmib=cmmib10 \skewchar\tencmmib='177
\font\sevencmmib=cmmib7 \skewchar\sevencmmib='177
\font\fivecmmib=cmmib5 \skewchar\fivecmmib='177
\newfam\cmmibfam
\textfont\cmmibfam=\tencmmib \scriptfont\cmmibfam=\sevencmmib
\scriptscriptfont\cmmibfam=\fivecmmib
\def\cmmib#1{{\fam\cmmibfam\relax#1}}
\numberwithin{equation}{section}

\def\d{\mathrm d}
\def\C{{\Bbb C}}
\def\Z{{\Bbb Z}}
\def\A{{\mathcal A}}
\def\bar{\overline}
\def\bar{\overline}
\def\tilde{\widetilde}
\def\R{{\Bbb{R}}}\def\Z{{\Bbb{Z}}}
\def\N{{\mathcal N}}
\def\hat{\widehat}
\font\teneurm=eurm10 \font\seveneurm=eurm7 \font\fiveeurm=eurm5
\newfam\eurmfam
\textfont\eurmfam=\teneurm \scriptfont\eurmfam=\seveneurm
\scriptscriptfont\eurmfam=\fiveeurm

 \font\teneusm=eusm10 \font\seveneusm=eusm7 \font\fiveeusm=eusm5
\newfam\eusmfam
\textfont\eusmfam=\teneusm \scriptfont\eusmfam=\seveneusm
\scriptscriptfont\eusmfam=\fiveeusm
\def\eusm#1{{\fam\eusmfam\relax#1}}
\font\tencmmib=cmmib10 \skewchar\tencmmib='177
\font\sevencmmib=cmmib7 \skewchar\sevencmmib='177
\font\fivecmmib=cmmib5 \skewchar\fivecmmib='177
\newfam\cmmibfam
\textfont\cmmibfam=\tencmmib \scriptfont\cmmibfam=\sevencmmib
\scriptscriptfont\cmmibfam=\fivecmmib
\def\cmmib#1{{\fam\cmmibfam\relax#1}}
\numberwithin{equation}{section}

\def\d{\mathrm d}
\def\frak{\mathfrak}
\begin{document}

\begin{titlepage}
\begin{flushright}
hep-th/yymm.nnnn
\end{flushright}
\vskip 1.5in
\begin{center}
{\bf\Large{The Omega Deformation,   }\vskip0cm \bf\Large{Branes,
Integrability, And Liouville Theory } }\vskip 0.5cm {Nikita
Nekrasov$^1$ and Edward Witten$^2$} \vskip 0.05in {\small{
\textit{$^1$Institut des Hautes Etudes Scientifiques,
Bures-sur-Yvette, France,}\vskip -.4 cm \textit{ and Simons
Center for Geometry and Physics, Stony Brook University, Stony Brook, New York,
11794}\vskip 0 cm \textit{$^2$School of Natural Sciences,
Institute for Advanced Study}\vskip -.4cm {\textit{Einstein Drive,
Princeton, NJ 08540 USA}}}}

\end{center}
\vskip 0.5in
 \baselineskip 16pt
\date{April, 2008}

\def\re{{\mathrm{Re}}}
\begin{abstract}
We reformulate the $\Omega$-deformation of four-dimensional gauge
theory in a way that is valid away from fixed points of the
associated group action.  We use this reformulation together with
the theory of coisotropic $A$-branes to explain recent results
linking the $\Omega$-deformation to integrable Hamiltonian systems
in one direction and Liouville theory of two-dimensional conformal
field theory in another direction.
\end{abstract}
\end{titlepage}
\vfill\eject \tableofcontents
\section{Introduction}

\def\Bcc{{\mathcal B}_{\mathrm{cc}}}
In this paper, we will apply methods of two-dimensional
sigma-models to two different problems in four-dimensional gauge
theory.

The first problem involves the relation of gauge theory to quantum
integrable systems.  Vacua of massive two-dimensional gauge
theories with $\N=2$ supersymmetry correspond unexpectedly
\cite{MNS,GS,GS2}
  to the quantum eigenstates of a
quantum integrable system.  This correspondence has recently been extended
 \cite{NS,NSmore} to a much
wider class of examples. In the  present paper, we will approach
this rather surprising relation from a new angle.  We focus on
what is perhaps the most challenging example of the correspondence
in question. This is the case \cite{NS2} that the two-dimensional
theory arises by reducing a four-dimensional $\N=2$ supersymmetric
gauge theory to two dimensions with the help of the $\Omega$
deformation \cite{Nek}. The resulting theories are associated to
quantum integrable systems that arise by quantizing a
finite-dimensional classical phase space. By contrast, many purely
two-dimensional examples
 are more directly described in terms of quantum
spin systems.

Our basic idea is to map this problem to a  brane construction in
two dimensions.  Under certain conditions, a two-dimensional
$A$-model admits unusual branes \cite{KO} whose existence brings
noncommutativity into the $A$-model in several related ways
\cite{Kap,AZ,KW,GW}. Of most direct relevance to us is an
$A$-brane construction that leads to quantization of
finite-dimensional classical phase spaces \cite{GW}. In this
construction, integrability is natural.  The construction involves
a pair of branes; one is an ordinary Lagrangian $A$-brane, and the
second, which has been called the canonical coisotropic $A$-brane
$\Bcc$, is the most simple example of the unusual $A$-branes
introduced in \cite{KO}.

In section 2, we describe some background that may be helpful.  We
recall the basic reason that massive supersymmetric gauge theories
in two dimensions are related to integrability. We review a
variety of facts about two-dimensional sigma-models and their
relation to four-dimensional gauge theories, including the brane
construction \cite{GW} that will be our main tool.

 In section 3, we describe how to study the
$\Omega$-deformation via this framework.  Here, as in \cite{NS2},
we consider the $\Omega$-deformation of $\N=2$ super Yang-Mills
theory on $\R^4$, defined by a $U(1)$ action that leaves fixed a
two-plane $\R^2\subset\R^4$.  There is an immediate problem, as it
appears that both the $\Omega$-deformed Lagrangian and the
supersymmetry preserved by the $\Omega$-deformation are not what
we need to make contact with the brane construction of \cite{GW}.
To overcome this difficulty, we give a new interpretation of the
$\Omega$-deformation. The $\Omega$-deformation is defined using a
vector field $V$ that generates a $U(1)$ symmetry of spacetime. As
originally defined in \cite{Nek}, building on \cite{MNS,LNS1,LNS2}, the $\Omega$ deformation involves a
deformation of the Lagrangian that preserves part of the
supersymmetry.  We will give an alternative description of the
$\Omega$-deformation that is valid (for our purposes, which do not
depend on the precise choice of a $U(1)$-invariant metric) away
from the zeroes of $V$: by a change of field variables, one can
remove the deformation from the Lagrangian while rotating the
unbroken supersymmetry.

Taking this into account, we show that $\Omega$-deformed
supersymmetric gauge theory, in the situation considered in
\cite{NS2}, reduces naturally to an $A$-model in two dimensions,
with precisely the brane setup of \cite{GW}. The most unusual part
of this construction is the exotic $A$-brane $\Bcc$.  It arises in
giving a two-dimensional interpretation to what in four dimensions
are simply the $U(1)$ fixed points.

Section 4 is devoted to applying our framework to a very
different-sounding problem. Here our aim is to make contact with
remarkable results \cite{AGT} linking the $\Omega$-deformation in
four dimensions with Liouville theory (and its higher rank
analogs) in two dimensions.  Our method can be applied to this
situation in an interesting way and answers some of the questions,
although many points are not yet clear.  In this application, the
important branes are all rotated or dual versions of the
coisotropic brane $\Bcc$.

Though our discussion is applicable to any $\N=2$ supersymmetric
gauge theory, we often specialize to a convenient and large class
of such theories which arise  by compactification  of the
six-dimensional $(0,2)$ model of type $G$ (here $G$ is a simple
and simply-laced Lie group) on a Riemann surface $C$.  For every
choice of a system of $A$-cycles on $C$, one gets in four
dimensions a gauge theory realization \cite{DG} in which the gauge
group is a product of copies of $G$. We call these generalized
quiver  theories. Compactification on a two-torus reduces a
generalized quiver theory to a sigma-model in two dimensions in
which the target space is $\M_H$, the moduli space of Higgs
bundles on $C$, endowed with a hyper-Kahler metric \cite{Hitchin}.

To keep things simple, a number of calculations in the body of
this paper are carried out only for compactification on a
rectangular two-torus and only for special values of the
parameters of the $\Omega$-deformation.  A more complete treatment
is given in the appendix.

Our strategy throughout this paper is qualitatively similar to
many applications of toric geometry in string theory.  For
example, see \cite{CN}.  Also, the brane construction we use in
section 3 to obtain an eigenvalue problem for the commuting
Hamiltonians of a quantum integrable system is a cousin of a
construction that has been analyzed in the literature on geometric
Langlands \cite{F1,FFR,F2,F3}.  Our (limited) understanding of the
relation is described in section \ref{eigand}. Finally, our
results in section 4 are qualitatively in agreement with previous
arguments suggesting a relation between duality of Liouville
theory and what is often called quantum geometric Langlands
\cite{FF,F,Teschner,Teschner2}. We hope it will prove possible to
make this connection more precise.

\def\L{\mathcal L}
\def\tW{\widetilde W}
\section{Some Background}\label{aspects}

The present section is devoted to describing some background that
may be helpful.  None of these results are new.

In section \ref{linko}, we will review the basic reason that there
is a link between two-dimensional massive supersymmetric gauge
theories and integrability. In section \ref{realsections}, we
discuss some generalities about what it might mean to quantize a
complex integrable system. In section \ref{coiso}, we review some
relevant facts about the two-dimensional topological $A$-model and
recall  how $A$-branes can be used for quantization.  In section
\ref{little}, we describe some pertinent differential geometry.

\subsection{The Basic Link Between Gauge Theory And
Integrability}\label{linko}

\def\d{{\mathrm d}}
\def\rm{\mathrm}
For illustrative purposes, we consider a two-dimensional $\N=2$
supersymmetric gauge theory with $(2,2)$ supersymmetry and a gauge
group $U(1)^r$.  (The general case of a gauge group $G$ of rank
$r$ can be treated similarly; it leads to a nonabelian but equally
tractable version of the Lagrangian (\ref{kerf}).)

The $U(1)^r$ gauge theory has $r$ vector multiplets.  Their
gauge-invariant content can be described by the twisted chiral
multiplets
\begin{equation}\label{twch}\Sigma_a=\sigma_a-i\sqrt 2\theta^+\bar\lambda_{a\,+}-
i\sqrt 2\bar\theta^-
\lambda_{a\,-}+\sqrt 2\theta^+\bar\theta^-(D_a-i\star F_a)+\dots,~a=1,\dots,r\end{equation}
where $\theta^\pm$ and $\bar\theta^\pm$ are superspace
coordinates, $\sigma_a$ and $\lambda_a$ are ordinary scalar and
fermi fields, $D_a$ are auxiliary fields,
 $F_a$ is the field strength of the $a^{th}$ $U(1)$ gauge field, and $\star$ is the Hodge
 star operator.    We are interested in theories in which all vacua are massive, that is, admit only excitations
 with positive mass.  For this to be the case, any chiral multiplets that are present in the theory
 must be massive; they can be integrated out, possibly making contributions to the effective twisted chiral superpotential
 $\tilde W(\Sigma_1,\dots,\Sigma_r)$ of the vector multiplets.  The theory is then massive if $\tilde W$
 is sufficiently generic.

 If so, the only important part of the effective action at low energies is the contribution of $\tilde W$:
\begin{equation}\label{kerf}I_{\tilde W}=\frac{1}{2\sqrt 2}\int d^2x\left(\d\theta^+\d\bar\theta^-\tilde W(\Sigma_1,\dots,
\Sigma_r)+\rm{c.c.}\right).\end{equation}
This is the action of a topological field theory; by dropping
higher derivative terms in the action, we have effectively taken
the masses of all excitations to infinity.  The quantum states
obtained by quantizing (\ref{kerf}) are the same as the vacua in
the infinite volume limit of an underlying physical theory whose
action consists of $I_{\tilde W}$ plus irrelevant terms of higher
dimension.

After performing the $\theta$ integrals, $I_{\tilde W}$ gives for
the fermions $\lambda_a$ a mass matrix $m_{ab}=\partial^2\tilde
W/\partial\sigma^a\partial\sigma^b$.  If this is nondegenerate
(invertible) in every vacuum, as we will assume, then the fermions
are ``massive'' (but nonpropagating in the approximation of
(\ref{kerf}), as they have no kinetic energy). Let us look more
closely at the bosonic part of $I_{\tilde W}$, which turns out to
be
\begin{equation}\label{zerf}I_{\tilde W,B}=\int \d^2x \,\sum_a\left( D_a\rm{Re}\frac{\partial \tilde W_a}
{\partial\sigma_a}+\star F_a \rm{Im}\frac{\partial\tilde
W_a}{\partial\sigma_a}\right).\end{equation} Since the part of
$I_{\tilde W,B}$ that involves the auxiliary fields $D_a$ has no
derivatives, the effect of the terms involving $D_a$ is simply to
impose constraints:
\begin{equation}\label{constraint}\rm{Re}\,\frac{\partial\tilde W}{\partial\sigma_a}=0.\end{equation}
Let $\cmmib P$ be the locus (in a copy of $\C^r$ parametrized by
the $\sigma_a$) defined by these constraints. If $\tilde W$ is
sufficiently generic, then $p_a=\rm{Im}\,\partial \tilde
W/\partial\sigma_a$ is a good system of coordinates on $\cmmib P$.

We will analyze the dynamics of the fields $p_a$ and the $U(1)$
gauge fields $A_a$ on a two-manifold $\R\times S^1$, with metric
$\d s^2=\d t^2-\d x^2$, where $t$ is real-valued and $x\cong x+R$
(for some constant $R$) parametrizes $S^1$.  We work in the gauge
in which the $t$ components of all gauge fields vanish, that is
$A_{t,a}=0$. The Gauss law constraint $\delta I_{\tilde
W,B}/\delta A_{t,a}=0$ is $\partial p_a/\partial\sigma=0$, so in
this gauge, the $p_a$ are functions of $t$ only.  We can
further fix the gauge so that the spatial parts of the gauge
fields are constants: $A_{x,a}=\phi_a/R$, where the $\phi_a$ are
angular variables that depend only on $t$. The action then reduces
to
\begin{equation}\label{polly}I'=\int \d t\sum_a p_a\frac{\d\phi_a}{\d t}.\end{equation}
This is an integrable system written in action-angle coordinates.
The $p_a$ are the action variables and the $\phi_a$ are the angle
variables.  The nonzero Poisson brackets are
$\{p_a,\phi_b\}=\delta_{ab}$.  In particular, the $p_a$ are
Poisson-commuting and quantum states can be labeled by the values
of the $p_a$.

However, not all values of the $p_a$ occur.  The reason for this
is that the space $T$ obtained by specifying the values of the
$p_a$ is compact; it is a torus, parametrized by the $\phi_a$.  In
the WKB approximation, the good values of the $p_a$ are those for
which the one-form $\omega=\sum_ap_a\d\phi_a$ has all its periods
integer multiples of $2\pi$. Each $\phi_a$ parametrizes a circle
$S_a$, and the basic periods are
$\int_{S_a}\d\phi_b=2\pi\delta_{ab}$.  So the condition for
$\omega$ to have all integer periods is simply that the $p_a$
should be integers.  Recalling the definition of the $p_a$, our
conclusion is that
\begin{equation}\label{polyp}\rm{Im}\,\frac{\partial\tilde W}{\partial\sigma_a}=n_a,~n_a\in\Z.\end{equation}
Evidently, this result can be combined with eqn.
(\ref{constraint}) to give the holomorphic relation
\begin{equation}\label{zolyp}\frac{\partial\tilde W}{\partial\sigma_a}=in_a, ~n_a\in\Z\end{equation}
or equivalently
\begin{equation}\label{olyp}\exp\left(2\pi \frac{\partial\tilde W}{\partial\sigma_a}\right)=1\end{equation}
which characterizes the quantum states of the topological field
theory with action $I_{\tilde W}$, or equivalently the vacua of an
underlying massive theory whose action differs from $I_{\tilde W}$
by irrelevant operators.

Apart from possibly orienting the reader to the results of
\cite{MNS,GS,GS2,NS,NSmore,NS2}, the reason that we have explained
these matters here is to draw a lesson that will be important for
our derivation in the rest of this paper.  The constraint
(\ref{olyp}) that determines the vacuum states is holomorphic in
$\tilde W$, but in the derivation the real and imaginary parts of
$\partial\tilde W/\partial\sigma_a$ have gone their separate ways.
Since this happens just in the two-dimensional derivation, it will
hopefully come as no surprise when something similar happens in
deriving this story from four dimensions.

\subsubsection{Validity Of The WKB Approximation}\label{wkb}

There remain two points to clarify about this derivation. First,
we have presented (\ref{zolyp}) or (\ref{olyp}) as the result of a
WKB approximation, but actually in the context of supersymmetric
gauge theory, these formulas are exact (modulo additive constants
that will be discussed). This is a standard result; for example,
see \cite{LNS} for a derivation via path integrals. The basic idea
of the derivation is to integrate over the gauge fields with the
action (\ref{zerf}). In the integral
\begin{equation}\label{kerfk}\int DA_a\,\exp\left(i\sum_a\int
F_a\,{\mathrm {Im}}\,\partial \tilde W/\partial
\sigma_a\right)\end{equation} one would like to change variables
from $A_a$ to $F_a$, using the fact that $F_a$ contains almost the
full gauge-invariant content of $A_a$ and that (as the gauge group
is abelian) the volume of the space of gauge fields with given
curvature $F_a$ is independent of $F_a$.  If the $F_a$ could be
treated as arbitrary two-forms, the resulting integral
\begin{equation}\label{omerf}\int DF_a\,\exp\left(i\sum_a\int F_a\,{\mathrm{Im}}\,\partial \tilde W/\partial
\sigma_a\right)\end{equation} would simply give a delta function
setting ${\mathrm{Im}}\,\partial \tilde W/\partial \sigma_a=0$.
Instead, $F_a$ is constrained by Dirac quantization, $\int
F_a/2\pi \in\Z$, which one can incorporate by including in the
path integral a factor $\sum_{n_a\in\Z}\exp(-i\sum_an_a\int F_a)$.
With this factor included, the integral over the $F_a$ now gives a
constraint ${\mathrm{Im}}\,\partial \tilde W/\partial \sigma_a=
n_a$.

The result $\partial\tilde W/\partial\sigma_a=i n_a$ has a more
elementary analog for chiral multiplets.  If $\Phi_i=\phi_i+\theta
\psi_i+\theta^2F_i+\dots$ are chiral multiplets with
superpotential $W(\Phi_1,\dots,\Phi_s)$, then a vacuum is
characterized by $\partial
W(\phi_1,\dots,\phi_s)/\partial\phi_i=0$.  No integer analogous to
$n_a$ enters because the auxiliary fields $F_i$ of the chiral
multiplet are independent complex fields.  The result
(\ref{zolyp}) for twisted chiral multiplets of gauge theory is
slightly more complicated because the ``auxiliary field'' $\tilde
F_a=D_a-i\star F_a$ in (\ref{twch}) is not an arbitrary complex
field; its imaginary part is subject to Dirac quantization.

In general, it would be unrealistic to expect the WKB
approximation to be always exact for an abstract integrable system
written in action-angle variables. Indeed, if $p_a,\phi_a$ are a
set of action-angle variables, then via a canonical
transformation, one can map them (in many different ways) to
another set of equally good action-angle variables $p'_a,\phi'_a$.
It is impossible for the WKB approximation to be valid in every
set of action-angle variables, so in general at best it will be
exact only for a system of action-angle variables that is in some
way distinguished. However, in the present context, the angle
variables $\phi_a$ are indeed distinguished, as they originate
from gauge fields; hence the conjugate variables $p_a$ are also
distinguished, modulo possible additive constants which will
indeed play a role.  This makes it possible for the WKB
approximation to be exact.

  The constraint (\ref{constraint}) is not holomorphic in $\tilde W$,
but the condition defining the vacuum states of the underlying
gauge theory must be holomorphic. The WKB formula (\ref{polyp})
has indeed combined with the constraint (\ref{constraint}) to give
the desired holomorphy in (\ref{olyp}). A quantum correction to
(\ref{polyp}), apart from the possibility of adding a constant to
the right hand side, would spoil holomorphy. By adding a constant,
we mean replacing (\ref{polyp}) with
\begin{equation}\label{rufus}\rm{Im}\,\frac{\partial\tilde W}{\partial\sigma_a}=n_a-\frac{\theta_a}{2\pi},
\end{equation}
for some angles $\theta_a$.  The $\theta_a$ must be constants,
since if they were nontrivial functions of the $p_a$, this would
spoil holomorphy.

\def\J{{\mathcal J}}
For a perhaps fuller explanation, let us consider adding
$\theta$-angles to the underlying gauge theory action:
\begin{equation}\label{tangle}I_{\tilde W}\to I_{\tilde W}+\sum_a\theta_a\int\frac{F_a}{2\pi}.\end{equation}
In the above derivation, the one-form $\omega=\sum_ap_a\,\d\phi_a$
is then replaced by $\omega'=\sum_a(p_a+\theta_a/2\pi)\d\phi_a$,
and this leads exactly to the generalized WKB condition
(\ref{rufus}).  So this generalization is indeed something that we have to consider.

On the other hand, two-dimensional gauge theory is in general
invariant under $\theta_a\to \theta_a+2\pi m_a$.  (This is usually
deduced from Dirac quantization of $\int F_a$, or alternatively
proved by observing that the unitary operator
$\exp(i\sum_am_a\oint_S A_a)$ brings about a shift
$\theta_a\to\theta_a+2\pi m_a$.) Allowing for arbitrary integers
$n_a$ in (\ref{rufus}) insures that the spectrum has this
invariance.

This discussion may raise the following question: is there a
renormalization of the angles $\theta_a$ in going from the
underlying Lagrangian to the exact condition (\ref{rufus}) that
characterizes the quantum states?  In fact, this question is
equivalent to asking whether the effective superpotential $\tilde
W$ has been identified correctly, since (as one can see by
comparing eqns. (\ref{zerf}) and (\ref{tangle})) introducing the
angles $\theta_a$ is equivalent to changing $\tilde W$ to $\tilde
W+i\sum_a \theta_a\sigma_a/2\pi$.   So, given a microscopic theory, the problem
of finding the right constants in the WKB formula is part of the problem of correctly
computing the effective twisted chiral superpotential $\tilde W$.

\subsubsection{Observables}\label{observables}

What are the natural observables of this integrable system?  Going
back to the underlying gauge theory description, the natural
operators are those of the twisted chiral ring.  The twisted
chiral ring of this theory is a polynomial ring generated by the
operators $\sigma_a$, $a=1,\dots,r$.  They commute, and the
existence of these commuting operators can be regarded as an
explanation of why a detailed analysis of this system has led to
an integrable description in action-angle variables.

The derivation sketched above has led to a basis of quantum
states, characterized by (\ref{olyp}),  in which the   $\sigma_a$
have definite values.   The chiral
 ring generators -- and therefore all the operators of the chiral ring -- are diagonal in this
 basis.

This was indeed the starting point of \cite{MNS,GS,GS2,NS,NSmore,
NS2}, where the conditions (\ref{olyp}) that determine the
eigenvalues of the chiral ring generators were interpreted in
terms of Bethe ansatz equations of an integrable system presented
not in action-angle variables but in some alternative and
physically interesting description.  The relation between the two
pictures is not well understood, and we will not shed light on it
in the present paper. Rather, in this paper, our goal is to
understand the analog of the above two-dimensional derivation in
the context of $\Omega$-deformed theories in four dimensions.

\subsection{Complex Integrable Systems And Their Real
Sections}\label{realsections}

\def\M{{\mathcal M}}
We want to put this in the context of $\N=2$ supersymmetric gauge
theories in four dimensions. Let us denote as $\cmmib B$ the
Coulomb branch of the moduli space of vacua of such a theory.
(Depending on the spectrum of hypermultiplets, there may also be
Higgs branches of vacua or mixed branches, but they will not be
important in the present paper.) $\cmmib B$ parameterizes a family
of abelian varieties. We denote as $X$ the total space of this
fibration.  $X$ is a complex symplectic manifold, with a
holomorphic symplectic form $\Omega$.  The fibers of the fibration
$X\to\cmmib B$ are holomorphic submanifolds that are Lagrangian
with respect to $\Omega$.  This means that $X$ is a completely
integrable Hamiltonian system in the complex sense \cite{DonW}.
The holomorphic functions on $\cmmib B$ are the action variables,
and the fiber coordinates are the angle variables.

An important class of examples \cite{DG} is derived by
compactification of the six-dimensional $(0,2)$ theory of type $G$
on a Riemann surface $C$.  In this case, $X$ is a moduli space of
Higgs bundles on $C$, with structure group $G$, and the fibration
$X\to\cmmib B$ is the Hitchin fibration. The integrable system is
that of Hitchin \cite{Hitchin} and we denote $X$ as $\M_H$.  It
may be that all $\N=2$ supersymmetric gauge theories in four
dimensions are related to Hitchin systems with suitable
singularities included. At any rate, we will continue the
discussion with this important class of examples in mind.  When
formulated as four-dimensional gauge theories, these theories have
a generalized quiver structure.

In what sense might one quantize the completely integrable system
$\M_H$?  An important property of $\M_H$ is that it is birational
to a cotangent bundle $T^*\M$, where $\M$ is the moduli space of
stable holomorphic $G$-bundles on $C$.  Such a ``birational''
equivalence holds after deleting complex submanifolds on both
sides.  The holomorphic symplectic form of $\M_H$ maps to the
standard one on $T^*\M$.

This equivalence means that one can hope to map the commuting
Hamiltonians of the integrable system -- in other words, the
polynomial functions on $\cmmib B$ -- to holomorphic differential
operators on $\M$. This was first done for rank 2 Higgs bundles
(in other words, for the case that the gauge group of the
generalized quiver is a product of $SU(2)$'s) in \cite{HitchA} and
in generality in \cite{BD}; these constructions were based on
conformal field theory on $C$. An alternative argument starting
from four-dimensional gauge theory was explained at the end of
section 11.1 of \cite{KW}.  An important fact is that the
differential operators in question act not on functions but on
sections of $K^{1/2}$, where $K$ is the canonical line bundle of
$\M$.  They commute with each other, just like the underlying
classical Hamiltonians.

\def\D{{\mathcal D}}
\def\B{{\mathcal B}}
For illustration, let us suppose that $\M$ is of complex dimension
1, which happens for rank 2 if $C$ is a Riemann surface of genus 1
with one marked point.  (This corresponds to an $\N=2$ system that
is known as the $\N=2^*$ theory, whose Coulomb branch was related
to the corresponding Hitchin system in \cite{DonW}. It is also the
example considered in \cite{NS2}.) Consider a holomorphic function
$f$ on $\M_H$ that when restricted to the cotangent bundle $T^*\M$
is quadratic on each fiber of the cotangent bundle.  (Such a
function can be derived from a Beltrami differential on $C$; in
gauge theory with a product of $SU(2)$ gauge groups, it
corresponds to a linear combination of the usual order parameters
of the Coulomb branch.) Letting $z$ be a local complex coordinate
on $\M$, and $p$ a fiber coordinate, $f$ is of the form
$f(z,p)=a(z)p^2$. In deformation quantization, $p$ maps to $\d/\d
z$ so $f$ will map to a second order holomorphic differential
operator $\D_f$ on $C$. Picking a local trivialization of
$K^{1/2}$, $\D_f$ is concretely given by a formula
\begin{equation}\label{givefor}\D_f=a(z)\frac{\d^2}{\d
z^2}+b(z)\frac{\d}{\d z}+c(z),\end{equation} with local
holomorphic functions $a,b,c$. Here, $a$ can be read off from the
classical function $f=a(z)p^2$, but the usual quantum mechanical
problem of operator ordering affects $b$ and $c$.  In principle,
$b$ and $c$ can be computed in $\sigma$-model perturbation theory
for the brane system described in section \ref{coiso}. It is a
non-trivial fact that unique $b$ and $c$ functions do exist such
that the operator $\D_f$ is a globally-defined holomorphic
differential operator acting on sections of $K^{1/2}$. (This is
not true if $K^{1/2}$ is replaced by any other line bundle.)  For
various explanations, see \cite{HitchA}, \cite{BD}, or section 11
of \cite{KW}.  The details are anyway not really pertinent to the
qualitative remarks we will make here.

 This construction has been formally called
quantization, but it is not what physicists usually  mean by
quantization.  A holomorphic differential operator $\D_f$ is
constructed, but there is no Hilbert space that it acts on.
Indeed, what could such a Hilbert space be?  $\D_f$ could act on
global holomorphic sections of $K^{1/2}$, but there are none.
$\D_f$ could also act on global $C^\infty$ sections of $K^{1/2}$,
or on meromorphic sections of $K^{1/2}$ with prescribed poles or
with arbitrary poles. But none of these spaces is a Hilbert space
in a natural way.

\def\H{{\mathcal H}}
To get a Hilbert space, we should pick\footnote{The actual construction made
in \cite{NS2} is more complicated than we are about to explain.  The reason is
that in their example, $\M_H$ is only birational to $T^*\M$, for a Riemann surface $\M$.
If one aproximates $\M_H$ as $T^*\M$, one introduces a bad point in $\M$ where the
Hamiltonians have poles.  The curves
$\gamma$ considered in \cite{NS2} pass through the bad point and this causes
the definition of the Hilbert space to be more involved.} a closed curve $\gamma\in
\M$. $K^{1/2}$ restricts on $\gamma$ to the bundle of
complex-valued half-densities on $\gamma$, and these form a
natural Hilbert space $\H_\gamma$. $\D_f$ acts naturally on
$\H_\gamma$ (as an unbounded operator that is densely defined) for
the following reason. Real-analytic sections of $K^{1/2}$ along
$\gamma$ form a dense subspace $W_\gamma\subset \H_\gamma$. By
definition, a real-analytic section $s$ of $K^{1/2}$ along
$\gamma$ can be extended to a holomorphic section of $K^{1/2}$ on
a small neighborhood of $\gamma$ in $\M$.  $\D_f$ then acts
naturally on $s$, and the restriction of $\D_f s$ to $\gamma$ is
again in $W_\gamma$.  So $\D_f$ acts naturally on the dense
subspace $W_\gamma$ of the Hilbert space $\H_\gamma$.  Under
suitable hypotheses on $\gamma$, $\D_f$ will be an elliptic
operator with a discrete spectrum.

Thus, the function $f$ has been mapped to an operator $\D_f$ that
acts on a Hilbert space $\H_\gamma$.  This is what we usually mean
by quantization, and it is what is meant by quantization in
\cite{NS2}. However, quantization in this physical sense is not a
property of the complex integrable system $\M_H$ alone. It depends
on a choice of the real curve $\gamma$. In fact, in \cite{NS2},
two different choices of $\gamma$ are considered. (We will derive
these two choices from gauge theory in section \ref{farend}.)

In an $n$-dimensional example, the analog of $\gamma$ is an
$n$-dimensional real subspace $N\subset \M$.  (A subspace $N$ is
called real if letting $TN$ be its tangent bundle and $I$ the
complex structure of $\M$, one has $TN\cap I(TN)=0$. This implies
that $\M$ is a sort of complexification of $N$.) Let
$L\subset\M_H$ be a completion of $T^*N\subset T^*\M$. Hitchin's
system restricted to $L$ is an ordinary real integrable system,
and the construction above can be regarded as  quantization in the
ordinary sense of this integrable system.

What we have described is a two-step process.  Hitchin's
integrable system can be quantized at a formal level by the
construction of certain holomorphic differential operators. To get
an actual Hilbert space requires a choice of a real cycle
$N\subset\M$.  We have described this for two reasons: to orient
the reader to the sense in which we aim to quantize $\M_H$, and
also to help motivate the two-step nature of the brane
construction to which we turn next.  In the brane construction,
the formal quantization is associated with a single brane, the
coisotropic $A$-brane that in the context of the
$\Omega$-deformation we will call $\B_\varepsilon$.  Construction
of an actual Hilbert space $\H$ depends on the choice of a second
brane, a Lagrangian $A$-brane $\B_L$.

\subsection{Coisotropic $A$-Branes}\label{coiso}

\def\I{{\mathcal I}}
The brane construction that we will need in this paper relies on
aspects of the two-dimensional topological $A$-model that are not
novel \cite{KO,Kap,AZ,KW,GW} but are perhaps also not well known.
We will here summarize the facts that will be used later in the
paper, without attempting full explanations.

Consider the $A$-model of a symplectic manifold $X$ with symplectic structure $\omega$.
An $A$-brane of the familiar sort is supported on a Lagrangian submanifold $L\subset X$ that is endowed with
a flat vector bundle.   Such an $L$  automatically has half the dimension of $X$.
However, in general the $A$-model admits
additional branes known as ``coisotropic'' branes.  (One reason that such branes are not
well known is that they do not arise for Calabi-Yau threefolds.)  The support of such an $A$-brane is
a submanifold $Y\subset X$ whose dimension exceeds half of the dimension of $X$.

The most basic new case is the case that $Y=X$.   This case will
suffice in the present paper. (The general case is a sort of
hybrid of this with the more familiar Lagrangian $A$-branes.)
Unlike Lagrangian $A$-branes, whose Chan-Paton bundle is flat, the
curvature $F$ of a coisotropic $A$-brane is necessarily nonzero.
Only the rank 1 case is understood; in this case $F$ is an
ordinary two-form. The condition for a rank 1 brane whose support
is precisely $X$ and whose Chan-Paton bundle has curvature $F$ to
be an $A$-brane is \cite{KO} that the linear transformation of the
tangent bundle defined by $\I=\omega^{-1}F$ should obey $\I^2=-1$.
It is then automatically true that $\I$ is an integrable complex
structure on $X$.  The condition $\I^2=-1$ implies that $F$ is not
only nonzero but is in fact  non-degenerate.  To generalize this
to include a $B$-field, we simply replace $F$ by $F+B$.

Reading this construction backwards, $X$ is a complex manifold and
$\Omega=F+i\omega$ is a  holomorphic $(2,0)$-form that is closed
and non-degenerate.  Thus, $X$ is a complex symplectic manifold.
However, neither the complex structure of $X$ nor
$F=\rm{Re}\,\Omega$ are part of the definition of the $A$-model of
$X$; rather they were used to define a  brane.  Only
$\omega=\rm{Im}\,\Omega$ is used in defining the $A$-model of $X$.
The same $X$ may have many different structures of complex
symplectic manifold (not related to each other by exact
symplectomorphisms with respect to $\omega$, which are trivial in
the $A$-model) each with a holomorphic two-form whose imaginary
part is $\omega$ and whose real part is the curvature of some line
bundle.  Each of these will lead to a different brane in the same
$A$-model of $X$.  Concrete examples can be constructed in the
hyper-Kahler case of section \ref{hyperkahler}.

\def\tN{{{\cmmib N}}}
\def\B{{\mathcal B}}
\def\RR{{\mathcal R}}
\def\R{{\mathbb R}}
\def\L{{\mathcal L}}
Though not yet part of the standard toolkit of physicists,
coisotropic $A$-branes have very interesting properties. If $\B$
is a Lagrangian $A$-brane, supported on a Lagrangian submanifold
$L$ and endowed with a flat bundle $E$, then the space of
$(\B,\B)$ strings is ordinarily finite-dimensional and not really
quantum mechanical in nature. Additively, it is the cohomology of
$L$ with values in the bundle $E\otimes E^*$ ($E^*$ is the dual
bundle to $E$).

By contrast, if $\B$ is a rank 1 coisotropic $A$-brane whose
support is $X$, then the space of $(\B,\B)$ strings of fermion
number zero is additively the space of complex-valued functions on
$X$ that are holomorphic in complex structure $\I$; allowing the
fermion number to vary, the space of  $(\B,\B)$ strings is the
$\bar\partial$ cohomology of $X$. This is quite an unusual answer;
we are accustomed to holomorphic functions and $\bar\partial$
cohomology in the $B$-model, but not in the $A$-model. Of course,
the size of the space of $(\B,\B)$ strings depends very much on
$X$.  At one extreme, if $X$ is compact, the only global
holomorphic functions are constants.  We will be interested in
cases in which $X$ admits many holomorphic functions.  The extreme
case is the case that $X$ is an affine variety, admitting in a
sense as many global holomorphic functions as there are on $\C^n$.
(In our applications, $X$ will or will not have this property
depending on the precise choice of coisotropic $A$-brane.)

Even more remarkable is the ring structure that arises from the
joining of $(\B,\B)$ strings (or equivalently, from the
multiplication of boundary vertex operators that represent such
strings).  This ring is a noncommutative deformation of the ring
of holomorphic functions on $X$ (or its extension to include the
higher $\bar\partial$ cohomology).  In the case that $X$ has lots
of holomorphic functions, the noncommutative ring $\RR$ of
$(\B,\B)$ strings is the ring  that can be obtained by deformation
quantization of the ring of holomorphic functions with respect to
the holomorphic symplectic form $\Omega$.  In other words, the
first order departure from commutativity is given by the Poisson
bracket $\{f,g\}=(\Omega^{-1})^{i j}
\partial_if\,\partial_jg$, with higher order corrections largely determined by associativity of the
operator product expansion.  See \cite{Kap} or in more detail
\cite{KW}, section 11; see also \cite{Pestun} for a related
analysis.

\def\O{{\mathcal O}}
To understand the noncommutative structure of the ring of
$(\B,\B)$ strings, recall that, in general, interactions of open
strings involve a noncommutative structure that is much more
complicated than deformation quantization of a finite-dimensional
manifold. It is exceptional to find a situation in which
noncommutativity survives but reduces to something as simple as
ordinary deformation quantization.  This occurs in somewhat
similar ways in the presence of a strong $B$-field \cite{CDS,SW}
or in the $A$-model with a coisotropic brane; either the strong
$B$-field or the topological symmetry of the $A$-model can
eliminate most of the open string modes, reducing to a
finite-dimensional but still noncommutative story. Let us briefly
describe how this comes about.

Suppose that on the target space $X$ of a sigma-model we have a
metric $g$ and  a $B$-field $B$. We consider a brane endowed with
Chan-Paton curvature $F$. We make no assumption in general that
$X$ is a complex manifold, and write $I,J=1,\dots,n$ for tangent
space indices to $X$.  In sigma-model, one encounters \cite{SW} an
effective inverse metric
\begin{equation}\label{overdu}G^{IJ}=\left(\frac{1}{g+2\pi\alpha' (F+B)}g\frac{1}{g-2\pi\alpha'(F+ B)}\right)^{IJ}\end{equation}
and a noncommutativity parameter
\begin{equation}\label{nerdu}\theta^{IJ}=-(2\pi\alpha')^2\left(\frac{1}{g+2\pi\alpha'(F+
B)}(F+B) \frac{1}{g-2\pi\alpha' (F+B)B}\right)^{IJ}.\end{equation}
In sigma-model perturbation theory, in computing the operator
product expansion of boundary operators $\O_u$ and $\O_v$
associated to functions $u$ and $v$ on $X$, one meets symmetric
contractions proportional to $G^{IJ}\partial_Iv
\partial_Jv$, and antisymmetric contractions proportional to
$\theta^{IJ}\partial_Iu\partial_Jv$.  For the sigma-model to
reduce to something as simple as deformation quantization, the
symmetric contraction must be negligible compared to the
antisymmetric one.  The most familiar way to achieve this result
is to take $F+B\to\infty$, noting that $G^{IJ}\sim 1/(F+B)^2$
while $\theta^{IJ}\sim 1/(F+B)$.  However, in the $A$-model, there
is another way to suppress the symmetric contractions.  In the
context of the coisotropic $A$-brane $\B$ just described, $u$ and
$v$ are required to be holomorphic functions in complex structure
$\I$. Moreover,  $g$ is of type $(1,1)$ while $F+B$ is of type
$(2,0)\oplus (0,2)$.  In this case, the symmetric contraction
vanishes, and the antisymmetric contraction is governed by
$\theta^{-1}=-(F+B)^{-1}_{2,0}$, that is, the $(2,0)$ part of
$-(F+B)^{-1}$. In our above presentation, we took $B=0$,
$F=\mathrm{Re}\,\Omega$, so the anticommutativity parameter is
$-\Omega^{-1}$.

\def\S{{\mathcal S}}

Now the question arises of whether we can use this framework to
see quantum mechanics, and not simply deformation quantization. To
do so, we need a Hilbert space on which the noncommutative ring
$\RR$ acts. For this, we consider a pair of $A$-branes -- a
coisotropic $A$-brane $\B$ of support $X$ and a rank 1 Lagrangian
$A$-brane $\B_L$ of support $L$.  We write $\L$ and $\S$ for the
Chan-Paton bundles of $\B$ and $\B_L$, respectively. $\L$ and $\S$
are both endowed with unitary connections.
 The curvature of $\L$ is a non-degenerate two-form $F$;  the curvature of $\S$ vanishes.
 Whatever $L$ is, the ring $\RR$ of $(\B,\B)$ strings will act on the space $\mathcal H$ of $(\B,\B_L)$
 strings, by the usual operation of joining strings.  The fact
 that we must introduce a second brane to define $\mathcal H$ has
 an obvious parallel with what we stated more naively in
 section\footnote{In our presentation, we are eliding a few key
 details that are described in \cite{GW}.  Though $\mathcal H$ can
 always be defined in the $A$-model, and has a Hilbert space structure
 because it is the space of ground states of the sigma-model, this Hilbert space structure is
 natural in the $A$-model only if $L$ is the fixed point set of an
 antiholomorphic involution of $X$.  And when this is the case,
 the Hilbert space structure on $\mathcal H$ that is natural in the
 $A$-model coincides with the naive one introduced in section
 \ref{realsections} only in the semiclassical limit, that is, to
 lowest order in sigma-model perturbation theory.}
 \ref{realsections}.
 Under certain conditions which we will now state, $\mathcal H$ can be interpreted in terms of
 quantization of $L$.

The fact that $L$ is Lagrangian means by definition that $\omega$
vanishes when restricted to $L$.  What about $F$? Let us consider
two contrasting cases. (There are also various intermediate cases,
but they will not be important for us.)  If $F$ vanishes when
restricted to $L$, then $L$ is actually a complex
submanifold\footnote{For a proof of this statement, see section
\ref{little} -- though a different notation is used there with
$I$, $J$, and $K$ cyclically permuted. As will become clear in
this paper, it is difficult to find a single and uniformly
convenient notation.} in complex structure $I$. This case does not
lead to what physicists usually understand as quantization, but
can lead to interesting and purely holomorphic constructions of
spaces on which $\RR$ acts, and it is important for geometric
Langlands \cite{KW}.

\def\H{{\mathcal H}}
The opposite case is the case that $F$ remains nondegenerate when
restricted to $L$. Thus, though  Lagrangian with respect to
$\omega$, $L$ is symplectic with respect to $F$. If unitarity is
desired, one also requires that $X$ should have an antiholomorphic
involution (a symmetry of order 2) with $L$ as a component of its
fixed point set. Under these conditions, as found in \cite{AZ} in
examples and discussed more systematically in \cite{GW}, the space
$\H$ of $(\B,\B_L)$ strings can be understood as a quantization of
$L$, with symplectic structure $F$ (and prequantum line bundle
$\L\otimes \S^{-1}$, whose curvature is $F$).  The basic reason
for this is easily explained.  As usual, the physical states of
the $A$-model are the string ground states, which can be found by
quantizing the zero-modes of the string. In the case of the
$(\B,\B_L)$ strings, one finds that there are no fermion
zero-modes.  The bosonic zero-modes describe the motion of the
string along $L$, and the relevant part of the action for these
modes is the Chan-Paton contribution. Writing $\cmmib A=\sum_i
p_i\, \d q^i$ for the connection on the prequantum line bundle
$\L\otimes \S^{-1}$ ($p_i$ and $q^ j$ are a local system of
canonically conjugate coordinates on $L$ viewed as a symplectic
manifold with symplectic form $\omega_J$), the relevant action is
$\int\d t \,p_i \,\d q^ i/\d t$; quantization of the zero-modes
with this action is usually called quantization of $L$.

 The functions on $L$ that can be most
naturally quantized as operators on $\H$ are
the functions that are restrictions to $L$ of holomorphic functions on $X$.
  Such functions
are quantized by identifying them with $(\B,\B)$ strings which then act naturally on $\H$.

If $X$ is an affine variety in complex structure $\I$,  functions
on $L$ that are restrictions of holomorphic functions on $X$ are
dense in the space of functions on $L$.  We consider two primary
applications in this paper.  In section \ref{confblocks}, where we
discuss the relation \cite{AGT} between four-dimensional gauge
theory and two-dimensional Liouville theory, $X$ is indeed  an
affine variety in complex structure $\I$.  The holomorphic
functions on $X$ are numerous and act irreducibly in the
quantization. In section \ref{compom}, where we discuss the
relation \cite{NS2} of the $\Omega$-deformation to quantization of
Hitchin's integrable system, $X$ is not affine in complex
structure $\I$ (but instead is a fibration by abelian varieties
over an affine base $\cmmib B$). The holomorphic functions are the
commuting Hamiltonians of Hitchin's integrable system.  Their
interpretation via $(\B,\B)$ strings amounts to their
interpretation as commuting holomorphic differential operators on
$\M$, as explained in detail in section 11 of \cite{KW}.  To get a
Hilbert space $\H$ on which these operators can act, we need to
pick a real section of $\M$, as explained heuristically in section
\ref{realsections}. More fundamentally, we need to pick a
Lagrangian brane $\B_L$ in the $A$-model of $\M_H$ in symplectic
structure $\omega_K$.

Our two applications will involve different sides of the same
coin. But to explain this, we must now specialize to the
hyper-Kahler situation.

\subsubsection{The Hyper-Kahler Case}\label{hyperkahler}

In our examples, $X$ will be the Coulomb branch of the moduli
space of vacua of a four-dimensional theory with $\N=2$
supersymmetry, after compactification on a circle (or a two-torus)
to three or two dimensions.  Thus $X$ will be hyper-Kahler.  An
important class of examples \cite{DG}, already considered for
illustration above, are the generalized quiver theories in which
$X$ is a moduli space $\M_H$ of Higgs bundles on a Riemann surface
$C$.

Such an $X$ has a distinguished complex structure $I$ (in which
the Hitchin fibration is holomorphic) and another distinguished
complex structure $J$ (in which it parametrizes complex-valued
flat connections on $C$).  We set $K=IJ$; $I$, $J$, and $K$ obey
the quaternion algebra. Together with a Riemannian metric $g$,
they define the hyper-Kahler structure of $X$.  Any linear
combination $\I=aI+bJ+cK$, where $a,b,c$ are real and
$a^2+b^2+c^2=1$, is an integrable complex structure. This family
of complex structures is parametrized by $\Bbb{CP}^1\cong S^2$.
The three real symplectic forms, which are Kahler in complex
structures $I,J$, or $K$ respectively, are $\omega_I=gI$,
$\omega_J=gJ$, $\omega_K=gK$. Similarly, the three holomorphic
symplectic forms, which are of type $(2,0)$ with respect to $I,J$,
or $K$, are $\Omega_I=\omega_J+i\omega_K$, $\Omega_J
=\omega_K+i\omega_I$, and $\Omega_K=\omega_I+i\omega_J$. For
simplicity, we will assume the $B$-field to vanish.

We will be studying the $A$-model of $X$ in the symplectic
structure $\omega=\omega_K$. To define a coisotropic brane in this
situation, we can take $F=\omega_I\sin p + \omega_J\cos p$ for
some angle $p$.  We must constrain $p$ and the hyper-Kahler metric
of $X$ so that $F/2\pi$ has integer periods and hence $F$ is the
curvature of some line bundle $\L\to X$. Given this, we get a
coisotropic $A$-brane with $\I=\omega^{-1}F=I\cos p -J\sin p $.

\def\Bcc{{\mathcal B}_{\rm{cc}}}
For our eventual application, the most important case will be that
$p=0$, so $F=\omega_J$ and $\I=I$.  (In examples arising by
compactification from $\N=2$ theories in four dimensions,
$\omega_J$ is cohomologically trivial, so there is no problem in
constructing a line bundle with curvature $F$.) The brane $\B$ so
obtained is a sufficiently basic example of a coisotropic
$A$-brane that it has been called the canonical coisotropic
$A$-brane $\Bcc$.  We have constructed it to be an $A$-brane in
the $A$-model of symplectic structure $\omega_K$. However, it has
additional supersymmetric properties and these will be important.
Suppose we take $\omega=\omega_I$; then $\omega^{-1}F=-K$, which
is again an integrable complex structure. So the same brane $\Bcc$
is also an $A$-brane for another $A$-model, the one with
$\omega=\omega_I$. Finally, as $\omega_J$ is of type $(1,1)$ in
complex structure $J$, we see that $\Bcc$ is a $B$-brane for the
$B$-model of complex structure $J$.

These three facts are related, in the following sense.  The
topological supercharges of the $A$-model of symplectic structure
$\omega_I$, the $B$-model of complex structure $J$, and the
$A$-model of symplectic structure $\omega_K$ obey one linear
relation.  They are linear combinations of two supercharges $Q$
and $Q'$ (this will be explained in detail in section
\ref{compom}). Any linear combination $uQ+vQ'$, with complex
coefficients $u,v$ that are not both zero, squares to zero and is
the topological supercharge of some topological field theory.
Varying the ratio $u/v$, this gives a family of topological field
theories, parametrized by $\Bbb{CP}^1$, which admit the same brane
$\Bcc$.  For brevity, we will describe this by saying that $\Bcc$
is a brane of type $(A,B,A)$.

As branes with multiple supersymmetric properties may be
unfamiliar, we will mention a much more obvious example that will
also be important in this paper.  This is the brane $\B_*$ whose
support is all of $X$ and whose Chan-Paton line bundle $\L_*$ is
trivial. The support of $\B_*$ (being all of $X$) and the line
bundle $\L_*$ (being flat) are both holomorphic in every complex
structure $\I=aI+bJ+cK$.  So the brane $\B_*$ is a $B$-brane in a
family of $B$-models parametrized by $\Bbb{CP}^1$.  We summarize
this by saying that
 $\B_*$ is a brane of type $(B,B,B)$.   Again, the various supercharges
are linear combinations of any two of them.

The differential geometry of branes with multiple supersymmetric
properties is further described in section \ref{little}.

\subsubsection{First Among Equals}\label{distinguished}

In this hyper-Kahler situation, we can define a plethora of
topological field theory structures -- the $B$-model in any
complex structure $aI+bJ+cK$, or the $A$-model in any symplectic
structure $a\omega_I+b\omega_J+c\omega_K$.  However, in the
important case
 \cite{DG} that $X$ is
actually a moduli space $\M_H$ of Higgs bundles on a Riemann
surface $C$, with some gauge group $G$, some of these structures
are more special than others.

\def\A{{\mathcal A}}
Almost all of these two-dimensional topological field theories
depend on the complex structure of $C$.  But some do not.  In one
of its complex structures, customarily called $J$, $\M_H$ is the
moduli space of flat connections on $C$ with values in the
complexification of $G$.  This is a complex symplectic manifold in
a completely natural way, independent of any choice of metric or
even complex structure  on $C$.  Thus, both the complex structure
$J$ and the holomorphic two-form $\Omega_J=\omega_K+i\omega_I$ do
not depend on any property of $C$ beyond its orientation. In fact,
$\Omega_J$ can be defined by the formula
\begin{equation}\label{zork}\Omega_J=\frac{i}{4\pi}\int_C\Tr\,\delta \A\wedge\delta\A,\end{equation}
where $\A=A+i\phi$ is a complex-valued flat connection; this
formula does not use a metric or complex structure on $C$, so it
makes clear the topological nature of $\Omega_J$.

Accordingly, the $B$-model of type $J$ and the $A$-models of types
$\omega_I$ and $\omega_K$ are special -- they do not depend on the
choice of a complex structure on $C$.  In that sense, branes of
type $(A,B,A)$ are special, compared to say branes of type
$(B,B,B)$, $(B,A,A)$, or $(A,A,B)$, all of which have a mixture of
properties that do or do not depend on a complex structure on $C$.

As explained earlier, a brane on $\M_H$  with multiple
supersymmetric properties is a brane in a whole family of
topological field theories, parametrized by $\Bbb{CP}^1$. In
general, this family depends on the complex structure of the
underlying Riemann surface $C$.  Precisely in the case of a brane
of type $(A,B,A)$, the relevant family of topological field
theories is independent of the complex structure of $C$.  Our
applications are based on this family, as is also the gauge theory
approach to geometric Langlands \cite{KW}.

In a four-dimensional $\N=2$ theory with $U(1)_R$ symmetry, the
forms $\omega_J$ and $\omega_K$ are rotated by the $U(1)_R$
symmetry, which ensures that their cohomology classes (which would
have to be rotation-invariant) must vanish.  This is actually true
even without $U(1)_R$ symmetry, as long as hypermultiplet bare
masses vanish.  For generalized quiver theories associated to a
Riemann surface $C$ without marked points, this can be shown in
terms of differential geometry as follows.
 Describing a Higgs bundle by a gauge
field $A$ with Higgs field $\phi$, the exactness of $\omega_J$ and
$\omega_K$ follows from an explicit formula
\begin{equation}\label{zobox}\omega_J+i\omega_K=\delta\left(\frac{1}{\pi}\int_C\Tr\phi_z\delta
A_{\bar z}\right).\end{equation} (In the presence of a
hypermultiplet bare mass, $\phi_z$ has a pole whose residue has an
eigenvalue proportional to the mass, and the above argument fails
because the one-form in parentheses is not gauge-invariant; that
is, it is not the pullback of a one-form on $\M_H$.) However, the
form $\omega_I$ is topologically non-trivial.  It is given by the
imaginary part of (\ref{zork}), or
\begin{equation}\label{obox}\omega_I=\frac{1}{4\pi}\int_C\Tr\,\left(\delta
A\wedge \delta A-\delta\phi\wedge\delta\phi\right).\end{equation}
One way to prove that $\omega_I$ has a non-zero cohomology class
is to observe that, when restricted to the locus $\phi=0$, it
becomes the Kahler form of the compact Kahler manifold
$\M\subset\M_H$ that parametrizes flat $G$-bundles on $C$.  The
Kahler form of a compact Kahler manifold always has a non-trivial
cohomology class.

\def\aalpha{{\cmmib u}\,}
\def\bbeta{{\cmmib v}\,}
Now let us discuss what two-dimensional $A$-models we can make
using the symplectic forms $\omega_I$ and $\omega_K$ that do not
depend on a choice of complex structure on $C$.  (We keep away
from $\omega_J$, since it does depend on the complex structure of
$C$.)  Superficially, we can introduce two complex parameters
$\aalpha,\bbeta$, since the complexified Kahler class $\hat\omega$
used in defining the $A$-model may be any complex linear
combination
\begin{equation}\label{oplo}\hat\omega=\aalpha
\omega_I+\bbeta\omega_K.\end{equation} We must take the real parts
of $\aalpha$ and $\bbeta$ to be not both zero, because to define
an $A$-model, ${\mathrm{Re}}\,\hat\omega$ must be a symplectic
form. However, in the absence of hypermultiplet bare masses,
because $\omega_K$ is exact, there is really only one complex
parameter that matters, namely $\aalpha$. If
$\mathrm{Re}\,\aalpha=0$, we must take
$\mathrm{Re}\,\bbeta\not=0$, but its magnitude and even sign are
not relevant.  (The sign of $\mathrm{Re}\,\bbeta$ can be reversed
by a rotation of the Higgs field.)  So really the only meaningful
parameter is $\aalpha$. Arbitrary values of $\aalpha\in\C$ make
sense, with $\bbeta$ turned on if necessary (or if desired). There
is actually also a limit as $\aalpha\to\infty$; this limit is the
$B$-model in complex structure $J$.  The fact that the limit of
$A$-models for $\hat\omega\to\infty$ exists and is a $B$-model is
most readily understood using generalized complex geometry and
will be reviewed in section \ref{rolegen}.

Both in interpreting geometric Langlands duality via gauge theory
\cite{KW} and in section \ref{compom} of the present paper, the
important special case of the $A$-model is the case $\aalpha=0$.
This is naturally called the $A$-model of type $\omega_K$; as we
have explained, in an important situation, this model has no
Kahler parameter. In an extended version of geometric Langlands
duality, analyzed in section 11 of \cite{KW} and encountered in
section \ref{confblocks} of the present paper, one requires the
generic model with variable $\aalpha$.  This model is naturally
called the $A$-model of type $\omega_I$, and of course, it does
always have a Kahler parameter, namely $\aalpha$.

We will make one last comment on branes of type $(A,B,A)$.
Although the conditions that characterize a brane of type
$(A,B,A)$ do not depend on a choice of complex structure on $C$, a
particular $(A,B,A)$-brane might be defined in a way that does
depend on that complex structure. Indeed, we have already
discussed a very important example. Given a complex structure on
$C$, $\M_H$ becomes hyper-Kahler, and in particular it acquires a
symplectic form $\omega_J$ that is Kahler with respect to $J$.
Unlike $\omega_I$ and $\omega_K$, $\omega_J$ does depend on the
metric of $C$ (as do $I=\omega_I^{-1}\omega_J$ and
$K=\omega_K^{-1}\omega_J$).  So the canonical coisotropic
$(A,B,A)$ brane does depend on the complex structure of $C$,
though it is a brane in a family of topological field theories
that do not depend on this metric.

\subsection{A Little Differential Geometry}\label{little}

On a hyper-Kahler manifold $X$, we have described some  branes
with multiple supersymmetric properties -- the brane $\B_*$ is of
type $(B,B,B)$, and its cousin $\Bcc$ is of type $(A,B,A)$.  These
are not the only examples. Another simple example of a brane of
type $(B,B,B)$ is a brane supported at a point in $X$ -- since a
point is a complex submanifold in any complex structure. The most
obvious branes of type $(A,B,A)$ are Lagrangian branes of this
type. Such a brane is supported on a middle-dimensional
submanifold $L\subset X$ that is holomorphic in complex structure
$J$, and is Lagrangian for the holomorphic symplectic form
$\Omega_J=\omega_K+i\omega_I$. As $L$ is Lagrangian for both
$\omega_K$ and $\omega_I$, a brane supported on $L$ with vanishing
Chan-Paton curvature is an $A$-brane of these types; as $L$ is
holomorphic in complex structure $J$, such a brane is of type
$(A,B,A)$.

The structures that we have described are redundant, in the
following sense.  A brane that is (for example) a $B$-brane of
type $I$ and a $B$-brane of type $J$ is automatically a $B$-brane
of type $K$ (and more generally, a $B$-brane in any complex
structure $aI+bJ+cK$, $a^2+b^2+c^2=1$), since the conserved
supercharges of these three $B$-models are linearly dependent.
This linear dependence will become very clear in section
\ref{compom}, but here we will briefly explain the redundancy
among the different supersymmetric structures from the point of
view of differential geometry.

In a hyper-Kahler manifold $X$, consider a brane $\B$ with support
$L\subset X$. One condition for $\B$ to be  a $B$-brane for
complex structures $I$ and $J$ is that $L$ must be holomorphic in
those complex structures.  If so, then $L$ is also holomorphic in
complex structure $K=IJ$.  Indeed, if the tangent space to $L$ is
invariant under the endomorphisms of the tangent bundle to $X$
corresponding to $I$ and $J$, it is certainly invariant under
$K=IJ$.  The other condition for $\B$ to be a $B$-brane for
complex structures $I$ and $J$ is that the Chan-Paton curvature
$F$ is of type $(1,1)$ with respect to both $I$ and $J$;
equivalently, $I^tFI=J^tFJ=F$.  Clearly this implies that
$K^tFK=F$, completing the argument that $\B$ is a $B$-brane of
type $K$ if it is one of types $I$ and $J$.

For an analogous argument for $A$-branes of type $(A,B,A)$, we
will consider just the case of a Lagrangian brane $\B$ supported
on a middle-dimensional submanifold $L$.  We will show that if
$\B$ is an $A$-brane for both $\omega_I$ and $\omega_K$, then it
is a $B$-brane in complex structure $J$.  (We leave it to the
reader to show that if $\B$ is an $A$-brane for $\omega_I$ and a
$B$-brane for $J$, then it is an $A$-brane for $\omega_K$.) Let
$TL$ be the tangent bundle to $L$, and let $N^*L$ be the subspace
of $T^*X|_L$ (the restriction to $L$ of the cotangent bundle of
$X$) consisting of cotangent vectors that annihilate $TL$. The
fact that $L$ is Lagrangian for both $\omega_I$ and $\omega_K$
means that $\omega_I$ establishes an isomorphism from $TL$ to
$N^*L$, and $\omega_K^{-1}$ is an isomorphism from $N^*L$ to $TL$.
So $J=\omega_K^{-1}\omega_I$ is an isomorphism from $TL$ to
itself, and thus $L$ is holomorphic in complex structure $J$ and
$\B$ is a $B$-brane.

In particular (though we have only shown this for Lagrangian
branes) there is no such thing as a brane of type $(A,A,A)$ -- if
$\B$ is an $A$-brane of type $I$ and $K$, then it is a $B$-brane
of type $J$. To illuminate the last statement further, and for
some further applications, we will give an overview of the
possible half-BPS supersymmetry conditions for a brane on a
hyper-Kahler manifold $X$.

\subsubsection{General Half-BPS Condition}\label{genhalf}

\def\J{{\mathcal J}}
On $X$, there is a family of complex structures parametrized by
$\Bbb{CP}^1$; a general element of this family is a complex
structure $\J=aI+bJ+cK$, with $a,b,c$ real and $a^2+b^2+c^2=1$.

Twisted topological field theories in two dimensions are
conveniently constructed by twisting a theory with $(2,2)$
supersymmetry.  In general, a sigma-model with target $X$ and
$(2,2)$ supersymmetry is constructed \cite{Rocek} in terms of a
pair of integrable complex structures $\J_+$ and $\J_-$, which
govern right- and left-moving excitations, respectively.  They
obey a certain compatibility condition which also involves the
metric and the curvature $H$ of the $B$-field.  Generalized
complex geometry \cite{Hitchin2} leads to the most elegant
interpretation of the compatibility condition \cite{Gualtieri}. We
use this viewpoint below.

If $X$ is a hyper-Kahler manifold, $\J_+$ and $\J_-$ can be chosen
to correspond to arbitrary points in $\Bbb{CP}^1$; the
compatibility condition is always obeyed, with $H=0$. Hence, the
sigma-model with hyper-Kahler target space has a twisted version
corresponding to an arbitrary pair $(\J_+,\J_-)\in
\Bbb{CP}^1_+\times \Bbb{CP}^1_-$, that is, in the product of two
copies of $\Bbb{CP}^1$.  A $B$-model corresponds to the case that
$\J_+=\J_-$, and an $A$-model corresponds to the case that
$\J_+=-\J_-$.

A supersymmetric boundary condition preserves the supersymmetries
associated to certain pairs $(\J_+,\J_-)$, but of course, not all
possible pairs.  In general, a half-BPS boundary condition
preserves the supersymmetries associated with pairs of the form
 $\J_-=h\J_+$, where $h\in
SO(3)$ is a rigid rotation of $\Bbb{CP}^1\cong S^2$ that gives a
holomorphic map from $\Bbb{CP}^1_+$ to $\Bbb{CP}^1_-$.  For
example, if $h=1$, we have $\J_+=\J_-$ for all $\J_+$. This is the
condition for a brane of type $(B,B,B)$. Since the antipodal map
on $\Bbb{CP}^1$ is not as $SO(3)$ rotation, it is not possible to
have $\J_-=-\J_+$ for all $\J_+$, and hence there is no such thing
as a brane of type $(A,A,A)$.

\def\acalpha{\varrho}
Actually, any $h\in SO(3)$ leaves fixed some axis in
$\Bbb{CP}^1\cong S^2$, so there is always some choice of $\J_+$
for which $h\J_+=\J_+$. Hence any half-BPS brane $\B$ is a
$B$-brane in some complex structure $\J=aI+bJ+cK$, with
$a^2+b^2+c^2=1$.  It is not true that there is always some $\J_+$
with $h\J_+=-\J_+$. Such a $\J_+$ exists if and only if $h$ is a
$\pi$ rotation around the appropriate axis. If so, then regarding
$h$ as a linear transformation of a copy of $\R^3$ in which
$S^2\cong\Bbb{CP}^1$ is embedded, $h$ has two eigenvalues $-1$,
and so a brane associated with such a $h$ is an $A$-brane in two
different ways. After a suitable rotation of the coordinate axes
(so that $h$ is a rotation around the $J$ axis), such a brane is
of type $(A,B,A)$.

\subsubsection{Role Of Generalized Complex
Geometry}\label{rolegen}

But what happens if $h$ is a rotation by an angle other than
$\pi$? In this case, although $\B$ is a $B$-brane in one complex
structure, its other supersymmetric properties appear unfamiliar.

To allow for the case of an arbitrary $h$, we consider the general
case of a brane $\B$ that conserves a topological supercharge $Q$
associated to a pair of independent complex structures $\J_+$ and
$\J_-$ for the right-moving and left-moving modes.  It turns out
\cite{Kap} that on a hyper-Kahler manifold, the topological field
theory associated to a pair $(\J_+,\J_-)$ with $\J_+\not=\J_-$ can
always be reduced to an $A$-model, even if $\J_+\not=\J_-$. The
reduction is made using the language of generalized complex
geometry \cite{Hitchin2,Gualtieri}.  (The requisite formulas are
summarized in section 5.2 of \cite{KW}, where they are applied to
geometric Langlands.)

\def\M{{\mathcal M}}
Rather than defining a topological twist by a pair of complex
structures with a metric and $B$-field obeying certain conditions,
a useful point of view is that such a twist can be determined by
the choice of a generalized complex structure $\I$. A generalized
complex structure is a linear transformation $\I$ of $TX\oplus
T^*X$ (the direct sum of the tangent and cotangent bundles of $X$)
that obeys $\I^2=-1$ as well as a certain integrability condition.
A $B$-model associated to a complex structure $J$ (whose transpose
we denote as $J^t$) corresponds to the case that
\begin{equation}\label{mlef}\I_J=\begin{pmatrix} J & 0 \cr 0 &
-J^t\end{pmatrix}.\end{equation}
The $A$-model with a symplectic structure $\omega$ and zero $B$-field
corresponds to the case that
\begin{equation}\label{nlef}\I_\omega=\begin{pmatrix}0 & -\omega^{-1}\cr \omega& 0\end{pmatrix}.\end{equation}
In general, to turn on a $B$-field, pick a closed two-form $B_0$ and set
\begin{equation}\label{plef}\M(B_0)=\begin{pmatrix} 1 & 0 \cr B_0 & 1\end{pmatrix}.\end{equation}
The transformation
\begin{equation}\label{qlef}\I\to \M(B_0)\I\M(B_0)^{-1}\end{equation}
is known as a $B$-field transform.  It
preserves the condition  $\I^2=-1$ and the integrability condition obeyed by $\I$, and has the effect
of shifting the $B$-field by $B_0$.  In particular, the generalization of $\I_\omega$ to include a $B$-field
is
\begin{equation}\label{mulef}\I_{\omega,B}=\M(B)\begin{pmatrix}0 & -\omega^{-1}\cr \omega& 0\end{pmatrix}
\M(B)^{-1}.\end{equation}

Now return to the case of a hyper-Kahler manifold $X$ with a pair
of complex structures $\J_\pm$ for right-movers and left-movers.
Denoting as $g$ the hyper-Kahler metric of $X$, let
$\omega_\pm=g\J_\pm$ be the Kahler forms associated to the complex
structures $\J_\pm$. The generalized complex structure associated
to this data is, according to eqn. 6.3 of \cite{Gualtieri},
\begin{equation}\label{nulef}\J=\frac{1}{2}\begin{pmatrix}\J_++\J_-
& -(\omega_+^{-1}-\omega_-^{-1})\cr
\omega_+-\omega_-&-(J_+^t+J_-^t)\end{pmatrix}.\end{equation}
(There is also a second generalized complex structure that we do
not need here; it is obtained  by reversing the sign of $\J_-$ and
$\omega_-$.) As long as $\J_+\not=\J_-$, this takes the form of
eqn. (\ref{mulef}) with
\begin{align}\label{golf}\omega^{-1}&=\frac{1}{2}\left(\omega_+^{-1}-\omega_-^{-1}\right)
\cr \omega^{-1}B&=\frac{1}{2}(\J_++\J_-).\end{align}
Hence, the model is equivalent to an $A$-model.

For our application, an important special case is that $\J_+$ is
very close to $\J_-$ -- so the model is almost a $B$-model.  If
$\J_+-\J_-$ is of order $\varepsilon$, where $\varepsilon$ is a
small parameter, then according to (\ref{golf}), $\omega$ and $B$
and therefore the complexified Kahler form $\hat\omega=\omega+iB$
are of order $\varepsilon^{-1}$.  This seems a little puzzling
because one expects the effects of rotating $\J_+$ slightly away
from $\J_-$ to be small.  However, as we have reviewed in section
\ref{coiso}, noncommutative effects in the $A$-model are of order
$\hat\omega^{-1}$, which in the present context means that these
effects are of order $\varepsilon$.  In the limit that $\J_+$
approaches $\J_-$, the noncommutative effects in the $A$-model
vanish and the $A$-model becomes an ordinary commutative
$B$-model.

Since $\hat\omega$ diverges as $\varepsilon\to 0$, one might not
expect the $A$-model with complexified symplectic form
$\hat\omega$ to have a limit as $\varepsilon\to 0$.  But in fact
this limit exists and is simply the $B$-model of type $\J_+$ or
equivalently $\J_-$.

\subsubsection{``Rotation'' Group}\label{rotgroup}

\def\G{{\Gamma}}
\def\CP{{\Bbb{CP}}}
Two-dimensional topological field theories of the class considered
here are labeled by the pair $(\J_+,\J_-)$, which parametrize what
we may call $\CP^1_+\times \CP^1_-$, with one copy of $\CP^1$ for
$\J_+$ and one for $\J_-$.  It is natural to introduce a group
$\G=SU(2)_+\times SU(2)_-$ that rotates $\CP^1_+\times \CP^1_-$,
with one factor of $SU(2)$ for each factor of $\CP^1$.  The group
that acts faithfully on $\CP^1_+\times \CP^1_-$ is actually
$SO(3)_+\times SO(3)_-$, where $SO(3)_\pm = SU(2)_\pm/\Z_2$.  $\G$
is a double cover of $SO(4)=(SU(2)_+\times SU(2)_-)/\Z_2$.

Consider a half-BPS brane characterized by a condition
$\J_-=h\J_+$, $h\in SU(2)$.  Obviously, if we transform
$(\J_+,\J_-)$ to $(g_+\J_+,g_-\J_-)$, then $h$ is transformed to
$h'=g_-^{-1}hg_+$.

For example, suppose that $h=1$, corresponding to a brane of type
$(B,B,B)$.  Then $h'=g_-^{-1}g_+$.  If (as will occur in our
application), $g_+$ is a rotation around some axis by an angle
$\vartheta$, and $g_-=g_+^{-1}$, then $h'$ is a rotation around
the given axis by an angle $2\vartheta$.

\section{Compactification And $\Omega$-Deformation}\label{compom}

Finally, we are prepared to consider our first application: the
relation of the $\Omega$-deformation to quantization.

The object of study in \cite{NS2} was a four-dimensional gauge
theory with $\N=2$ supersymmetry, ``compactified'' to two
dimensions on $\R^2_\varepsilon$.  Here $\R^2_\varepsilon$ is simply
$\R^2$, endowed with a $U(1)$ rotation symmetry that leaves the
origin fixed; the gauge theory on $\R^2_\varepsilon$ is deformed via
the $\Omega$-deformation \cite{Nek} with parameter $\varepsilon$.

The precise metric on $\R^2$ is not essential, as long as it is
$U(1)$-invariant.  We will find it helpful to place on $\R^2$ a
``cigar-like'' metric
\begin{equation}\label{cigar}\d s^2=\d r^2+
f(r)\,\d\theta^2,~0\leq r<\infty,~~0\leq\theta\leq
2\pi,\end{equation} with $f(r)\sim r^2$ for $r\to 0$ and $f(r)\to
\rho^2$ for $r\to\infty$. Thus $\rho$ is the asymptotic radius of
the circle parametrized by $\theta$.  We can assume that $f(r)$ is
identically equal to $\rho$ for sufficiently large $r$ (say $r\geq
r_0$).  We write $D$ for $\R^2$ endowed with this kind of metric.
We also write $D_R$ for $D$ restricted to $r\leq R$, where we
choose $R$ so that $R>>\rho,r_0$.

We will compactify to two dimensions on $D_R$,
 with an $\Omega$-deformation and a
suitable supersymmetric boundary condition at $r=R$.  However,
first we will need to understand what happens in the absence of
the $\Omega$-deformation.

To make contact with the explanation of integrability in section \ref{linko}, we take
the two-manifold to which we compactify on $D_R$ to be $\R\times S^1$.
So overall, we will be doing gauge theory on $M=\R\times S^1\times D_R$.
Since we take the cutoff $R$ very large, $D_R$ looks macroscopically like $I\times
 \tilde S^1$, where $\tilde S^1$ is a second circle, parametrized by $\theta$, and the interval
 $ I$ is parametrized by $r$, $0\leq
r\leq R$.  Macroscopically, $M$ is a two-torus fibration, that is
an $S^1\times \tilde S^1$ fibration, over $\R\times I$.  We should
be able to reduce to an effective description in a sigma-model on
$\R\times I$.

\def\BB{{\cmmib B}}
\def\M{{\mathcal M}}
The appropriate sigma-model is obtained by compactification of our
four-dimensional gauge theory to two dimensions on a two-torus
$T^2$.  For orientation, we consider the generalized quiver
theories that are obtained \cite{DG} by compactifying the
six-dimensional $(0,2)$ theory on a Riemann surface $C$, perhaps
with surface operators supported at marked points on $C$.  In this
case, our $\N=2$ gauge theory on $M=\R\times S^1\times D_R$ will
reduce at long distances to the sigma-model on $\Sigma=\R\times I$
with target $\M_H$, the moduli space of Higgs bundles on $C$.

To complete this description, we need to specify two branes,
supplying boundary conditions at the two ends of $I$. The brane at
$r=0$ will in some sense arise purely from geometry, as $r=0$ is
not really a boundary point in the more microscopic description on
$M$.  So one of our questions will be to identify the brane that
is generated by geometry. The second boundary condition in the
two-dimensional description, the one at $r=R$, will descend from a
choice of a boundary condition in the four-dimensional gauge
theory.

To account for the results of \cite{NS2}, we want quantization of
the sigma-model on $\Sigma=\R\times I$ to give quantization in the
ordinary sense of a middle-dimensional real subspace of $\M_H$.
From section \ref{coiso}, we know how this might happen: the
effective model on $\Sigma$ should be a two-dimensional $A$-model;
one brane should be a canonical coisotropic brane $\Bcc$, with
support all of $\M_H$, while the other should be an ordinary
Lagrangian brane $\B_L$, with support a Lagrangian submanifold
$L\subset\M_H$.  It will turn out that the brane that arises from
geometry will be $\Bcc$, while the Lagrangian brane $\B_L$ will
depend on a choice of boundary condition at $r=R$.

In section \ref{undeformed}, we study compactification on $D_R$ in
the absence of the $\Omega$-deformation.  We identify the brane
that arises at $r=0$ in the effective two-dimensional description.
The support of this brane is all of $\M_H$; however, it is not a
coisotropic brane, but the more elementary brane $\B_*$ of type
$(B,B,B)$ described at the end of section \ref{hyperkahler}. In
view of \cite{NS2}, the way to remedy this must be to incorporate
the $\Omega$-deformation. In section \ref{rethink}, we reformulate
the $\Omega$-deformation in a way suitable for our purposes.  In
section \ref{deformed}, we consider the $\Omega$-deformed theory
on $M=\R\times I\times D_R$ and explain why the
$\Omega$-deformation has the desired effects.  In section
\ref{farend}, we describe boundary conditions at the far end of
$D_R$.

\subsection{The Undeformed Case}\label{undeformed}

Two different $2+2$-dimensional splits of $M=\R\times S^1\times D_R$
will be important in this paper.  The first is the obvious decomposition of $M$
as the product of two two-manifolds $\R\times S^1$ and $D_R$.  The second
involves using the fact that $D_R$ is asymptotic to $I\times \tilde S^1$ and viewing $M$ as an $S^1\times \tilde S^1$ fibration over $\R\times I$.
Unfortunately, it is difficult to find a notation that is well-adapted to both decompositions.
What we will do is simply to number the coordinates as $0,1,2,3$ for $\R$, $S^1$,
$I$, and $\tilde S^1$, respectively.

The  bosonic part of the four-dimensional vector multiplet
comprises a gauge field $A_\mu$, $\mu=0,\dots,3$ and a complex
scalar $\phi$ in the adjoint representation. It is convenient to
adopt a six-dimensional notation in which $A_\mu$ and $\phi$
combine to a six-dimensional gauge field $A_I$, $I=0,\dots,5$
which is independent of the last two coordinates; from this point
of view, $\phi=(A_4-iA_5)/\sqrt 2$.  This is useful because,
although there is not really an $SO(6)$ symmetry rotating the six
components of $A_I$, many key equations can conveniently be
written in $SO(6)$ notation. For example, the supersymmetry
generator $\eta$ is a spinor of definite chirality, so if we
introduce gamma matrices $\Gamma_I$, $I=0,\dots,6$, obeying (in
Euclidean signature) $\{\Gamma_I,\Gamma_J\}=2\delta_{IJ}$, then
\begin{equation}\label{turmok}\Gamma_0\Gamma_1\cdots\Gamma_5\eta
= i\eta.\end{equation} The fermions $\Psi$ of the vector multiplet
are similarly a Weyl spinor in the adjoint representation of the
gauge group, with the same chirality as $\eta$.   Apart from being
chiral spinors of $SO(6)$, $\eta$ and $\Psi$ are also spinors of
the $SU(2)_R$ group of $R$-symmetries.  We denote as $\sigma_i$,
$i=1,2,3$ the analogs for $SU(2)_R$ of the gamma matrices, obeying
$\sigma_i\sigma_j=\delta_{ij}+i\epsilon_{ijk}\sigma_k$. We will
use standard abbreviations such as $\Gamma_{IJ}=\Gamma_I\Gamma_J$,
$I\not=J$, and
$\sigma_{ij}=\sigma_i\sigma_j=i\epsilon_{ijk}\sigma_k$, $i\not=j$.
An example of the usefulness of the six-dimensional notation is
that the supersymmetry transformations for the vector multiplet
are simply written:
\begin{align}\label{urkok}\delta A_I& = i\bar\eta\Gamma_I\Psi     \cr   \delta\Psi & = \frac{1}{2}\Gamma^{IJ}F_{IJ}\eta.\end{align}

Now consider an $\N=2$ gauge theory on the four-manifold
$M=\R\times S^1\times D_R$.  Away from the tip of the cigar (that
is, the region near $r=0$), $D_R$ is equivalent to $I\times
\tilde S^1$ and so $M$ reduces to the flat manifold $\R\times
S^1\times I\times \tilde S^1$.  On this flat manifold, there are
eight unbroken supersymmetries corresponding to all eight
components of $\eta$.

The curvature near the tip of the cigar inevitably breaks some of
the supersymmetries, in fact at least half of them.  (Any set of
at least five supersymmetries would include one whose square would
generate in the asymptotic region a translation along the first
factor of $D_R\sim I\times \tilde S^1$, but such a translation
cannot be extended to a symmetry of $D_R$.)  There is a standard
way  \cite{WD} to make a topological twist so that half of the asymptotic
supersymmetries are preserved in the exact $D_R$ geometry.  The
supersymmetries that are preserved are the ones that are invariant
under a rotation of the tangent space of $I\times \tilde S^1$
together with an $SU(2)_R$ rotation.  The rotation of the tangent
space of $I\times \tilde S^1$ is generated by $\Gamma_{23}$, and
up to conjugation in $SU(2)_R$, we can assume that the $SU(2)_R$
rotation in question is generated by $\sigma_{23}$.  So with the
standard topological twist, the four supersymmetries that are
preserved are the ones that can be characterized, in the
asymptotic region of $D_R$, by
\begin{equation}\label{zomely}\left(\Gamma_{23}+\sigma_{23}\right)\eta=0.\end{equation}

Let us look at this from the point of view of toroidal
compactification, on $S^1\times \tilde S^1$, to $\Sigma=\R\times
I$.  The tip of the cigar at $r=0$ gives a boundary condition at
one end of $I$. This boundary condition preserves half of the
supersymmetry.  In other words, in the effective two-dimensional
sigma-model, the tip, with the standard topological twist,
determines a half-BPS brane $\B_*$. We would like to interpret in
two-dimensional terms the unbroken supersymmetry and the brane
that carries this symmetry. A convenient way to do this is to
understand what topological properties this brane possesses. What
structures of twisted topological field theory on $\Sigma$ are
preserved by this brane?

Any topological field theory structure on $\Sigma$ is associated
with a supersymmetry generator $\eta$ that is invariant under
a rotation of the tangent space to $\Sigma$ together with an
$SU(2)_R$ transformation.  The rotation of the tangent space is
generated by $\Gamma_{02}$, and as this anticommutes with the
matrix $\Gamma_{23}$ that appears on the left of eqn.
(\ref{zomely}), we must pick an $SU(2)_R$ generator that
anticommutes with $\sigma_{23}$ or we will reach a contradiction.
With no essential loss of generality, we can look for a
supersymmetry generator that obeys
\begin{equation}\label{pomely}\left(\Gamma_{02}+\sigma_{31}\right)\eta=0.\end{equation}

The two equations (\ref{zomely}) and (\ref{pomely}) characterize a
two-dimensional space of $\eta$'s.  To determine a particular
topological field theory structure on $\Sigma$, we need one more
condition restricting to a one-dimensional space of $\eta$'s.
Any condition will do, so there is a $\Bbb{CP}^1$ family of
topological field theories on $\Sigma$ that are all compatible
with the same brane $\B_*$.

For one convenient choice, we supplement (\ref{zomely}) and
(\ref{pomely}) with the additional condition
\begin{equation}\label{tomely}\left(\Gamma_{01}+\sigma_{23}\right)\eta=0.\end{equation}
Although presented here in a non-invariant way, these three
conditions combine to something that can be described invariantly.
By commuting the operators appearing on the left hand sides of the
three equations, one learns that a spinor obeying the three
equations  actually obeys
\begin{align} \label{fomely}
\left(\Gamma_{ij}+\sigma_{ij}\right)\eta& = 0\cr
\left(\Gamma_{0i}+\sigma_{i+1,i-1}\right)\eta& = 0,
\end{align}
for $i,j=1,2,3$.  By subtracting (\ref{tomely}) from (\ref{zomely}), we find that
$\left(\Gamma_{01}-\Gamma_{23}\right)\eta=0$, which together with (\ref{turmok})
implies that
\begin{equation}\label{gelmok}\Gamma_{0123}\eta=-\eta,~~\Gamma_{45}\eta=-i\eta.\end{equation}
 The conditions (\ref{fomely}) are the standard conditions
that characterize the supersymmetry generator of a twisted
four-dimensional topological field theory \cite{WD} that (in the
case of $SU(2)$ gauge theory without hypermultiplets) is related
to Donaldson theory.  They can be characterized in group-theoretic terms.
 Let $SO(4)$ be the group of rotations of the
tangent space to $M$; denote its double cover as $SU(2)_l\times
SU(2)_r$. Then the above conditions mean that $\eta$ is invariant under
$SU(2)_l\times SU(2)'_r$, where $SU(2)'_r$ is a diagonal subgroup
of $SU(2)_r\times SU(2)_R$.

We can get two more useful choices of supersymmetry parameter by
observing that $\Gamma_1$, $\Gamma_4$, and $\Gamma_5$ all commute
with the operators on the left hand sides of our first two
conditions (\ref{zomely}) and (\ref{pomely}).  So these conditions
commute with a group $SO(3)_{145}$ that rotates those three gamma
matrices.  Making an $SO(3)_{145}$ transformation that rotates
$\Gamma_1$ to $\Gamma_4$ or $\Gamma_5$, we replace (\ref{tomely})
by
\begin{equation}\label{domely}\left(\Gamma_{04}+\sigma_{23}\right)\eta=0\end{equation}
or
\begin{equation}\label{bomely}\left(\Gamma_{05}+\sigma_{23}\right)\eta=0.\end{equation}

\subsubsection{Support Of The Brane}\label{otto}

Before trying to interpret the supersymmetries in two dimensional
terms, let us first determine the support of the brane $\B_*$.

In four dimensions, a single vector multiplet contains a complex
scalar field $\phi$ or equivalently a pair of real scalars.  When
we compactify to three dimensions on a circle $S^1$, two more
scalars come from the gauge field $A$ -- one is the holonomy of
$A$ around $S^1$, and the second is the dual photon. All four
scalars combine to a three-dimensional hypermultiplet. The key
point here is that to arrive at this hypermultiplet, a duality
transformation was needed, converting the photon to a scalar.

The geometry of the hypermultiplets that arise in compactification
can be described as follows.  Let $\cmmib B$ be the Coulomb branch
of vacua  of the four-dimensional gauge theory.  It parametrizes a
family of abelian varieties.  We denote the total space of this
family as $\M_H$  because in a large class of examples \cite{DG},
this total space is a moduli space of Higgs bundles on a Riemann
surface $C$.  After compactification on a circle and dualization
of the photons (one in each vector multiplet), $\M_H$ becomes
endowed with a hyper-Kahler metric, and one gets \cite{SW2} a low
energy description by a sigma-model of maps from three-dimensional
spacetime to $\M_H$.

Compactification to two dimensions on $S^1\times \tilde S^1$ is a
little different.  In this case, a four-dimensional gauge field
leads to two scalars -- its holonomies around the two circles --
without  dualization of any kind.  This actually gives a
description by linear multiplets rather than hypermultiplets
\cite{Rocek}; this description is inconvenient as there is not a
powerful theory of nonlinear models built from linear multiplets.
A $T$-duality transformation for the scalar fields that arise from
holonomies around one circle or the other is useful because it
leads to a description by hypermultiplets, and here hyper-Kahler
geometry is an effective tool for studying nonlinear models.  (For
more on this, see the end of this subsection as well as section
\ref{mixed}.)

From a two-dimensional point of view, in compactification on
$S^1\times \tilde S^1$, there is no natural choice of which of the
two sets of scalars should be $T$-dualized.  A description in
which we $T$-dualize one set of scalars differs from a description
in which we $T$-dualize the other set of scalars  by a combined
$T$-duality on both sets of scalars. The combined operation is a
$T$-duality on all the scalars that come from gauge fields, so it
can be described simply: it is the $T$-duality on the fibers of
the fibration $\M_H\to \cmmib B$. This particular instance of
$T$-duality is related to $S$-duality in another description of
the same models \cite{HM,BJV}, and is the basic geometric
Langlands duality \cite{KW}.  This $T$-duality transforms $\M_H$
to an analogous moduli space of Higgs bundles for the Langlands
dual gauge group.

Now let us specialize to our problem with $M=\R\times S^1\times
D_R\sim\R\times S^1\times I\times \tilde S^1$.  The symmetry
between $S^1$ and $\tilde S^1$ is broken by the fact that $\tilde
S^1$, and not $S^1$, is capped off at the tip of the cigar. It
turns out that to explain the results of \cite{NS2}, it is better
to $T$-dualize the holonomies of the gauge field around $\tilde
S^1$.  (In section \ref{confblocks}, we will explore another
problem in which the two circles enter symmetrically.)

Recalling that we have labeled the four dimensions of $M\sim
\R\times S^1\times I\times \tilde S^1$ consecutively as 0123, we
write simply $A_1$ or $A_3$ for scalars arising from the holonomy
of a gauge field $A$ around $S^1$ or $\tilde S^1$, respectively.
 The boundary conditions on $\phi$, $A_1$, and $A_3$ at the
tip of the cigar are uniquely determined, since in
four-dimensional terms there is no boundary at all. There is no
reason for $\phi$ or $A_1$ to vanish at the tip of the cigar, so
in the two-dimensional description on $\Sigma=\R\times I$, they
obey Neumann boundary conditions.  On the other hand, $A_3$ must
vanish because it is the holonomy of the gauge field around a
circle that shrinks to a point at the tip. So $A_3$ obeys
Dirichlet boundary conditions.

However, to get a description by a two-dimensional sigma-model
with target $\M_H$, we are supposed to $T$-dualize $A_3$,
replacing it by another scalar that we will call $\acalpha$.  For
future reference,\footnote{\label{firstone} Here we assume the
circles $S^1$ and $\tilde S^1$ are orthogonal; otherwise $A_1$
enters the formulas. See the appendix for a much more complete
treatment.} we write the equations describing this $T$-duality:
\begin{align}\label{confo}\partial_0 \acalpha &=- i\partial_2 A_3\cr
                          \partial_2\acalpha & =  i\partial_0 A_3.\end{align}
After $T$-duality, $\acalpha$ obeys Neumann boundary conditions at
the tip of the cigar. In fact, at this stage all scalars
$\phi,A_1,$ and $\acalpha$ obey Neumann boundary conditions at the
tip.  So the tip of the cigar corresponds in two-dimensional terms
to a brane $\B_*$ whose support is all of $\M_H$.

The topological twist that was used to preserve supersymmetry on
$\R\times S^1\times D_R$ does not generate at the tip of $D_R$ any
couplings that look obviously like Chan-Paton couplings.  So it is
natural to think that the Chan-Paton bundle of $\B_*$ may be
trivial.  If so, as the support of $\B_*$ (being all of $\M_H$) is
holomorphic in every complex structure on $\M_H$, $\B_*$ will be a
brane of type $(B,B,B)$ -- a $B$-brane for every complex structure
that makes up the hyper-Kahler structure of $\M_H$.  We will show
in section \ref{twod} that this is the case.

We will add a word on the more naive description by scalars $\phi,
A_1$, and $A_3$ without any $T$-duality.  In this description,
precisely one scalar in each hypermultiplet (namely $A_3$) obeys
Dirichlet boundary conditions.  Since the usual supersymmetric
branes have Dirichlet boundary conditions for an even number of
scalars in each multiplet, this is another indication that a
simple description requires $T$-duality for either $A_1$ or $A_3$.
(Without such a $T$-duality, a four-dimensional vector multiplet
reduces in two dimensions to a linear multiplet rather than a
hypermultiplet.)

\subsubsection{Two-Dimensional Interpretation Of The
Supersymmetries}\label{twod}

Now we want to determine the two-dimensional interpretation of the
supercharges whose generator $\eta$ obeys  (\ref{zomely}) and
(\ref{pomely}) plus one of the three supplementary conditions (\ref{tomely}),
(\ref{domely}), or (\ref{bomely}).

We first consider the case of a spinor $\eta$ obeying
(\ref{tomely}). We claim that $\eta$ generates the topological
symmetry of the $B$-model in complex structure $I$ -- the complex
structure in which the Hitchin fibration is holomorphic.

The basic functions on $\M_H$ that are holomorphic in complex
structure $I$ are $\phi$ and $A_1+i\acalpha$.  So, if $Q$ is the
supersymmetry generated by $\eta$, we must show that
$[Q,\phi]=[Q,A_1+i\acalpha]=0$.

It is straightforward to show that $[Q,\phi]=0$; this is actually
a standard fact in the context of applications to Donaldson theory
\cite{WD}.  As $\phi=(A_4-iA_5)/\sqrt 2$ and $\delta
A_I=i\bar\eta\Gamma_I \Psi$, what we need to show is that
$\bar\eta(\Gamma_4-i\Gamma_5)\Psi=0$, which will follow if
$(\Gamma_4-i\Gamma_5)\eta=0$.   This is equivalent to
$\Gamma_{45}\eta =-i\eta$,  which was deduced in (\ref{gelmok}).

The other condition $[Q,A_1+i\acalpha]=0$ is more subtle, because
$\acalpha$ is defined via a $T$-duality that only makes sense
after reduction to two dimensions. So in analyzing this condition,
we work in the effective two-dimensional theory. Thus, we discard
terms involving derivatives in the 1 or 3 directions, and for
example that means that $F_{01}$ reduces to $\partial_0 A_1$. The
zero mode of a scalar field such as $\acalpha$ that is defined via
$T$-duality is subtle to understand.  However, there are
straightforward formulas (\ref{confo}) for the derivatives of
$\acalpha$.  So we will content ourselves with showing the
vanishing of the derivatives along $\R\times I$ of
$[Q,A_1+i\acalpha]$ in the effective two-dimensional theory. For
example, the derivative in the 0 direction is
$[Q,\partial_0A_1+i\partial_0\acalpha]=[Q,\partial_0
A_1+\partial_2A_3] =[Q,F_{01}+F_{23}]$.  This vanishes; indeed,
the combination $F_{01}+F_{23}$ is self-dual and therefore is
$Q$-exact and in particular $Q$-closed in the topological field
theory related to Donaldson theory.  The derivative of
$[Q,A_1+i\acalpha]$ in the 2 direction vanishes similarly.

What we have learned then is that if we select a spinor $\eta$
using the supplementary condition (\ref{tomely}), we get the
topological supercharge of the $B$-model of $\M_H$ in complex
structure $I$.  This is the complex structure in which the Hitchin
fibration $\M_H\to \cmmib B$ is holomorphic, and the scalar fields
 that are functions on $\cmmib B$ are likewise holomorphic.

The other conditions that we want to analyze, namely
(\ref{domely}) and (\ref{bomely}), can be formally obtained from
(\ref{tomely}),  which we have just analyzed, by an $SO(3)_{145}$
transformation that exchanges $A_1$ with $A_4$ or $A_5$.  This
fact can be used to interpret the supersymmetries in the low
energy theory without any computation.  From each vector
multiplet, we get four scalars in the effective two-dimensional
sigma-model. Schematically we call them $A_1$, $A_4$, $A_5$, and
$\acalpha$.  Each set of four scalars forms a hypermultiplet whose
tangent space admits an action of the quaternion units $I,J,$ and
$K$.  A formal $SO(3)_{145}$ rotation that exchanges $A_1$ with
$A_4$ or $A_5$, so as to map (\ref{tomely}) to (\ref{domely}) or
(\ref{bomely}), maps $I$ to $J$ or $K$.  So while the auxiliary
condition (\ref{tomely}) determines $\eta$ to be the generator of
the $B$-model in complex structure $I$, (\ref{domely}) or
(\ref{bomely}) similarly determines $\eta$ to be the generator of
the $B$-model in complex structure $J$ or $K$, respectively.

At this level of generality, it is in part a convention which of the complex structures on $\M_H$
is called $J$ rather than $K$.   In a large class of models \cite{DG} in which $\M_H$ actually is a moduli
space of Higgs bundles on a Riemann surface $C$, we can fix the definition of $A_4$ and $A_5$
and the conventions in the Higgs bundle equations so that
$J$ is the complex structure in which $\M_H$ parametrizes flat bundles on $C$ with complex
structure group.  $K=IJ$ is then distinct from $J$ but equivalent to it by a $U(1)_R$ rotation, provided hypermultiplet
bare masses (which violate $U(1)_R$) are absent.
The notation just described is in accord with that
of \cite{Hitchin}.

\subsubsection{A Mixed $AB$-Model}\label{mixed}

Now we can be more precise about what would happen if we describe
the two-dimensional effective field theory with the naive set of
scalars $\phi$, $A_1$, and $A_3$, without any $T$-duality.

For example, let us consider the supercharge $Q$ of Donaldson
theory, the one that we associated with the $B$-model of complex
structure $I$.  Making or not making a $T$-duality on $A_1$ or
$A_3$ does not affect the fact that $\phi$ obeys the $B$-model
condition $[Q,\phi]=0$.  But in the absence of any $T$-duality,
the conditions obeyed by $A_1$ and $A_3$ are $A$-model conditions,
not $B$-model conditions.

Indeed,
in our derivation, we used the fact that $\partial_0A_1+\partial_2A_3$ is $Q$-exact in the
effective two-dimensional theory;
similarly, the same is true of $\partial_0 A_3-\partial_2 A_1$.  We can combine these
statements into the assertion that $(\partial_0+i\partial_2)(A_3+iA_1)$ is $Q$-exact.
In the complex structure on $\R\times I$ in which $z=x^0+ix^2$ is holomorphic, this says
that $A_3+iA_1$ is holomorphic modulo $\{Q,\cdot\}$.

Thus the model under discussion, in terms of the obvious variables
without any $T$-duality, is from the point of view of the
supercharge $Q$ a mixed $AB$ model, with $B$-model conditions on
$\phi$ and $A$-model conditions on $A_3+iA_1$.  A similar story
holds if $Q$ is replaced by one of the other supercharges
considered above.

\subsubsection{More On Compactification}\label{morec}
\def\Cm{{\eusm C}}

We will conclude this discussion with a few more observations
about compactification of $\N=2$ gauge theories from four to two
dimensions.  The goal is to review some points made in
\cite{HM,BJV} and derive some formulas that will be used later.
(See the appendix for a much more complete treatment.)

We will here consider only compactification of a $U(1)$ vector
multiplet on a rectangular torus $T^2$.  We write $\Cm,\Cm'$ for
the circumferences of the two circles. Also, for simplicity, we
take the four-dimensional $\theta$-angle to vanish.

The scalar $\phi$ in the vector multiplet reduces to a scalar in two dimensions that we denote
as $a$.    Its kinetic energy is
\begin{equation}\label{boken}  \frac{ \Cm \Cm'}{e^2}\int \d^2x\,|\nabla a|^2.\end{equation}
The components of the gauge field along $T^2$ have zero modes that reduce in two dimensions
to angle-valued field $b,b'$.  Their kinetic energy is
\begin{equation}\label{zoken}\frac{1}{e^2}\int \d^2x\,\left(\frac{ \Cm'}{ \Cm}|\nabla b|^2+
\frac{ \Cm}{ \Cm'}|\nabla b'|^2\right).\end{equation} However, we
want a description obtained by $T$-duality on one of these
scalars. Making a $T$-duality that replaces $b'$ by another
angle-valued field $\acalpha$, (\ref{zoken})  is replaced by
\begin{equation}\label{token}\frac{ \Cm'}{ \Cm}\int \d^2x\,\left(\frac{1}{e^2}|\nabla b|^2+
\frac{e^2}{16\pi^2}|\nabla \acalpha|^2\right).\end{equation}

In this simple model, the angles $b,\acalpha$ parametrize the
fibers of the Hitchin fibration. Clearly, the area of a fiber
$\cmmib F$ of the Hitchin fibration is independent of $e^2$ and is
proportional to $ \Cm'/ \Cm$.  So $ \Cm'/ \Cm$ is a Kahler
parameter, as in \cite{HM,BJV}.  If we rescale $a$ to $\tilde a=a
\Cm/e$, then the metric on $\M_H$ become $ \Cm'/ \Cm$ times a
metric that depends only on $e$ and not on $ \Cm$ or $ \Cm'$:
\begin{equation}\label{zongo}\d s^2=\frac{ \Cm'}{ \Cm}\left(|\d \tilde a|^2+\frac{1}{e^2}\d b^2
+\frac{e^2}{16\pi^2}\d \acalpha^2\right).\end{equation}

Conversely, the complex structure of $\cmmib F$ is determined by
$e^2$ and is independent of $ \Cm$ and $ \Cm'$.  In fact, the
$\tau$ parameter of $\cmmib F$ is that of the underlying gauge
theory. If one thinks of the $U(1)$ vector multiplet as arising
from compactification of the abelian version of the
six-dimensional $(0,2)$ theory on a two-torus $C$, then $\M_H$ is
the moduli space of $U(1)$ Higgs bundles on $C$.  This picture
extends to more interesting examples  with $U(1)$ replaced by a
nonabelian group and $C$ by a more general Riemann surface.  The
effective two-dimensional description in an example of this type
is obtained by compactifying the six-dimensional $(0,2)$ theory to
two dimensions on $T^2\times C$.  If one compactifies first on
$C$, one gets a generalized quiver theory \cite{DG} in four
dimensions in which the complex structure of $C$ is encoded in the
gauge coupling parameters (generalizing what we called $e^2$ in
the abelian theory). If one compactifies first on $T^2$, one gets
$\N=4$ super Yang-Mills theory with coupling parameter
$\tau=i\Cm_1/\Cm_2$. Further compactification on $C$ gives the
situation studied in \cite{KW}: supersymmetric vacua correspond to
Higgs bundles on $C$, so the moduli space of vacua is $\M_H$, and
the metric on $\M_H$ is scaled by ${\mathrm
{Im}}\,\tau=\Cm_1/\Cm_2$, generalizing the abelian result of
(\ref{zongo}).

Since the metric has a factor of $\Cm'/\Cm$, the Kahler forms all
contain this factor as well. Writing
$(\omega_I^*,\omega_J^*,\omega_K^*)$ for the Kahler forms computed
relative to the metric (\ref{zongo}), these Kahler forms are
$\Cm'/\Cm$ times the standard ones defined in eqns. (\ref{zobox}),
(\ref{obox}):
\begin{equation}\label{retso}(\omega_I^*,\omega_J^*,\omega_K^*)
=\frac{\Cm'}{\Cm}(\omega_I,\omega_J,\omega_K).\end{equation}
Of course, this factor will also appear in the holomorphic two-forms.  For example, the holomorphic
two-form in complex structure $I$ is
\begin{equation}\label{metso}\Omega_I^*=
\omega_J^*+i\omega_K^*=\frac{\Cm'}{\Cm}\left(\omega_J+i\omega_K\right).\end{equation}
In fact, $\Omega_I^*=(e\,\Cm'/2\pi\Cm )\d \tilde
a\wedge(\d\acalpha-(4\pi i/e^2)\d b)=(\Cm'/2\pi)\d a\wedge(\d
\acalpha-(4\pi i/e^2)\d b)$.  It is convenient to introduce
$a_D=\tau a$, with\footnote{Here $\tau$ is imaginary since we took
the four-dimensional $\theta$-angle to vanish.  For a derivation
including the $\theta$-angle, see eqns. (2.34) and (3.16) of
\cite{GMN}.} $\tau=4\pi i/e^2$, in terms of which
\begin{equation}\label{poln}\Omega_I^*=\frac{\Cm'}{2\pi}
\bigl(\d a\wedge\d\acalpha -\d a_D\wedge \d b\bigr).\end{equation}
The expression in parentheses is a standard formula for the
holomorphic two-form of the total space of the Seiberg-Witten
fibration over the Coulomb branch.  (This holomorphic two-form is
unaffected by compactification.)  The main purpose of computing
$\Omega_I^*$ here has been  to explain the factor of $\Cm'$.

The formula (\ref{poln}) has an analog for the general case of a
semi-simple gauge group of rank $r$.  Pick a local description of
the Coulomb branch in terms of $r$ vector multiplets with scalar
components $a_i$ and dual scalar fields $a_D^i=\partial{\mathcal
F}/\partial a_i$ (here $\mathcal F$ is the prepotential of the
gauge theory). The gauge field of the $i^{th}$ vector field gives
two angle-valued fields $\acalpha^i$ and $b_i$ (as in the $U(1)$
case, $b_i$ is a holonomy and $\acalpha_i$ is the dual of a
holonomy).  The analog of (\ref{poln}) is then
\begin{equation}\label{zpoln}\Omega_I^*=\frac{\Cm'}{2\pi}\sum_i
\bigl(\d a_i\wedge\d\acalpha^- -\d a_D^i\wedge \d
b_i\bigr).\end{equation} Because it requires the choice of a
duality frame, this formula is valid only locally on the Coulomb
branch.  Given the choice of a duality frame, the formula can be
derived by the same steps as in the $U(1)$ case, starting with the
low energy effective action on the Coulomb branch.

\subsection{Rethinking The $\Omega$-Deformation}\label{rethink}

By now, we have learned that the brane associated with the tip of the cigar manifold $D_R$
is a brane of type $(B,B,B)$ whose support is all of $\M_H$.  On the other hand, we know
from section \ref{coiso} that this is not what we need in order to generate the quantization of
a real slice in $\M_H$.  What we need is a brane of support $\M_H$ and of type $(A,B,A)$.

It is also clear from \cite{NS2} what we need to do in order to get the desired result: we need
to implement an $\Omega$-deformation -- that is, we need to replace $D_R$ by its
$\Omega$-deformed version
$D_{R,\varepsilon}$, where $\varepsilon$ is the parameter of
the $\Omega$-deformation.

\subsubsection{The Standard Formulation}\label{standard}

First, let us recall a standard way of introducing the
$\Omega$-deformation. One uses the fact that the scalar fields
$A_4$ and $A_5$ can be viewed as components of the gauge field
$A_I$ in two extra dimensions.  For simplicity, let us focus on
$A_4$. We think of $x^4$ as an angular variable. Instead of taking
spacetime to be a product $\R^4\times S^1$, with the second factor
parametrized by $x^4$, we take it to be an $\R^4$ bundle over
$S^1$, with the monodromy around $S^1$ being an element $g\in
SO(4)$.  The monodromy action on $\R^4$ can be accompanied by an
$R$-symmetry transformation $r\in SU(2)_R$ acting on the fermions,
and then the unbroken supersymmetries are those that are invariant
under the product $gr$. For our present purposes, we take $g$ to
rotate a two-plane in $\R^4$, so $g$ actually lies in an $SO(2)$
subgroup of $SO(4)$.  The element $r$ is chosen in the usual way
so that the product $gr$ preserves one-half of the supersymmetry.

Now instead of taking the metric on $\R^4\times S^1$ to be a
simple product, we consider a fiber bundle metric in which the
monodromy around $S^1$ is the element $gr\in SO(4)\times SU(2)_R$
(in other words, the geometric monodromy is $g$, and we also make
an $R$-symmetry twist by $r$).  In formulas, the metric is
\begin{equation}\label{frog}\d s^2=\sum_{\mu=0}^3(\d x^\mu-\varepsilon V^\mu\d x^4)^2 +(\d x^4)^2,\end{equation}
The $\Omega$-deformed four-dimensional theory is defined by starting with five-dimensional
super Yang-Mills theory in this spacetime, and then taking the fields to be independent of $x^4$.

\def\W{{\mathcal W}}
 The $\Omega$-deformation has its name because it actually is a
deformation of the action. A basis of orthornormal vector fields
for the metric (\ref{frog}) is given by $u_\mu=\partial/\partial
x^\mu$, $\mu=0,\dots,3$, $u_4=\partial/\partial x^4 +\varepsilon
V^\mu\partial/\partial x^\mu$.  The only one that is unusual is
$u_4$, and the extra term in $u_4$ means that the
$\varepsilon$-dependence of the Lagrangian can be computed roughly
by a substitution $A_4\to A_4+\varepsilon V^\mu D_\mu$. More
precisely, the deformation of the bosonic part of the action can
be computed by the following substitutions
\begin{align}\label{helpo}[A_4,A_5] & \to [A_4,A_5]+\varepsilon V^\mu D_\mu A_5\cr
      D_\nu A_4& \to D_\nu A_4+\varepsilon V^\mu F_{\nu\mu}.\end{align}
The Yukawa couplings containing $A_4$ are modified in a similar way.

The $\Omega$-deformation preserves whatever supersymmetry commutes
with $gr$.   However, the supersymmetry algebra is modified.  The
reason for this is that usually the supersymmetry algebra of
$\N=2$ super Yang-Mills theory closes modulo gauge transformations
generated by $A_4$ and $A_5$.  But in the present situation, $A_4$
is effectively replaced by $A_4+\varepsilon V^\mu D_\mu$, and so
wherever a gauge transformation generated by $A_4$ would usually
appear, there is now an additional term that is $\varepsilon$
times the conserved charge $\W_V$ associated to the Killing vector
field $ V^\mu$. The most important special case \cite{Nek}
concerns the supercharge $Q$ that is associated to Donaldson
theory and the counting of instantons.  In the undeformed theory,
its square is a gauge transformation; after $\Omega$-deformation,
it obeys $Q^2=\varepsilon \W_V$ modulo a gauge transformation.

\subsubsection{An Alternative Description}\label{alternative}

In constructing the $\Omega$-deformation,
we can replace $\R^4\times S^1$ by $M\times S^1$, where $M$ is any  Riemannian
four-manifold and $g$ is an isometry of $M$.  Our application will be to the case that
$M=\R\times S^1\times D_R$, with $g$ a rotation of the cigar metric $D_R$ that
leaves fixed the tip of the cigar.

In its asymptotic region, $D_R$ is simply a product $I\times \tilde S^1$, with the circumference
of $\tilde S^1$ being $2\pi \rho$.   Rotations of $D_R$
act in the asymptotic region by rotations of  $\tilde S^1$.  Such rotations of the flat metric
$\R\times S^1\times I\times \tilde S^1$ preserve all supersymmetry, and are not accompanied
by an $R$-symmetry transformation.\footnote{This may be understood as follows.
To preserve half of the supersymmetry on $\R\times S^1\times D_R$, in the absence of
the $\Omega$-deformation, one twists the fermions, changing their spins.  This
modifies how they transform under rigid rotations of $D_R$ in the curved region near
the tip of the cigar, but not in the asymptotic region.}

In principle, we could now proceed to study the claim of \cite{NS2} in the context of the
$\Omega$-deformed theory.  The only problem is that the $\Omega$-deformed theory
is a different theory from $\N=2$ super Yang-Mills theory, and we have much less experience
with its dynamics.

Here we will avoid having to understand the dynamics of $\Omega$-deformed
theories, via the following device.  We will show that if $M$ is of the form
$W\times\tilde S^1$, with a product metric,  then the $\Omega$-deformation associated
to a rotation of $\tilde S^1$ can be eliminated by a change
of variables.  Of course, our $M$ is of the stated form.

We proceed as follows.  In the product situation, the $\Omega$-deformation can be described
formally as a substitution
\begin{equation}\label{turkey} A_4\to \tilde A_4=A_4+\varepsilon\rho\frac{D}{Dx^3}.
\end{equation}
(The reason for the factor of $\rho$ is that, as $x^3$ has period
$2\pi\rho$, the normalized generator of the rotation of $\tilde
S^1$ is the vector field $\rho\partial/\partial x^3$.) This is a
substitution only in a very formal sense; because  the right hand
side of (\ref{turkey}) is a differential operator of degree 1
rather than a field, this operation does not make sense as a
change of variables in any standard sense. However, the
substitution (\ref{turkey}) does make sense as a formal device to
generate a deformation of the action of $\N=2$ super Yang-Mills
theory. The reason for this is that $A_4$ enters the action only
via commutators such as $[A_4,A_5]$ and covariant derivatives
$D_\nu A_4$.  Under the substitution (\ref{turkey}), these
transform in a sensible way as indicated in eqn. (\ref{helpo}).

Now, consider the transformation
\begin{equation}\label{durkey}\frac{D}{Dx ^3}\to \frac{D}{Dx ^3}-\varepsilon \rho A_4.\end{equation}
Though superficially similar, this does make sense as an ordinary change of variables
in the quantum field theory, since it is equivalent to
\begin{equation}\label{yurkey}A_3\to \tilde A_3=A_3-\varepsilon\rho  A_4, \end{equation}
and the right hand is a field rather than a differential operator of positive degree.

Since the distinction between the formal operation (\ref{turkey})
and an ordinary change of variables such as (\ref{durkey}) will be
important, we will belabor the point slightly.  If the ordinary
change of variables (\ref{durkey}) is applied to a given quantum
field theory, we simply get an equivalent theory written in terms
of a different set of variables. By contrast, the operation
(\ref{turkey}) is not a change of variables in any ordinary sense,
and when it is applied to a theory to which it can be applied, it
gives a theory that in general is inequivalent.  In general, the
$\Omega$-deformation really is a non-trivial deformation. However,
we will show here  that in the special case of a product $W\times
\tilde S^1$ and a deformation involving a rotation of the second
factor, the $\Omega$-deformation is trivial, in the sense that it
can be removed by an ordinary change of variables.  The
identification of the deformed theory with the undeformed one
will, however, involve a non-trivial transformation of the
observables, and that is why the analysis will lead to something
useful.

The combination of (\ref{turkey}) and (\ref{durkey}) can be
viewed, formally, to first order in $\varepsilon$, as a rotation
of the $34$ plane.  As such, it leaves fixed, to first order, the
bosonic part of the Lagrangian that contains $A_3$ and $A_4$. For
example, the contributions
\begin{equation}\label{luckey} \int_{W\times \tilde S^1}\d^4x\sum_{\mu=0,1,2}\Tr\,\left(F_{\mu 3}^2
+D_\mu A_4^2\right)\end{equation} are invariant to first
nontrivial order in $\varepsilon$ under the combination of the
formal operation (\ref{turkey}), which we implement via
(\ref{helpo}), and the actual change of variables (\ref{durkey}).
The same is true for the other bosonic terms involving $A_3$ and
$A_4$, namely
\begin{equation}\label{plucky}\int_{W\times\tilde S^1} \d^4x \,\Tr\left((D_3A_5)^2+[A_4,A_5]^2\right)
\end{equation}
and
\begin{equation}\label{zucky}\int_{W\times\tilde S^1}\d^4x \,\Tr \,(D_3A_4)^2.\end{equation}

To compensate  for the $\Omega$-deformation of the terms in the
action involving fermions, we must make a similar rotation of the
fermions
\begin{equation}\label{fermrot}\Psi\to \exp\left(\frac{\varepsilon\rho}{2}\Gamma_{34}\right)\Psi.\end{equation}

The formulas we have considered so far compensate for the
$\Omega$-deformation only to first non-trivial order in
$\varepsilon$, because the formulas (\ref{turkey}) and
(\ref{durkey}) represent a rotation of the 34 plane only to that
order.  (By contrast, for the fermions we have used an exact
rotation matrix in (\ref{fermrot}).)  How can we improve our
formulas to represent a rotation of the 34 plane beyond first
order?

For $A_4$, it is fairly clear what to do.  We compose the
$\Omega$-deformation with an ordinary change of variables $ A_4\to
\hat A_4=A_4/\sqrt{1+\varepsilon^2\rho^2}$.  So the modified $A_4$
becomes
\begin{equation}\label{lanky}\hat A_4=\frac{1}{\sqrt{1+\varepsilon^2\rho^2}}
\left(A_4+\varepsilon\rho\, \frac{D}{Dx^3}\right).\end{equation}
Now it is more or less clear what the remaining formula for the
redefinition of $A_3$ ought to be.  To get a rotation of the 34
plane, we would like the formula to be in some sense
\begin{equation}\label{zanky}\frac{D}{Dx^3}\to \frac{1}{\sqrt{1+\varepsilon^2\rho^2}}
\left( \frac{D}{Dx_3}-\varepsilon\rho A_4\right).\end{equation}
The only problem is to explain what this formula means, as we are not free to
rescale the derivative $\partial/\partial x^3$.

We should interpret rescaling of $D/Dx^3$ as rescaling of this
covariant derivative referred to an orthonormal frame, which will
result from rescaling of the metric in the $x^3$ direction.
Instead of saying that $x^3$ has period $2\pi\rho$, we write
$x^3=\rho w$ where $w$ is an ordinary angular variable of period
$2\pi$.  Thus the metric on $\R\times S^1\times I\times \tilde
S^1$ becomes
\begin{equation}\label{oggo} \d s^2 =\sum_{\mu=0}^2(\d
x^\mu)^2+\rho^2\, \d w^2.\end{equation} Let us now rewrite the contribution (\ref{luckey}) to the action
 making explicit the dependence on
$g_{ww}=\rho^2$ and also on the gauge coupling constant $e^2$:
\begin{equation}\label{tuckey}\frac{1}{e^2}\int_{W\times\tilde
S^1} \d^3x\,\d w \sqrt{g_{ww}}
\sum_{\mu=0,1,2}\Tr\,\left(g_{ww}^{-1}F_{\mu w}^2 +D_\mu
A_4^2\right).\end{equation} For this to be invariant, we should
rotate $A_4$ into $g_{ww}^{-1/2}D/D w$, which is the covariant
derivative referred to an orthonormal frame. This will entail
rescaling of $g_{ww}=\rho^2$. At the same time we will rescale
$e^2$ to ensure that
\begin{equation}\label{gensu}\frac{\sqrt{g_{ww}}}{e^2}\end{equation}
remains fixed.  Given this, a rotation of the $A_4-X$ plane, where
 $X=g_{ww}^{-1/2}D/Dw$  (which in the undeformed theory is the
same as $ D/Dx^3$), will leave (\ref{tuckey}) invariant. The
equations for the rotation should thus be
\begin{align}\label{talign} A_4&=\frac{1}{\sqrt{1+\varepsilon^2\rho^2}}
\biggl(\hat A_4-\varepsilon\rho\, \hat X\biggr)\cr
 X &=\frac{1}{\sqrt{1+\varepsilon^2\rho^2}} \biggl(\hat
X+\varepsilon\rho\, \hat A_4\biggr),\end{align} where it is
convenient to express the objects $A_4, X$ of the undeformed
description in terms of corresponding deformed objects $\hat A_4$,
$\hat X$. Interpreting $\hat X$ as $\hat g_{ww}^{-1/2}D/Dw$, we
require
\begin{equation}\label{gwelf}
\frac{1}{\sqrt{g_{ww}}}=\frac{1}{\sqrt{1+\varepsilon^2\rho^2}}\frac{1}{\sqrt{\hat g_{ww}}}
\end{equation}
or more simply
\begin{equation}\label{onelf}
\hat g_{ww}=\frac{g_{ww}}{1+\varepsilon^2\rho^2}=\frac{\rho^2}{1+\varepsilon^2\rho^2}.\end{equation}

To keep (\ref{gensu}) fixed, we also need to change the gauge
coupling so that $\sqrt{\hat g_{ww}}/\hat
e^2=\sqrt{g_{ww}}/{e^2}$, or
\begin{equation}\label{nelf}\hat e ^2=
\frac{e^2}{\sqrt{1+\varepsilon^2\rho^2}}.\end{equation}

The parameter $\varepsilon$ is a physical parameter of the
$\Omega$-deformation, and for example it enters the results of
\cite{NS2} that we aim to understand.  However, the asymptotic
radius $\rho$ in the $\Omega$-deformed description is not an
important parameter for the topological field theory observables
of interest.

Accordingly, we could proceed  with any choice of $\rho$. However,
it turns out that we get a particularly simple description if we
take $\rho$ to be very large, $\rho>>1/\varepsilon$, so we will
consider this case first. (In fact, in discussions of the
$\Omega$-deformation, a flat metric on $\R^2$ is typically
assumed, rather than our cigar metric $D$. This corresponds to
taking $\rho=\infty$.)

The above formulas simplify in the limit $\rho\to\infty$. The
formulas (\ref{talign}) expressing the objects $X,A_4$ of the
$\Omega$-deformed description in terms of analogous objects $\hat
X,\hat A_4$ in an undeformed description reduce to
\begin{align}\label{zlign} A_4&=-\hat X\cr
                            X& =\hat A_4.\end{align}
Thus, the transformation from the $\Omega$-deformed description to
an equivalent undeformed description is a simple $\pi/2$ rotation
of the 34 plane.  The limiting value of $\hat g_{ww}$ is
$1/\varepsilon^2$, which means that in undeformed language, we
take the radius of $\tilde S^1$ to be $1/\varepsilon$:
\begin{equation}\label{blign}
\hat\rho=\frac{1}{\varepsilon}.\end{equation}

Finally, for $\rho\to\infty$, we  have $\hat e\to 0$ and the
theory becomes weakly coupled.  Since we can  make the coupling
arbitrarily weak by taking $\rho$ large, one may wonder how the
theory can do anything interesting at all. The answer to this
question is that our formulas such as (\ref{nelf}) relating the
$\Omega$-deformed theory to an ordinary one are only valid away
from the tip of the cigar. For $\rho\to\infty$, quantum effects
remain, but they are localized near the tip of the cigar. Though
our explanation here is different in detail, our conclusion is the
same as in \cite{Nek}: when the $\Omega$-deformation is defined
using a vector field $V$, quantum effects are localized near
zeroes of $V$, because the effective gauge coupling becomes small
far away from the zeroes.

Thus, as long as we consider quantities that do not depend on
$\rho$, so that we can take $\rho$ large, the $\Omega$-deformed
theory on $\R\times S^1\times D_R$ is equivalent, away from the
tip of $D_R$, to a weakly coupled and undeformed theory in which
the asymptotic radius of $\tilde S^1$ is $1/\varepsilon$. However,
in going from the $\Omega$-deformed description to this weakly
coupled and undeformed description, physical observables undergo a
$\pi/2$ rotation in the tangent space to the 34 plane.  Such a
rotation acts on fermions as multiplication by
$(1+\Gamma_{34})/\sqrt 2$, so that is the transformation of the
supersymmetry generator $\eta$ in comparing the deformed
description to the weakly coupled and undeformed one:
\begin{equation}\label{kelmit}\eta\to\frac{1+\Gamma_{34}}{\sqrt 2}\eta.\end{equation}
Conjugation by this matrix maps $\Gamma_3$ to $\Gamma_4$ and
$\Gamma_4$ to $-\Gamma_3$.

\subsection{The Deformed Brane}\label{deformed}

We aim to generalize our result of section \ref{undeformed} and
describe in two-dimensional terms the half-BPS brane
$\B_\varepsilon$ that arises by compactification on
$D_{R,\varepsilon}$, with the $\Omega$-deformation.  In
$\Omega$-deformed language, the supersymmetries preserved by this
brane are those whose generator obeys eqn. (\ref{zomely}):
\begin{equation}\label{umok}\left(\Gamma_{23}+\sigma_{23}\right)\eta=0.\end{equation}
Since the symmetry that we used in making the $\Omega$-deformation
commutes with the matrix on the left hand side of (\ref{umok}),
the same condition also characterizes the unbroken supersymmetries
of the $\Omega$-deformed brane.  However, to describe the
$\Omega$-deformed brane in conventional language (in the simplest
case, the limit of large $\rho$), we must make pick a rotation of
the 34 plane.  The rotation angle $\vartheta$ can be read off from
eqn. (\ref{talign}):
\begin{equation}\label{zerro}\cos\vartheta
=\frac{1}{\sqrt{1+\varepsilon^2\rho^2}}.\end{equation}

\subsubsection{The Case
$\varepsilon\rho\to\infty$}\label{largerho}

Neither the $\Omega$-deformation nor the change of variables that
maps us back to the undeformed theory affects the support of the
brane $\B_\varepsilon$ that comes from the tip of the cigar. So
$\B_\varepsilon$ is a half-BPS brane whose support is all of
$\M_H$, just like the brane $\B_*$ that arises in the absence of
the $\Omega$-deformation. However, the supersymmetry preserved by
$\B_\varepsilon$ is different from that preserved by $\B_*$.

We first consider the simplest case, namely the limit of large
$\rho$. In this limit, $\vartheta=\pi/2$ and the necessary
rotation of supersymmetry generators in going from
$\Omega$-deformed variables to standard ones was given in eqn.
(\ref{kelmit}). The rotation in question just transforms
$\Gamma_3$ to $\Gamma_4$, so in a formalism in which the
Lagrangian away from the tip of the cigar is standard, the
supersymmetries preserved by the $\Omega$-deformed brane
$\B_\varepsilon$  are characterized not by (\ref{umok}) but by
\begin{equation}\label{zumok}\left(\Gamma_{24}+\sigma_{23}\right)\eta=0.\end{equation}

Just as in section \ref{undeformed},  it is convenient to view
$M=\R\times S^1\times D_{R,\varepsilon}$ as an asymptotic $T^2$
bundle over $\R\times I$.  We want to understand what topological
field theory structures  $\B_\varepsilon$ preserves in the
effective two-dimensional theory.

As in section \ref{undeformed}, any two-dimensional topological
field theory structure is associated to a supersymmetry generator
$\eta$ that (modulo a possible $SU(2)_R$ transformation) obeys
\begin{equation}\label{urft}\left(\Gamma_{02}+\sigma_{31}\right)\eta=0.\end{equation}

The conditions (\ref{zumok}) and (\ref{urft}) select a
two-dimensional space of $\eta$'s.  Any non-zero $\eta$ in this
two-dimensional space determines a topological field theory, so
just like $\B_*$, $\B_{\varepsilon}$ is compatible with a family
of topological field theories parametrized by $\Bbb{CP}^1$.  To
select a particular member of this family, we need to place an
additional condition on $\eta$; any condition will do.

By analogy with section \ref{undeformed}, we consider three simple
conditions.  One of these will be
\begin{equation}\label{plumkin}\left(\Gamma_{23}+\sigma_{12}\right)\eta=0.\end{equation}
What we get with this choice is easily determined.  The three
conditions (\ref{zumok}), (\ref{urft}), and (\ref{plumkin}) are
equivalent modulo an $SU(2)_R$ symmetry to the three conditions
used in section \ref{undeformed} to characterize the $B$-model in
complex structure $J$, namely eqns.  (\ref{zomely}),
(\ref{tomely}), and (\ref{domely}).  (To compare our current set
of three conditions to the previous three, one uses the fact that
if $A\eta=B\eta=0$, then $[A,B]\eta=0$, and  one also makes an
$SU(2)_R$ transformation $(\sigma_1,\sigma_2,\sigma_3)\to
(\sigma_3,\sigma_2,-\sigma_1)$.)  So the brane $\B_\varepsilon$ is
a $B$-brane in complex structure $J$.

That is the only complex structure in which $\B_\varepsilon$  is a
$B$-brane. It turns out that the other simple properties of
$\B_\varepsilon$ are conveniently understood by imposing on $\eta$
the additional condition
\begin{equation}\label{guro}\left(\Gamma_{21}+\sigma_{12}\right)\eta=0\end{equation}
or
\begin{equation}\label{puro}\left(\Gamma_{25}+\sigma_{12}\right)\eta=0\end{equation}
together with (\ref{zumok}) and (\ref{urft}):
\begin{align}\label{holf}\left(\Gamma_{24}+\sigma_{23}\right)\eta&=0\cr
\left(\Gamma_{02}+\sigma_{31}\right)\eta&=0.\end{align}

Either condition (\ref{guro}) or (\ref{puro}) leads to an
$A$-model for the effective topological field theory on $\R\times
I$.  For example, let us consider (\ref{puro}).  We will show that
it leads to an $A$-model for the symplectic structure $\omega_I$.
This means that the fields $A_4-iA_5$ and $A_1+i\acalpha$ that are
holomorphic in complex structure $I$ on $\M_H$ will both obey
$A$-model conditions; in other words, they depend holomorphically
on $z=x^0+ix^2$, modulo exact terms.  We will show this by
repeating the analysis of section (\ref{twod}), but now, instead
of the $T$-duality $A_3\leftrightarrow\acalpha$ converting a mixed
$AB$-model to a $B$-model, it will convert a slightly different
mixed $AB$-model to an $A$-model.

If one places all $\sigma$-matrices on the right in equations
(\ref{puro}) and (\ref{holf}), and then multiplies the left and
right hand sides of the three equations, one learns that
\begin{equation}\label{prof}
\Gamma_{0245}\eta=-\eta,\end{equation} as a result of which the
six-dimensional chirality condition implies also that
\begin{equation}\label{trof} \Gamma_{13}\eta=i\eta.\end{equation}
Now, consider the group $SO(4)_{0245}$ that rotates the tangent
spaces to the bosonic fields $A_0,A_2,A_4,A_5$.  Decomposing its
double cover as $SU(2)_{\ell '}\times SU(2)_{r'}$, (\ref{prof})
implies that $\eta$ is invariant under $SU(2)_{\ell'}$, and eqns.
(\ref{puro}) and (\ref{holf}) further imply that $\eta$ is
invariant under a diagonal subgroup of $SU(2)_{r'}\times SU(2)_R$.

In short, $\eta$ is characterized by precisely the conditions that
characterize the supersymmetry generator of Donaldson theory,
modulo a rotation of the tangent space that exchanges $A_3$ with
$A_4$ and $A_1$ with $A_5$.  This means that we can borrow the
analysis of section (\ref{twod}), except that the roles of
$A_1+iA_3$ and $A_4-iA_5$ are now reversed.  In section
(\ref{twod}), $A_4-iA_5$ obeyed a $B$-model condition, and
$A_1+iA_3$ obeyed an $A$-model condition that was converted to a
$B$-model condition by the $T$-duality
$A_3\leftrightarrow\acalpha$. So now, $A_4-iA_5$ obeys an
$A$-model condition, and $A_1+iA_3$ obeys a $B$-model condition
that is converted to an $A$-model condition by the $T$-duality on
$A_3$.

We can analyze the consequence of eqn. (\ref{guro}) in the same
way.  This equation differs from (\ref{puro}) by the exchange
$A_1\leftrightarrow A_5$.  So now, mimicking the last sentence of
the last paragraph, $A_4-iA_1$ obeys an $A$-model condition, and
$A_5+iA_3$ obeys a $B$-model condition that is converted to an
$A$-model condition by the $T$-duality on $A_3$.   So we get an
$A$-model in which $A_4-iA_1$ and $A_5+i\acalpha$ are holomorphic
modulo exact terms.

Given our conventions for what is $J$ and what is $K$, the
functions $A_4-iA_1$ and $A_5+i\acalpha$ are holomorphic in
complex structure $K$, and this means that the $A$-model that we
have just arrived at is the $A$-model of symplectic form
$\omega_K$.

In summary, the brane $\B_\varepsilon$ is -- in the limit of large
$\rho\varepsilon$ -- a brane of type $(A,B,A)$. For understanding
the results of \cite{NS2}, the most important property is that it
is an $A$-brane of type $\omega_K$.    Indeed, as we see if we
rotate back to the deformed language, the topological supercharge
of the $A$-model of type $\omega_K$ is the one that corresponds to
the usual supercharge of the $\Omega$-deformed theory (or
equivalently of Donaldson theory).  So this is the supercharge
that we must use to make contact with the results of \cite{NS2}.
It is also the right one for a different reason described in
section \ref{coiso}: the coisotropic brane with Chan-Paton
curvature $F=\omega_J$ in the $A$-model with symplectic structure
$\omega_K$ is the right tool if we want to study quantization of a
real section of the integrable system $\M_H$.

By contrast, in section \ref{confblocks}, where we make contact
with the relation \cite{AGT} of four-dimensional gauge theory to
two-dimensional Liouville theory, the important fact will be that
the same brane $\B_\varepsilon$ is an $A$-brane for the symplectic
structure $\omega_I$.

\subsubsection{The General Case}\label{gencase}

We have by now interpreted the supersymmetry preserved by the tip
of the cigar in the $\Omega$-deformed theory in undeformed
language for $\varepsilon\rho>>1$. However, since the
$\Omega$-deformation is most directly understood as a deformation
of the standard theory, with $\varepsilon$ as a small deformation
parameter, one would naturally like to analyze the opposite limit
$\varepsilon\rho<<1$. In fact, it is not hard to analyze the
general case using tools described in section \ref{little}.

We recall that a general topological twist in two-dimensions is
determined by a pair of complex structures $(\J_+,\J_-)\in
\CP^1_+\times \CP^1_-$.  A brane of type $(B,B,B)$ is
characterized by $\J_-=\J_+$.  On the other hand, as explained in
section \ref{genhalf}, a brane of type $(A,B,A)$ is characterized
by $\J_-=h\J_+$, where $h$ is a $\pi$ rotation around the $J$
axis.

We get from one to the other using the rotation (\ref{zlign}).  We
view this rotation as an element $g$  of a group $SO(4)_{1345}$
that acts on the tangent space to dimensions 1345.  As such, $g$
is a $\pi/2$ rotation of the $34$ plane. The double cover of
$SO(4)_{1345}$ is the group $\G=SU(2)_+\times SU(2)_-$, introduced
in section \ref{rotgroup}, that acts on $\CP^1_+\times\CP^1_-$. As
a $\pi/2$ rotation of one plane, $g$ corresponds to a pair
$(g_+,g_-)\in SU(2)_+\times SU(2)_-$, where each of $g_\pm$
corresponds to a $\pi/2$ rotation in $SO(3)_\pm=SU(2)_\pm/\Z_2$.
The $SU(2)$ element $h$ of the last paragraph is then
$h=g_-^{-1}g_+$, as noted in section \ref{rotgroup}.  Given that
$h$ is a $\pi$ rotation around the $J$ axis, evidently  $g_+$ and
$g_-$ are rotations around the $J$ axis by respective angles
$\pi/2$ and $-\pi/2$.

Next let us consider the case of general $\varepsilon\rho$.  $g$
is now a rotation of the 34 plane by a more general angle
$\vartheta$. So $g_+$ and $g_-$ are rotations around the $J$ axis
by angles $\vartheta$ and $-\vartheta$, respectively.

Now, starting with the fact that in the undeformed theory the tip
of the cigar preserves supersymmetry of type $(B,B,B$), we  can
understand the supersymmetry preserved by the brane
$\B_\varepsilon$ for general $\varepsilon\rho$. First of all,
complex structure $J$ is invariant under rotations around the $J$
axis.  So $\B_\varepsilon$ is a $B$-brane of type $J$ for all
$\varepsilon\rho$.  One can verify this explicitly by observing
that the conditions (\ref{plumkin}) and (\ref{holf}) are invariant
under a rotation of the 34 plane, modulo an $SU(2)_R$
transformation generated by $\sigma_{31}$.

More pressing for our application is to interpret in undeformed
variables the usual supercharge of the $\Omega$-deformed theory,
which in $\Omega$-deformed variables corresponds to $\J_+=\J_-=I$.
After rotating $\J_+$ by an angle $\vartheta$ and $\J_-$ by an
angle $-\vartheta$ in the $IK$ plane, they map to
\begin{align}\label{porz} \J_+& =I \cos\vartheta+ K\sin\vartheta \cr
                          \J_-&= I\cos\vartheta - K\sin\vartheta. \cr\end{align}
According to (\ref{golf}), the topological field theory with this
pair $(\J_+,\J_-)$ is equivalent to an $A$-model with
\begin{align}\label{ogolf}\omega^{-1}&=\frac{1}{2}\left(\omega_+^{-1}-\omega_-^{-1}\right)
\cr \omega^{-1}B&=\frac{1}{2}(\J_++\J_-).\end{align} Here, writing
$g$ for the hyper-Kahler metric on $\M_H$, we have
$\omega_\pm=g\J_\pm$.

To make (\ref{ogolf}) more explicit, we use
$\omega_+=\omega_I^*\cos\vartheta+\omega_K^*\sin\vartheta$,
$\omega_-=\omega_I^*\cos\vartheta-\omega_K^*\sin\vartheta$. (The
meaning of the asterisks was explained in eqn. (\ref{retso}):
$\omega_I^*$ and $\omega_K^*$ are the Kahler forms for the actual
Kahler metric on $\M_H$, which depends on the radii of the
compactification.)  Evaluating (\ref{golf}), we find that the
symplectic form of the equivalent $A$-model is
\begin{equation}\label{dorf}\omega=\frac{\omega_K^*}{\sin\vartheta}.\end{equation}
So the supersymmetry is that of the $A$-model with symplectic
structure a multiple of $\omega_K$. The multiple is inessential as
explained in the discussion of eqn. (\ref{oplo}). The result
agrees, as expected, with what we found for large
$\varepsilon\rho$ in section (\ref{deformed}).

Eqn. (\ref{golf}) also tells us how to determine the $B$-field:
\begin{equation}\label{torf}\omega^{-1}B=I\cos\vartheta,\end{equation}
which implies that $B$ is a multiple of $\omega_J$.

\subsubsection{A Canonical Coisotropic Brane}\label{canco}

Since $\omega_J$ is cohomologically trivial in a large class of
models, as explained in section \ref{distinguished}, one may
wonder what we actually learn by determining $B$.  The answer is
that we do not learn much until we also consider the couplings
generated at the tip of the cigar. Those couplings  generate a
Chan-Paton curvature $F$ for the brane in the effective
two-dimensional description, and only the combination $F+B$ is
invariant under $B$-field gauge transformations.

In this paper, we will not attempt to compute the couplings at the
tip of the cigar.  But we can make a simple observation.  Let us
consider the limit in which $\varepsilon\rho$ is very small.  We
approach this limit by keeping $\rho$ fixed and taking
$\varepsilon$ to zero. In this limit, the $\varepsilon$-dependent
corrections at the tip of the cigar are negligible, and $F$
vanishes. Since also $\vartheta\to 0$ for $\varepsilon\rho\to 0$,
eqn. (\ref{torf}) becomes $\omega^{-1}(F+B)=I$.

In other words, for $\varepsilon\rho$ very small, the brane
derived from the tip of the cigar approaches the coisotropic brane
of type $(A,B,A)$ in which the complex structure
$\I=\omega^{-1}(F+B)$ is equal to $I$.  Since
$\omega=\omega_K^*/\sin\vartheta$, this brane has
$F+B=\omega_J^*/\sin\vartheta$.  It is convenient to evaluate this
using (\ref{zpoln}), using the fact that $\omega_J^*={\mathrm
{Re}}\,\Omega_I^*$, and identifying $\Cm'$ as $2\pi\hat\rho$
(which for small $\varepsilon\rho$ is the same as $2\pi\rho$). For
small $\varepsilon\rho$, we can replace $\sin\vartheta$ by
$\varepsilon\rho$, whereupon $\rho$ cancels out and we get
\begin{equation}\label{zum}F+B={\mathrm{Re}}\,\frac{1
}{\varepsilon}\sum_i\biggl(\d a_i\wedge \d\acalpha^i-\d a_D^i\wedge
\d b_i\biggr)\end{equation} along with
\begin{equation}\label{um}\omega={\mathrm{Im}}\,\frac{1
}{\varepsilon}\sum_i\biggl(\d a_i\wedge \d\acalpha^i-\d a_D^i\wedge
\d b_i\biggr).\end{equation}

Recalling the formula (\ref{poln}) for
$\Omega_I=\omega_J+i\omega_K$, it follows that at least for small
$\varepsilon$, the brane coming from the tip of the cigar is a
space-filling brane in which the complex structure
$\I=\omega^{-1}(F+B)$ is equal to $\omega_K^{-1}\omega_J=I$. Since
$\varepsilon\rho$ is small in this computation,
 the difference
between the gauge couplings $e$ and $\hat e$ in $\Omega$-deformed
and undeformed variables is not important and $I$ is simply the
complex structure in which the Hitchin fibration of the moduli
space $\M_H$ of vacua is holomorphic.

In the case of a generalized quiver theory \cite{DG} that arises
by compactification from six dimensions on a Riemann surface $C$,
the brane is the canonical coisotropic brane of type $(A,B,A)$
that is naturally defined using the complex structure on $C$ that
prevails at the tip of the cigar -- the one that is used in the
$\Omega$-deformed description.  This is the right brane for any
$\rho$, since $\rho$ is irrelevant in the twisted topological
field theory.

Changing $\rho$ changes $\hat e$ and therefore changes the
effective complex structure on $C$.  However, the brane that we
have found makes sense for any choice of the complex structure on
$C$, because the concept of a brane of type $(A,B,A)$ is
independent of that complex structure. This was explained in
section \ref{distinguished}.

Our determination that the complex structure $\I=\omega^{-1}(F+B)$
is equal to $I$ can be tested in the following way.  Let
$Q_\varepsilon$ be the topological supercharge of the
$\Omega$-deformed theory.  For $\varepsilon=0$, away from the tip
of the cigar, there are no non-trivial $Q_\varepsilon$-invariant
local operators.  But at the tip of the cigar, $Q_\varepsilon$
reduces to the topological supercharge $Q$ of Donaldson theory,
and its cohomology in the space of gauge-invariant local operators
is generated by the gauge-invariant polynomials $\Tr\,\phi^n$ in
the scalar field $\phi$; these are the usual local operators in
Donaldson theory. In the $\sigma$-model with target $\M_H$, the
gauge-invariant polynomials in $\phi$ become the holomorphic
functions on the base of the Hitchin fibration (and thus, the
commuting Hamiltonians of the integrable system). Let us compare
this to the answer in the $A$-model with symplectic structure
$\omega$. In bulk, the $A$-model admits no local observables of
ghost number zero; however, on a boundary labeled by a
space-filling coisotropic brane with complex structure $\I$, the
observables are the holomorphic functions in that complex
structure.  Thus, the complex structure $\I$ on $\M_H$ must be one
in which the Hitchin fibration $\pi:\M_H\to\cmmib B$ is
holomorphic -- so that the holomorphic functions on $\cmmib B$
pull back to holomorphic functions on $\M_H$.  Our result $\I=I$
is consistent with this, since $I$ has the required property.

For this reason and because the deformation theory of coisotropic
branes is rather rigid, we believe that the result $\I=I$ or
equivalently the determination (\ref{zum}) of $F+B $ is exact,
even though our derivation was only valid for small $\varepsilon$.

\subsection{Boundary Conditions At The Far End}\label{farend}

Hitherto, in the effective field theory on $\R\times I$, we have
analyzed a distinguished brane $\B_\varepsilon$ that arises from
the tip of the cigar. It is a space-filling coisotropic brane in
an $A$-model with target $\M_H$.  The symplectic form of the
$A$-model is a multiple of $\omega_K$, and the gauge-invariant
field $F+B$ of the brane is the corresponding multiple of
$\omega_J$.

By contrast, what happens at the other end of the cigar is largely
up to us. We can pick any $A$-model boundary condition we want at
the second boundary $r=R$ of $\R\times S^1\times D_R$.   If the
aim is to study quantization of a real section of $\M_H$, we
should pick a boundary condition that determines a Lagrangian
$A$-brane with support  $L\subset \M_H$.  Moreover, while
Lagrangian for $\omega_K$, $L$ should be symplectic for
$\omega_J$.  We pick a flat line bundle $\S$ over $L$, and write
$\B_L$ for the corresponding $A$-brane.  The space of
$(\B_\varepsilon,\B_L)$ strings will be a quantization of $L$ with
symplectic form $F+B$.

To orient ourselves, we will begin by constructing some choices of
$L$ directly by hand, before describing them in four-dimensional
gauge theory language.  Also, we will begin with the free $U(1)$
vector multiplet, previously considered in section \ref{morec}.

We will consider two simple choices of Lagrangian submanifold.
Since the symplectic form in the abelian case is
\begin{equation}\label{umbo}\omega={\mathrm{Im}}\,\frac{1
}{\varepsilon}\biggl(\d a\wedge \d\acalpha-\d a_D\wedge
\d b\biggr),\end{equation} we can define a Lagrangian
submanifold $L_1$ by\footnote{Throughout our derivation, we have
taken $\varepsilon$ to be real.  However, a generalization to
complex $\varepsilon$ may be trivially made by a $U(1)_R$ rotation
(which may also act on bare masses and scale parameters of the
gauge theory). In this generalization, eqn. (\ref{umbo}) is still
valid, and the second conditions in (\ref{krumbo}) and
(\ref{rumbo}) should read ${\mathrm{Im}}\,(a_D/\varepsilon)=0$ and
${\mathrm{Im}}\,(a/\varepsilon)=0$. }
\begin{equation}\label{krumbo}\acalpha=0={\mathrm{Im}}\,a_D.\end{equation}
And we can define a second Lagrangian submanifold $L_2$ by
\begin{equation}\label{rumbo} b=0={\mathrm
{Im}} \,a.\end{equation}

Restricted to $L_1$, we have $F+B=\d\lambda$ with
\begin{equation}\label{tumbo}\lambda=-\frac{{\mathrm{Re}}\,a_D}{\varepsilon}\wedge \d  b. \end{equation}
 In the WKB approximation, quantum states
correspond to values of $\mathrm{Re}\,a_D$ for which
$\oint\lambda$ is an integer multiple of $2\pi$.  Since $ b$ is an
angular variable, this condition says that
$\mathrm{Re}\,a_D/\varepsilon$ should be an integer. We also set
$\mathrm{Im}\,a_D/\varepsilon$ to zero as part of the definition
of $L_1$.  The two conditions combine to
\begin{equation}\label{porf}
\frac{a_D}{\varepsilon}\in\Z,\end{equation} or equivalently
\begin{equation}\label{uporf}\exp(2\pi i
a_D/\varepsilon)=1.\end{equation} Both the result and also the
fact that the real and imaginary parts of $a_D$ have gone their
separate ways in the derivation are hopefully reminiscent of
section \ref{linko}.  In the case of the free abelian vector
multiplet, the WKB approximation is valid.

We can quantize $L_2$ similarly.  In this case, $F+B=\d\lambda$
with
\begin{equation}\label{phumbo}\lambda=\frac{{\mathrm
{Re}}\,a}{\varepsilon}\wedge\d\acalpha.\end{equation} Quantization gives
\begin{equation}\label{zporf}
\frac{a}{\varepsilon}\in\Z,\end{equation} or equivalently
\begin{equation}\label{zuporf}\exp(2\pi i
a/\varepsilon)=1.\end{equation}

How can we generalize this to the case of a non-abelian gauge
group?  Given a choice of duality frame, the symplectic form is
the obvious generalization of (\ref{umbo}),
\begin{equation}\label{gumbo}\omega={\mathrm{Im}}\,\frac{1
}{\varepsilon}\sum_i\biggl(\d a_i\wedge \d\acalpha^i-\d
a_D^i\wedge \d b_i\biggr).\end{equation} So it looks like we can
define an analog of $L_1$ by setting $\acalpha^i=0={\mathrm
{Im}}(a_D^i/\varepsilon)$.  Similarly we can imitate the
definition of $L_2$.  The only problem with these definitions is
that they only  make sense locally, since they require a choice of
duality frame.

To find a definition that makes sense globally in the nonabelian
case, we need a better point of view.  Let us first return to the
$U(1)$ problem and express the definitions of $L_1$ and $L_2$ in a
different way.  For gauge group $U(1)$, the moduli space of vacua
is the moduli space $\M_H$ of $U(1)$ Higgs bundles on a Riemann
surface $T$ that is simply a two-torus parametrized by the angles
$\acalpha$ and $b$.  In complex structure $J$, $\M_H$ parametrizes
flat $\C^*$ bundles over $T$ ($\C^*$ is the complexification of
$U(1)$).  Such a flat bundle is labeled by its holonomies.
Choosing an $A$-cycle and a $B$-cycle that correspond to the
$\acalpha$ and $b$ directions on $T$, the two holonomies are
$U=\exp( i(\acalpha+i\,{\mathrm{Im}}\,a_D/\varepsilon))$ and
$V=\exp( i(b+i\,{\mathrm{Im}}\,a/\varepsilon))$.  Thus, our two
Lagrangian submanifolds $L_1$ and $L_2$ can be described by the
conditions $U=1$ and $V=1$, respectively.

This definition can be adapted to the nonabelian case, if one
considers generalized quiver theories \cite{DG} which arise by
compactifying from six dimensions on a Riemann surface $C$.  In
fact, there are multiple ways to do this.

One approach, assuming that $C$ has genus $g$ (and for brevity no
punctures), is to pick a set of $g$ $A$-cycles and $g$ $B$-cycles
on $C$, generating the first homology of $C$. Then one can define
a complex Lagrangian submanifold $L\subset\M_H$ for complex
structure $J$ by requiring the holonomies of the flat bundle
around all $A$-cycles (or around all $B$-cycles) to be trivial.
This can lead to a construction in which formulas like
(\ref{uporf}) and (\ref{zuporf}) are valid at least
asymptotically.

For $G=SU(2)$, so that $G_\C=SL(2,\C)$, another construction of a
complex Lagrangian submanifold that in some sense generalizes what
we did for $U(1)$ is as follows.  In relation to generalized
quiver gauge theories \cite{DG}, it is convenient to build $C$
from $3g-3$ cylinders or tubes that are glued together via
trinions or ``pairs of pants.''  Topologically, such a
construction gives us $3g-3$ circles, one for each cylinder.  A
complex Lagrangian submanifold $L\subset\M_H$ can be defined by
requiring that the holonomy of the flat $SL(2,\C)$ bundle around
each circle is unipotent, in other words conjugate to a triangular
matrix
\begin{equation}\label{morz} \begin{pmatrix}1 & \star \cr 0 &
1\end{pmatrix}.\end{equation}  A
generalization of this Lagrangian submanifold for $G$ of higher
rank is not immediately apparent.

 To learn more, we now consider the construction of branes on
$\M_H$ of type $(A,B,A)$ from the point of view of
four-dimensional gauge theory.

\subsubsection{Gauge Theory Realization}\label{gaugereal}

To construct branes of type $(A,B,A)$ in gauge theory, we need
half-BPS boundary conditions that preserve those  supersymmetries
whose generator $\eta$  obeys
\begin{equation}\label{orf}M\eta =\eta\end{equation}
or equivalently \begin{equation}\label{oorf}\bar\eta M=\bar
\eta,\end{equation}
 where
\begin{equation}\label{torfox}M=\Gamma_{24}\sigma_{23}.\end{equation}
This choice of $M$ corresponds to the case
$\varepsilon\rho\to\infty$, so in this analysis we are restricted
to that case.

We will consider only the most elementary examples of half-BPS
boundary conditions.  The full classification and analysis of
half-BPS boundary conditions with $\N=2$ supersymmetry is likely
to be quite rich, just as in the case  of $\N=4$ supersymmetry
\cite{GW2}.  Here we will only scratch the surface.

In any local boundary conditions, half of the fermion fields will
vanish at the boundary.   The obvious condition that will
accomplish this, while preserving all the symmetries of the matrix
$M$, is to place on the fermions $\Psi$ in the vector multiplet a
condition analogous to (\ref{orf}) but possibly with the opposite
sign:
\begin{equation}\label{yorf}M\Psi =z\Psi,~~z=\pm 1.\end{equation}
For either choice of the sign $z$, we will describe a boundary
condition on the bosons in the vector multiplet that preserves the
supersymmetries with $M\eta=\eta$.

The condition that a boundary condition preserves supersymmetry is
that it ensures the vanishing of the normal component of the
supercurrent at the boundary.  The supercurrent is
$J_I=\Gamma^{JK}\mathrm{Tr}\,F_{JK}\Gamma_I\Psi$, and so, bearing
in mind that in our notation the normal coordinate is $x^2$, the
condition that the normal part of $\bar\eta J_I$ vanishes at the
boundary is that at the boundary we must have
\begin{equation}\label{porft}\bar\eta
\mathrm{Tr}\,\Gamma^{JK}F_{JK}\Gamma_2\Psi = 0. \end{equation}  To
ensure this condition, we must pick a judicious boundary condition
on the bosons $A_I$.

We find the condition we need by replacing $\bar\eta$ on the left
hand side of (\ref{porft}) by $\bar\eta M$, moving $M$ to the
right, and then using $M\Psi=z\Psi$. If $z=1$, then, since
$M\Gamma_2=-\Gamma_2M$, the condition that we want is that the
boundary values should obey
\begin{equation}\label{tommy}
[M,\Gamma^{JK}F_{JK}]=0.\end{equation} If instead we take $z=-1$,
then we need instead
\begin{equation}\label{zommy} \{M,\Gamma^{JK}F_{JK}\}=0\end{equation}
at the boundary.

To obey (\ref{tommy}), we impose Dirichlet boundary conditions on
$A_4$ and Neumann boundary conditions on the other
components\footnote{From this and the other statements below, we
omit $A_2$, the normal component of $A$, as it can be set to zero
near the boundary by a gauge transformation.} of $A$. (The former
implies that $F_{I4}=0$, $I\not=2$, while the latter implies
$F_{I2}=0$, $I\not=4$; together these imply (\ref{tommy}).) And
conversely, to obey (\ref{zommy}), by an argument similar to the
one just indicated, we place Neumann boundary conditions on $A_4$
and Dirichlet boundary conditions on the other components of $A$.

There is an important detail to mention here.  There is never a
problem with Dirichlet boundary conditions on gauge fields, but in
four dimensions, Neumann boundary conditions on gauge fields are
only possible if the gauge theory $\theta$-angles vanish. (Neumann
boundary conditions in our notation would say that the boundary
value of $F_{2I}$ vanishes, $I=0,1,3$  while at non-zero $\theta$,
one must add to this a multiple of $\theta\epsilon_{IJK}F_{JK}$.
Here $I,J,K$ take values 013.)  So at $\theta\not=0$, we are
limited to $z=-1$ and (\ref{zommy}).

We can make these two boundary conditions more concrete as
follows:

(I) In the first case (which as just explained requires the
$\theta$-angles to vanish), $A_4$ vanishes on the boundary and
$A_1,A_3,$ and $A_5$ do not.

(II)  In our second case, $A_1$, $A_3$, and $A_5$ vanish on the
boundary and $A_4$ does not.

We have described these boundary conditions for vector multiplets,
but they extend to hypermultiplets. In the case of a generalized
quiver theory obtained by compactification from six dimensions on
a Riemann surface $C$, the gauge group is semi-simple rather than
simple. We pick the same type of boundary condition -- type I or
type II -- for each factor.

Now let us describe what these boundary conditions look like after
toroidal compactification to two dimensions.  First we consider a
free vector multiplet with gauge group $U(1)$.  As usual, we
$T$-dualize the holonomy of $A_3$ to a scalar $\acalpha$.
$\acalpha$ obeys Dirichlet boundary conditions if $A_3$ obeys
Neumann boundary conditions, and vice-versa.

So in case I, the fields that vanish at the boundary are
$\acalpha$ and $A_4$.  We can think of $A_4$ as $\mathrm{Im}
\,a_D$ where $a=A_4-iA_5$ and $a_D=(4\pi i/e^2)a$.  (We recall
that for Type I, the gauge theory $\theta$-angle vanishes and so
the gauge coupling parameter $\tau$ reduces to $4\pi i/e^2$.)
Thus, the Type I boundary condition leads to a brane supported on
the Lagrangian submanifold $L_1$.

In case II, the fields that vanish on the boundary are $A_1$ and
$A_5$.  In the two-dimensional description, the fields that vanish
on the boundary are $b$, which is the holonomy of $A_1$, and
${\mathrm{Im}}\,a$, which is a multiple of the holonomy of $A_5$.
So the Type II boundary condition leads to a brane supported on
the Lagrangian submanifold $L_2$.

What about the nonabelian case?  We cannot so easily interpret the
boundary conditions I and II in the low energy theory.  But we can
do so asymptotically on the Coulomb branch.  Generically on the
Coulomb branch, the gauge group is broken to an abelian subgroup;
the only massless particles are $r$ vector multiplets, with $r$
the rank of the gauge group. Near infinity on the Coulomb branch
(and far from the locus on which additional massless particles
appear), an abelian treatment along the above lines is a good
approximation.

\def\I{{\mathrm{I}}}
\def\II{{\mathrm{II}}}
So the brane $\B_\I$ coming from a Type I boundary condition can
be described near infinity by the familiar conditions
$\acalpha^i={\mathrm {Im}}\,a_D^i=0$.  Similarly the brane
$\B_\II$ coming from a Type II boundary condition can be described
near infinity by $b_i={\mathrm{Im}}\,a_i=0$.

There are two reasons that we cannot simply imitate the derivation
of eqns. (\ref{porf}) and (\ref{zporf}) and determine the spectrum
by setting $a_i/\varepsilon$ or $a_D^i/\varepsilon$ to integers.
First, the branes $\B_\I$ and $\B_\II$ are described only near
infinity by the conditions mentioned in the last paragraph.
Second, in general the WKB approximation, which was used to arrive
at (\ref{porf}) and (\ref{zporf}), is not exact.

In section \ref{wkb}, we have described a situation in which the
WKB approximation is exact.  This involved a two-dimensional
theory with $(2,2)$ supersymmetry, formulated on $\R\times S^1$.
We have been studying here, after toroidal compactification from
four dimensions, a theory with $(4,4)$ supersymmetry on $\R\times
I$, with half-BPS boundary conditions at the ends.  As far as we
know, the WKB approximation is not exact in this situation.

\subsubsection{Alternative Compactification To Two
Dimensions}\label{altcom}

However, we can look at the same problem in another way. Our
starting point has been a four-dimensional gauge theory on
$\R\times S^1\times D_{R,\varepsilon}$, understood in terms of
$S^1\times \tilde S^1$ compactification to two dimensions.   This
leads to a problem on $\R\times I$ that we have by now discussed
at length.

However,  a more obvious way to reduce the same problem to
two dimensions is to simply view it as a compactification on
$D_{R,\varepsilon}$ down to $\R\times S^1$.  In this fashion, we
arrive at a $(2,2)$ theory on $\R\times S^1$.

Let us consider the two cases of a Type I or Type II boundary
condition.  In the first case, all vector multiplets obey Neumann
boundary condition.  A four-dimensional vector multiplet with
$\N=2$ supersymmetry reduces to a two-dimensional vector multiplet
with $(2,2)$ supersymmetry.  Four-dimensional hypermultiplets
reduce to two-dimensional chiral multiplets.  The net effect is
that a four-dimensional generalized quiver theory reduces to a
two-dimensional theory of $(2,2)$ supersymmetry based on the same
generalized quiver.

This is precisely the setting for section \ref{wkb}, and the WKB
approximation to the quantization will be exact when expressed in
terms of the effective twisted chiral superpotential $\tilde W$,
which in general will depend on both the scalar fields in the
vector multiplet and the deformation parameter $\varepsilon$.  This twisted chiral
superpotential has been analyzed in \cite{Nek} and \cite{NS2}.  It receives contributions from
perturbative effects and instanton effects  near the tip of the
cigar. General arguments concerning the $\Omega$-deformation show
that all contributions come from the region near the tip. As far
as we know, the analysis of $\tilde W$ may as well be carried out
in $\Omega$-deformed variables; we know of no advantage to the
rotation to undeformed variables that has been exploited in the
present paper.

Since the effective twisted chiral superpotential is the same as
the one used in \cite{NS2}, the spectrum is also the same.  All
that we have gained by our approach here is the understanding of
why this spectrum can be understood as coming from the
quantization of a real slice of Hitchin's integrable system.

The case of a Type II boundary condition is a little different.
All gauge fields obey Dirichlet boundary conditions.  This
completely breaks the gauge symmetry, so there are no gauge fields
in the effective description on $\R\times S^1$. Both vector
multiplets and hypermultiplets reduce to chiral multiplets in that
effective description.

It seems that the clearest picture emerges if we rotate back from
the ordinary variables that we have used in describing the
boundary conditions to $\Omega$-deformed variables.  This has the
effect of exchanging $A_4$, which is a scalar field (the real part
of $\phi$) with $A_3$, the component of the gauge field around
$\tilde S^1$, the boundary of $D_R$.  So in this description, the
boundary condition does not constrain the holonomy of $A_3$ around
$\partial D_R$.

However, for the purposes of constructing a supersymmetric ground
state, the holonomy around $\partial D_R$ must vanish, or
supersymmetry will be violated by curvature in the interior of
$D_R$.  Treating $A_3$ as a constant in the low energy theory, the
holonomy is $\exp(2\pi \rho A_3)$, where $\rho$ is the radius of
$\tilde S^1$ in $\Omega$-deformed variables, so the condition is
\begin{equation}\label{kornz}\exp(2\pi\rho A_3)=1.\end{equation}
We want to rotate this condition back to undeformed variables.  In
doing so, we take $\rho\to\infty$, $\hat\rho\to 1/\varepsilon$,
since this choice was built into our construction of boundary
conditions in section \ref{gaugereal}. In the undeformed
variables, $\rho A_3$ can be replaced by $A_4/\varepsilon$. So the
condition becomes $\exp(2\pi  A_4/\varepsilon)=1$.  The boundary
condition also set $A_5=0$.  We can combine the two statements to
\begin{equation}\label{zelk}\exp(2\pi
\phi/\varepsilon)=1.\end{equation}

In the low energy description, we rotate $\phi$ to a maximal torus and denote it as $a$.
Thus, (\ref{zelk}) seems to show that the quantization of $a$ for a Type II boundary
condition takes its most naive form, though this is not so for Type I.

\subsubsection{Eigenvalues And Opers}\label{eigand}

Finally, we will describe what may ultimately -- after some future
developments -- be the most powerful way to get detailed results
about quantization of these integrable systems. Something close to
what we will describe momentarily has actually been carried out in
the mathematical literature \cite{F1,FFR,F2,F3} in the context of
an integrable system known as the Gaudin model. (This work has
been done in a language much closer to geometric Langlands than to
gauge theory, and at the moment we do not know precisely how to
reformulate it in terms of gauge theory or even two-dimensional
sigma-models.)

Our starting point in this section has been to choose a Lorentz-invariant boundary condition
at the far end of $D_R$ that preserves supersymmetry of type $(A,B,A)$.
In a description that arises by $T$-duality on scalars that represent holonomies around
$\tilde S^1$, this boundary condition determines a Lagrangian submanifold $L\subset \M_H$
and a corresponding Lagrangian brane $\B_L$.  Quantization of $L$ is carried out
by taking the space of $(\B_\epsilon,\B_L)$ strings.

However, at  the far
end of $D_R$, the two circles $S^1$ and $\tilde S^1$ are on an equivalent footing.
Hence, had we made the $T$-duality on scalars associated to $\tilde S^1$ rather than
$S^1$,  the same boundary condition would have given a very similar Lagrangian
submanifold $\tilde L$ again of type $(A,B,A)$.  Actually $\tilde L$ is a Lagrangian submanifold
not of $\M_H$ but of a dual moduli space $\tilde\M_H$ obtained from $\M_H$ by $T$-duality on the
fibers of the Hitchin fibration.\footnote{In the context of generalized quiver gauge theories,
if $\M_H$ is a moduli space of Higgs bundles on a Riemann surface $C$ with gauge
group $G$, then \cite{HT}  $\tilde \M_H$ is a corresponding moduli space of Higgs bundles on
$C$ with gauge group $G^\vee$, the group dual to $G$.  If $G$ is simply-laced, which is the
case that arises most simply by reduction from the $(0,2)$ model in six dimensions,
then $G$ and $G^\vee$ have the same universal cover, and so do $\M_H$ and $\tilde\M_H$.
In this situation, $L$ and $\tilde L$ should be equivalent if lifted to the universal cover.}

What happens to the brane $\B_\varepsilon$ if we use a description
obtained by $T$-duality on scalars coming from holonomy around
$S^1$ rather than $\tilde S^1$?  In this case, $\B_\varepsilon$ is
replaced by a new brane, also of type $(A,B,A)$, that arises from
$\B_\varepsilon$ by $T$-duality on the fibers of the Hitchin
fibration.

\def\hh{{\mathcal H}}
 We will postpone a fuller explanation to section
\ref{operbrane},  but in brief the dual of $\B_\varepsilon$ is a
Lagrangian brane $\B_\tN$ supported on a Lagrangian submanifold
$\tN$ that is known as the variety of opers. We can use this dual
description to describe the space $\hh$ of $(\B_\varepsilon,\B_L)$
strings; it is the same as the space of $(\B_\tN,\B_{\tilde L})$
strings.   This space is easily described, assuming that the two
Lagrangian submanifolds $\tN$ and $\tilde L$ have generic
(transverse) intersections.    $\hh$ has a basis\footnote{In
general, in the $A$-model of a symplectic manifold $X$,
world-sheet instanton effects can remove from the cohomology the
states corresponding to some intersections.  In the present
context, both branes $\B_\tN$ and $\B_{\tilde L}$ are of type
$(A,B,A)$, so they have extra supersymmetry in common beyond what
is typical in an $A$-model.  This extra supersymmetry generates an
extra fermion zero mode in the field of an instanton, ensuring
that instanton effects do not alter the cohomology that one reads
off classically from the intersections of the two Lagrangian
submanifolds.} with one basis vector for every intersection point
of $\tN$ and $\tilde L$.

The commuting Hamiltonians of the integrable system correspond to
holomorphic functions on $\tN$.  The eigenvalues of the commuting
Hamiltonians are simply the values of the corresponding functions
at the points on $\tN$ at which $\tN$ intersects $\tilde L$.
Differently put, the joint eigenstates of the commuting
Hamiltonians correspond to opers (points in $\tN$) that obey a
certain system of equations stating that they lie in $\tilde L$.
As remarked above, there is an example \cite{F1,FFR,F2,F3} of an
integrable system whose spectrum has been described in just such a
fashion.

\section{Conformal Blocks From Four Dimensions}\label{confblocks}

In this section, we will use similar methods to study a different
problem -- the relation \cite{AGT} of four-dimensional gauge
theory to Liouville theory and like theories in two dimensions.

\subsection{Gauge Theory And Liouville Theory}\label{gl}

We consider an $\N=2$ supersymmetric gauge theory in four
dimensions on the four-manifold $M'=\R\times S^3$.   Because there
are now three curved dimensions in spacetime (as opposed to two in
section \ref{compom}, where we worked on $M=\R\times S^1\times
D_R$), topological twisting now leaves only two unbroken
supercharges. This is so for any  product metric on $\R\times
S^3$, regardless of the choice of metric on $S^3$. One unbroken
supercharge is the usual supercharge $Q$ that is related to
Donaldson theory (in the case of $SU(2)$ gauge theory without
hypermultiplets).  In Euclidean signature, the second supercharge
$\bar Q$ can be obtained from the first by an
orientation-reversing reflection of the first factor $\R$ of $M'$.
In Lorentz signature, $\bar Q$ is the hermitian adjoint of $Q$.

The twisting that preserves $Q$ and $\bar Q$ is defined in the
usual way, by identifying the $SU(2)_R$ global symmetry group with
the structure group of the spin bundle of $S^3$.  If one modifies
the  geometry of $M'=\R\times S^3$ so as to be no longer a
product, then the positive and negative chirality spin bundles of
$M'$ become distinct.  For twisting, we must then decide whether
we want to identify $SU(2)_R$ with the structure group of the
positive or negative chirality spin bundle. As a result, in the
twisted theory with a non-product metric, we can conserve either
$Q$ or $\bar Q$ but not both.

The most important special case of this arises if we ``cap off''
the metric on $M'$.  Topologically, we do this by viewing $S^3$ as
the boundary of a ball $B^4$.  Instead of a flat metric on $B^4$,
we pick a metric that looks near the boundary like a product
$\R^-\times S^3$, where $\R^-$ is a half-line $t\leq 0$, but such
that $S^3$ shrinks to a point at some $t=t_0<0$.  In defining a
topologically twisted theory on $B^4$ with such a metric, we can
preserve $Q$ or $\bar Q$ but not both.

The basic property of $S^3$ that we will use is that it admits a
$U(1)\times U(1)$ action.  If $S^3$ is viewed as the locus
$\sum_{i=1}^4 y_i^2=1$ in $\R^4$, then we introduce polar
coordinates $y_1+iy_2=ue^{i\alpha}$, $y_3+iy_4=ve^{i\beta}$ and
finally $(u,v)=(\sin w,\cos w)$.  $U(1)\times U(1)$ acts by shifts
of $\alpha$ and $\beta$.  The round metric on $S^3$ is $\d
w^2+\sin^2 w \,\d\alpha^2+\cos^2 w\,\d\beta^2$.  It will be
convenient for us to use a more general $U(1)\times
U(1)$-invariant metric on $S^3$.  We let $w$ run over an interval
$0\leq w\leq \ell$, and we take
\begin{equation}\label{pokk}\d s_{S^3}^2=\d w^2+ f(w) \d\alpha^2
+g(w)\d\beta^2,\end{equation} where $f(w)=\rho_1^2$ except very
near the left end-point $w=0$ where it vanishes quadratically, and
$g(w)=\rho_2^2$ except very near the right end-point $w=\ell$
where it vanishes quadratically. Here $\rho_1$ and $\rho_2$ are
constants.  In describing $S^3$ as in (\ref{pokk}), we are in
effect viewing it as a ``warped'' $T^2$ fibration over a
one-dimensional base.  On $M'=\R\times S^3$, we take the product
metric $\d s^2=\d t^2+\d s_{S^3}^2$.

\subsection{$\Omega$-Deformation}\label{omegadef}

What we want to do with this theory is to $\Omega$-deform it,
using the $U(1)\times U(1)$ symmetry.  We follow the same basic
procedure as in (\ref{frog}), but now we use the fact that
four-dimensional gauge theories with $\N=2$ supersymmetry can
arise by dimensional reduction from a six-dimensional theory with
two more coordinates that we will here call $x^4$ and $x^5$. Apart
from adding $(\d x^4)^2+(\d x^5)^2$ to the metric $\d s^2$, the
only change that we make is to modify the terms that involve $\d
\alpha$ and $\d \beta$.  A simple case is a modification in which
$\d\alpha$ ``mixes'' only with $\d x^4$ and $\d\beta$ ``mixes''
only with $\d x^5$, in the sense that the relevant part of the
six-dimensional metric is
\begin{equation}\label{gunn} f(w)\left(\d\alpha -\varepsilon_1 \d
x^4)^2+g(w)\right(\d\beta-\varepsilon_2 \d x^5)^2+(\d x^4)^2+(\d x^5)^2.\end{equation}

We have considered here a ``diagonal'' deformation in which
$\alpha$ and $\beta$ mix only with $x^4$ and $x^5$, respectively.
We take the deformation parameters $\varepsilon_1$ and
$\varepsilon_2$ to be real and positive. In greater generality, as
in \cite{Nek}, one can introduce complex deformation parameters
$\epsilon_1$, $\epsilon_2$, set $x=x^4+ix^5$ and generalize
(\ref{gunn}) to
\begin{equation}\label{punn}f(w)\left(\d\alpha-{\mathrm{Re}}(\bar\epsilon_1
\d x)^2\right)+g(w)\left(\d\beta-{\mathrm{Re}}(\bar\epsilon_2\d
x)^2\right)+(\d x^4)^2+(\d x^5)^2.\end{equation} Evidently, the special case
(\ref{gunn}) corresponds to \begin{equation}\label{potz}
\epsilon_1=\varepsilon_1,~~\epsilon_2=i\varepsilon_2,\end{equation}
and so
\begin{equation}\label{unn}\frac{\epsilon_2}{\epsilon_1}=i\frac{\varepsilon_2}{\varepsilon_1}.\end{equation}
We will analyze here the special case of a diagonal
deformation, and then in appendix \ref{gencompgeom}, we will
analyze the general case using generalized complex geometry.

\def\sdef{S^3_{\varepsilon_1,\varepsilon_2}}
\def\hdef{\H_{\varepsilon_1,\varepsilon_2}}
\def\bdef{B^4_{\varepsilon_1,\varepsilon_2}}
\def\ssdef{S^4_{\varepsilon_1,\varepsilon_2}}
\def\rdef{\R^4_{\varepsilon_1,\varepsilon_2}}
Let $S^3_{\varepsilon_1,\varepsilon_2}$ be $S^3$ with the
$\Omega$-deformation with the indicated parameters.  We let
$\H_{\varepsilon_1,\varepsilon_2}$ be the space of supersymmetric
ground states of the $\Omega$-deformed theory on $\R\times \sdef$.
We interpret the results of \cite{AGT} to mean that $\hdef$ can be
identified, for $G=SU(2)$, with the space of Virasoro conformal
blocks  on the Riemann surface $C$. (Virasoro conformal blocks are
the conformal blocks of Liouville theory. If $SU(2)$ is replaced
by another simply-laced group $G$, then \cite{Wyllard,MMMM} the
relevant objects are the conformal blocks of the corresponding
Toda field theory.) Understanding this claim and certain related
facts will be our goal in the rest of this paper.

Our first step will be to follow the logic of section \ref{alternative}
and compare the $\Omega$-deformed description to a standard one.
  In the region in which $f$ and $g$ are
constants, the $\Omega$-deformation can be removed by an ordinary
change of variables.  The combined operation of
$\Omega$-deformation plus ordinary change of variables amounts to
a rotation of the $\alpha 4$ plane times a rotation of the $\beta
5$ plane, each of them precisely analogous to (\ref{talign}). (The
more general deformation (\ref{punn}) can also be removed by a
change of variables in the region in which $f$ and $g$ are
constants, though the necessary formulas are more complicated.)

Just as in section \ref{alternative}, the simplest situation is
that in which the radii $\rho_1$, $\rho_2$ in the
$\Omega$-deformed description are taken to infinity, keeping the
deformation parameters $\varepsilon_i$ fixed.  Then the rotation
that goes from $\Omega$-deformed description to the one by
conventional variables is simply the product of $\pi/2$ rotations
in the $\alpha4$ and $\beta5$ planes.  In the undeformed language,
the radii of the circles parametrized by $\alpha$ and $\beta$ are
\begin{equation}\label{properly}\hat\rho_1=\frac{1}{\varepsilon_1},~~\hat\rho_2
=\frac{1}{\varepsilon_2}.\end{equation}
This means that the $\tau$ parameter of the torus parametrized by
$\alpha,\beta$ is in the undeformed description
\begin{equation}\label{roperly}\hat\tau=i\frac{\varepsilon_2}{\varepsilon_1}
=\frac{\epsilon_2}{\epsilon_1}.\end{equation} (We denote this
$\tau$ parameter as $\hat\tau$ to avoid confusion with generic
coupling parameters of an $\N=2$ supersymmetric gauge theory, or
complex structure parameters of a Riemann surface $C$, which we
have called $\tau$.)

Extending (\ref{nelf}), the gauge coupling in the ordinary
description is
\begin{equation}\label{troperly}\hat
e^2=\frac{e^2}{\sqrt{(1+\varepsilon_1^2\rho_1^2)(1+\varepsilon_2^2\rho_2^2)}}.\end{equation}
In particular, $\hat e^2\to 0$ as $\rho_1,\rho_2\to\infty$.

The $\pi/2$ rotations means that in transforming from the
$\Omega$-deformed description to the ordinary description, the
supersymmetry generator $\eta$ is transformed by
\begin{equation}\label{zoperly}\eta\to \frac{1+\Gamma_{\alpha 4}}{\sqrt 2}\frac{1+\Gamma_{\beta
5}}{\sqrt 2}\eta.\end{equation}

\subsection{Two-Dimensional Description}\label{zoth}

Just as in section \ref{compom}, we want to think of $M'=\R\times
S^3$ as an $S^1\times \tilde S^1$ fibration, with fiber
parametrized by $\alpha,\beta$, over $\R\times I$, parametrized by
$t,w$.  As in section \ref{compom}, there are branes at both ends
of $I$.  In  section \ref{compom}, just one of these branes
originated from the geometry at the tip of a cigar.  But in our
present problem, both branes have such an origin. In fact, near
either end of $I$, $M'$ resembles, up to a fairly obvious change
of variables, the cigar geometry that led to our friend
$\B_\varepsilon$.

This will enable us to borrow the analysis of section
\ref{compom}, but we do have to keep in mind that the isomorphism
with section \ref{compom} is different at the two ends of $I$.
Near $w=0$, the $\alpha$ circle is shrinking and thus corresponds
to $\tilde S^1$ in section \ref{compom}.  The coordinates
$(t,w,\alpha,\beta)$ should be matched near $w=0$ with
$(x^0,x^2,x^3,x^1)$ in section \ref{compom}.  Near $w=\ell$, the
$\beta$ circle shrinks and corresponds to $\tilde S^1$; the roles
of $\alpha$ and $\beta$ are exchanged.

\def\var{\varepsilon}
\def\varone{{\alpha}}
\def\vartwo{{\beta}}
We will denote as $\B_\varone$ the brane at $w=0$ where the
$\alpha$ circle shrinks, and as $\B_\vartwo$ the brane at $w=\ell$
where the $\beta$ circle shrinks. Let us first understand the
brane $\B_\alpha$ in terms of ordinary variables.

Since the geometry near $w=0$ is the same as the geometry near
$x^2=0$ in section \ref{compom}, the only reason that the brane
$\B_\alpha$ is not trivially equivalent to the brane
$\B_\varepsilon$ studied in section \ref{compom}  is that a
rotation of observables must be made to compare the two. For
$\B_\alpha$, the rotation that we want to make from
$\Omega$-deformed to ordinary variables is by the product
\begin{equation}\label{omurx}\frac{1+\Gamma_{\alpha 4}}{\sqrt 2}\frac{1+\Gamma_{\beta 5}}{\sqrt 2},\end{equation}
 which in the
notation of section \ref{compom} corresponds to
\begin{equation}\label{zomurx}\frac{1+\Gamma_{34}}{\sqrt 2}\frac{1+\Gamma_{15}}{\sqrt 2},\end{equation}
 By contrast, in section \ref{compom}, we
made the rotation by $(1+\Gamma_{34})/\sqrt 2$.  So in comparing
$\B_\varepsilon$ to $\B_\alpha$, we need to  make an additional
rotation of the supersymmetry parameter $\eta$ (and all other
operators and observables) by $(1+\Gamma_{15})/\sqrt 2$, whose
effect is to exchange $\Gamma_1$ and $\Gamma_5$.

With this in mind, we re-examine some of the key equations of
section \ref{deformed}.  The supersymmetries preserved by
$\B_\varepsilon$ were characterized by eqn. (\ref{zumok}), and as
the matrix on the left hand side commutes with $\Gamma_{15}$, the
brane
 $\B_\alpha$ preserves the same supersymmetries as
$\B_\varepsilon$.  Likewise, the condition (\ref{urft}) that
characterizes which supersymmetries of a given brane can be
interpreted in two-dimensional topological field theory is again
defined with a matrix that commutes with $\Gamma_{15}$. So
$\B_\alpha$ is a brane of type $(A,B,A)$, just like
$\B_\varepsilon$.

The only difference between the two branes is that the $15$
rotation acts non-trivially on the $\Bbb{CP}^1$ that parametrizes
the possible topological field theory structure. Eqn.
(\ref{plumkin}), which singles out the supersymmetry generator of
the $B$-model in complex structure $J$, is again invariant under
the 15 rotation. But the other two conditions (\ref{guro}) and
(\ref{puro}), which characterize the $A$-models of types
$\omega_K$ and $\omega_I$, are exchanged under this
rotation.\footnote{\label{urtz} To be more precise, there is a
minus sign here, and the mapping takes
$(\omega_I,\omega_K)\to(\omega_K,-\omega_I)$. Similarly, two
paragraphs below, the exchange of $J$ and $K$ is really $(J,K)\to
(K,-J)$.}

The important consequence of this is that the supersymmetry charge
$Q$ that is related to instanton counting and Donaldson theory,
and which in section \ref{compom} was interpreted in undeformed
variables as the generator of the $A$-model supersymmetry of type
$\omega_K$, will now have to be interpreted as the generator of
the $A$-model supersymmetry of type $\omega_I$.  This is the
supercharge we care about for our present purposes, since it is
the only one (apart from its adjoint) that is conserved in the
twisted theory on the full $\R\times S^3$ geometry.

Now let us consider our other brane $\B_\beta$, the one at
$w=\ell$. All the same reasoning applies except that $x^4$ is
exchanged with $x^5$.  This exchanges the scalar field $A_4$ with
$A_5$.  That exchange can be carried out by a $U(1)_R$
transformation which also rotates complex structure $J$ into $K$
(with a sign mentioned in footnote \ref{urtz}) while fixing $I$.
So $\B_\vartwo$ is a brane of type $(A,A,B)$.  (The $U(1)_R$
transformation that maps $A_4$ to $A_5$ may not be a symmetry, as
it also acts on mass parameters in an underlying four-dimensional
Lagrangian. But still it can be used to determine the type of
supersymmetry preserved by $\B_\vartwo$, given that we know this
for $\B_\varone$.) The same reasoning as before shows that the
usual supercharge $Q$ of instanton counting and Donaldson theory
is associated to the $A$-model of type $\omega_I$.

Obviously, the only structure that a brane $\B_\alpha$ of type
$(A,B,A)$ and a brane $\B_\beta$ of type $(A,A,B)$ have in common
is the $A$-model of type $\omega_I$.  Everything hangs together,
since the supersymmetry generator of this $A$-model is indeed the
supercharge $Q$ that is conserved in the $\R\times S^3$ geometry.

\subsection{The Kahler Parameter Of The Sigma Model}\label{zelf}

Of the three Kahler classes on $\M_H$, $\omega_I$ is topologically
non-trivial, but $\omega_J$ and $\omega_K$ are topologically
trivial.  This has been explained in section \ref{distinguished}.

Accordingly, the $A$-models of type $\omega_J$ or $\omega_K$ have
no coupling parameter.  But the $A$-model of type $\omega_I$ does
have such a parameter, which we can think of as the Kahler
parameter in complex structure $I$.  According to a standard
result \cite{HM,BJV} that was explained in section \ref{morec},
this parameter, assuming its imaginary part is positive, is the
modular parameter $\hat\tau$ of the two-torus $T^2$ that is
parametrized by the angles $\alpha,\beta$, measured in the metric
that is appropriate for the undeformed description.\footnote{As
was explained in section \ref{distinguished} and as we further
explain at the end of the present subsection, this Kahler
parameter is complex-valued, with no condition of positivity, but
we have to use a different description when its imaginary part is
not positive.} That modular parameter was computed in eqn.
(\ref{roperly}):
\begin{equation}\label{zongor}\hat\tau=i\frac{\varepsilon_2}{\varepsilon_1}
=\frac{\epsilon_2}{\epsilon_1}.\end{equation} So far we have only
deduced that the model is the $A$-model of type $I$ with this
$\hat\tau$ parameter for real $\varepsilon_1, \varepsilon_2$.
However, this must be true in general by holomorphy. This will be
shown more explicitly in appendix \ref{gencompgeom} using
generalized complex geometry.

In particular, the exchange of the $\alpha$ and $\beta$ circles
corresponds to $\hat\tau\to 1/\hat\tau$.  One might expect that
there would be a minus sign in this formula, so let us examine
this point closely.  A simple exchange
$\alpha\leftrightarrow\beta$ (with no minus signs) gives an
orientation-preserving automorphism of the $S^3$ metric
(\ref{pokk}) if accompanied by $w\to\ell-w$.  A look at
(\ref{punn}) shows that to extend this operation to a symmetry of
the $\Omega$-deformed theory, we must take
$\epsilon_1\leftrightarrow\epsilon_2$, again with no minus signs.
To get a symmetry that preserves the overall orientation of
$\R\times S^3_{\varone,\vartwo}$, we take it to act trivially on
$\R$.  We call the combined operation $\cmmib W$.  It acts on
$\hat\tau$ by $\hat\tau\to +1/\hat\tau$.

$\cmmib W$ has an interesting interpretation in the context of
generalized quiver gauge theories. Consider a four-dimensional
$\N=2$ theory obtained as in \cite{DG} by compactifying the
six-dimensional $(0,2)$ theory on a Riemann surface $C$.  Then
compactify further to two dimensions on the $\alpha\beta$ torus
$T^2$.

Altogether, we are compactifying from six to two dimensions on
$T^2\times C$. Introducing an interval $I$ parametrized by $w$,
the full spacetime, away from the boundaries of $I$, is $\R\times
I\times C\times T^2$. If we carry out first the compactification
on $T^2$, we get $\N=4$ super Yang-Mills theory on $\R\times
I\times C$. The orientation-preserving mapping class group of
$T^2$ is $SL(2,\Z)$, and this is usually called the
electric-magnetic duality group in four dimensions.  However, if
we allow diffeomorphisms of $T^2$ that reverse its orientation, we
get an extended mapping class group $GL(2,\Z)$.
Orientation-reversing symmetries of $ T^2$ lead to symmetries of
$\N=4$ super Yang-Mills theory if accompanied by
orientation-reversing symmetries of spacetime.  In the present
case, $\cmmib W$ acts by $\alpha\leftrightarrow\beta$, which
reverses the orientation of $T^2$, and is accompanied by a
transformation $w\to\ell-w$ which reverses the orientation of
$\R\times I\times C$.

We can relate what we have found to a more familiar symmetry
$\hat\tau\to -1/\hat\tau$. Let $\cmmib T$ be the combination
$(\alpha,\beta)\to(-\alpha,\beta)$ together with time-reversal
acting on $\R$ and the identity on $w$.  Then $\cmmib T$ preserves
the orientation of $\R\times S^3_{\varone,\vartwo}$.  It acts on
the deformation parameters by $(\epsilon_1,\epsilon_2)\to
(-\epsilon_1,\epsilon_2)$.  So the combination $\cmmib T \cmmib W$
 acts on the deformation parameters by
$(\epsilon_1,\epsilon_2)\to (\epsilon_2,-\epsilon_1)$, or
$\hat\tau\to -1/\hat\tau$.

In the description by $\N=4$ super Yang-Mills theory on $\R\times
I\times C$, $\cmmib T$ is time-reversal, acting as $-1$ on $\R$
and trivially on $I\times C$.  The composition $\cmmib{TW}$
preserves the orientation of $\R\times I\times C$, and acts on
$\hat\tau$ as a standard electric-magnetic duality transformation
$\hat\tau\to-1/\hat\tau$.

According to \cite{KW}, in the context of compactification from
four to two dimensions on a Riemann surface $C$, the symmetry
$\hat\tau\to -1/\hat\tau$ is the basis for geometric Langlands
duality. The most basic application of this duality to the
geometric Langlands program involves the duality between the
$B$-model of $\M_H$ in complex structure $J$ and the $A$-model of
type $\omega_K$. Section \ref{compom} of the present paper was
based on the $A$-model of $\M_H$ of type $\omega_K$, but its
duality with the $B$-model of type $J$ did not play an important
role.

However, geometric Langlands duality has an extension\footnote{In
the mathematical literature, this extension is often called
quantum geometric Langlands, but we will avoid this terminology
because we interpret also ``classical'' geometric Langlands via
quantum field theory.} involving a complex parameter that was
called $\Psi$ in \cite{KW}. As explained in section 5.2 of
\cite{KW}, extended geometric Langlands can be naturally described
by the $A$-model of $\M_H$ in symplectic structure $\omega_I$,
with the modulus $\hat\tau$ of that $A$-model equal to $\Psi$. The
operation $\cmmib{TW}$ amounts to the extended geometric Langlands
duality $\Psi\to -1/\Psi$.  Thus, the exchange of the two circles,
which certainly will be important in our analysis in the rest of
this paper, is the basis for extended geometric Langlands duality.

In the $A$-model of type $\omega_I$ as usually defined,
$\hat\tau=\Psi$ is a Kahler parameter that takes values in the
upper half of the complex plane. However, according to \cite{KW},
and as we explained in relation to eqn. (\ref{oplo}), in a more
complete description, $\Psi$ actually parametrizes $\Bbb{CP}^1$.
Values of $\Psi$ in the lower half plane can be reached by an
$A$-model of symplectic structure $-\omega_I$.  The cases that
$\Psi$ is real or $\infty$ are more subtle.  The case of real
$\Psi$ can be studied as an $A$-model with symplectic form
$\omega_K$ and a $B$-field proportional to $\omega_I$,  while
$\Psi=\infty$ is the $B$-model of type $J$.

\subsection{More About The Branes}\label{onko}

Now let us discuss in more detail the branes that appear in our
$T^2$ compactification to $\R\times I$.

To arrive at a conventional sigma-model on $\R\times I$, we must
make a $T$-duality on the scalars that arise from gauge field
holonomies on either the $\alpha$ circle or the $\beta$ circle.
Suppose that we make the $T$-duality associated to the $\alpha$
circle.  Then as explained in section \ref{zoth}, the brane
$\B_{\alpha}$ is a rotated version of the brane of type $(A,B,A)$
studied in section \ref{compom}.

If we make the opposite $T$-duality on the $\beta$ circle, the
other brane $\B_{\beta}$ can be described similarly; it is the
analogous space-filling rank 1 brane of type $(A,A,B)$.  By a
$U(1)_R$ chiral rotation (which in general will transform the mass
 and scale parameters of the theory), we can rotate $K$ back to
$J$ and arrive at a description just like the one in the last
paragraph.  Thus, $\B_\beta$ in one description is equivalent to
$\B_\alpha$ in the other description, up to the chiral rotation.

However, our goal is to understand the physical Hilbert space
$\hdef$ of an underlying four-dimensional $\N=2$ theory
compactified on $S^3$ with an $\Omega$-deformation.  From a
two-dimensional viewpoint, $\H$ is the space of
$(\B_\alpha,\B_\beta)$ strings. To describe this space, we need to
describe both branes in the same language.

We have two options.  We can describe $\B_\alpha$  via a
$T$-duality associated to the $\beta$ circle rather than the
$\alpha$ circle. This will turn $\B_\alpha$ into a Lagrangian
brane $\B_{\tN}$ of type $(A,B,A)$ that we will describe.

Or conversely we can describe $\B_\beta$ via a $T$-duality on the
$\alpha$ circle.   In this description, $\B_\beta$ turns into a
Lagrangian brane $\B_{\tN'}$ of type $(A,A,B)$ that can be reached
from $\B_\tN$ by a $U(1)_R$ rotation.

The space $\H$ has two dual descriptions.  It is the space of
$(\B_\tN,\B_\beta)$ strings in the $A$-model of type $\omega_I$
with coupling parameter $\hat\tau=i\varepsilon_1/\varepsilon_2$,
or the space of $(\B_\alpha,\B_{\tN'})$ strings in the ``same''
model but with $\hat\tau=i\varepsilon_2/\varepsilon_1$.

We have put quotes around the word ``same'' because there is
actually a duality involved. The two descriptions differ by the
combined $T$-duality associated to the $\alpha$ and $\beta$
circles (which we recall is the basic geometric Langlands
duality); this is the same as a $T$-duality on the fibers of the
Seiberg-Witten fibration. In the simplest generalized quiver
theories that originate in six dimensions, the Seiberg-Witten
fibration is the Hitchin fibration for Higgs bundles with a
simply-laced gauge group $G$. Under duality on the fibers of the
Hitchin fibration, the structure group $G$ of the Hitchin
fibration is mapped to $G^\vee$. (This was first shown in
\cite{HT}.)  Since $G$ is simply-laced, $G$ and $G^\vee$ have the
same universal cover; if one is $SU(2)$, the other is $SO(3)$.
Still, the distinction between $G$ and $G^\vee$ is significant in
a very precise description.

\begin{table}
\begin{center}
\begin{tabular}{c|c}
Model&Dual\\
\hline
 ${ I_B}$&${ I_B}$\\
${I_A}$&${ I_A}$\\ ${ J_B}$&${ K_A}$\\ ${
J_A}$&${ K_B}$\\ ${ K_B}$&${ J_A}$\\ ${
K_A}$&${ J_B}$\\
\end{tabular}
\end{center}
\begin{caption}\noindent\small{
Listed here are the $A$- and $B$-models to which a given model
transforms under $T$-duality on the fibers of the Hitchin
fibration.  For example, the last row asserts that $K_A$, the
$A$-model of type $\omega_K$, is mapped to $J_B$, the $B$-model of
type $J$.}
\end{caption}
\end{table}

The combined $T$-duality on the $\alpha$ and $\beta$ circles maps
$\B_\alpha$  to $\B_\tN$ and $\B_\beta$ to $\B_{\tN'}$. How it
acts on the supersymmetries is summarized in the table.  For
example, the first row of the table asserts that the $B$-model of
type $I$ -- denoted in the table as $I_B$ -- is mapped to itself
by the $T$-duality on the fibers of the Hitchin fibration. This
reflects the fact that the Hitchin fibration is holomorphic in
complex structure $I$.  Similarly, the $B$-model of type $J$ --
denoted $J_B$ -- is exchanged with the $A$-model of type
$\omega_K$ -- denoted $K_A$.  The table is explained in section 5
of \cite{KW}.

\subsection{A Practice Case}\label{practice}

\def\ww{{v}}
Before discussing the duals of the branes $\B_\alpha$ and
$\B_\beta$, we will first practice with a simpler but still subtle
case. This simpler case is the brane $\B_*$ of type $(B,B,B)$
whose support is all of $\M_H$ and whose Chan-Paton bundle is
trivial, by which we mean in particular that it is flat.

Since the Chan-Paton bundle of $\B_*$ is flat, it is certainly
flat when restricted to each fiber of the Hitchin fibration. So a
fiber of the Hitchin fibration is mapped by the duality to a
point, and hence the $T$-dual $\B_{\tN^*}$ of $\B_*$ is supported
on a section $\tN^*$ of the Hitchin fibration (of the dual group
$G^\vee$). $\tN^*$ is automatically middle-dimensional, and as
$\B_{\tN^*}$ is a brane of type $(B,A,A)$, it is holomorphic in
complex structure $I$ and Lagrangian with respect to
$\Omega_I=\omega_J+i\omega_K$.

The section $\tN^*$ of the Hitchin fibration is actually almost
uniquely determined (up to a choice in what follows of a spin
structure on $C$) by the fact that it is holomorphic in complex
structure $I$.  How to construct such a holomorphic section was
shown in \cite{Hitchin}.  We will describe the construction only
for $G=SU(2)$.  We think of $\M_H$ as a moduli space of stable
Higgs bundles, that is pairs $(E,\varphi)$ where $E$ is a rank two
holomorphic bundle vector over $C$ of trivial determinant, and
$\varphi$ is a holomorphic section of $K_C\otimes
{\mathrm{ad}}(E)$. To describe a holomorphic section of the
Hitchin fibration, we pick a square root $K_C^{1/2}$, and take
$E=K_C^{-1/2}\oplus K_C^{1/2}$.  This is the most unstable bundle
$E$ for which there exists a stable Higgs bundle $(E,\varphi)$.
 Requiring $(E,\varphi)$ to be stable, the most general choice of
$\varphi$ up to an automorphism of $E$ is
\begin{equation}\label{telf}\varphi=\begin{pmatrix} 0 & 1\cr \ww & 0
\end{pmatrix},\end{equation} where $\ww$ is a quadratic differential on $C$ and we regard
$\varphi$ as a matrix acting on $\begin{pmatrix}K_C^{-1/2} \cr
K_C^{1/2}\end{pmatrix}$. For $G=SU(2)$, the fibers of the Hitchin
fibration are parametrized by the value of $\Tr\,\varphi^2$. Since
$\Tr\,\varphi^2=2\ww$, there is one choice of $\ww$ for any
desired value of $\Tr\,\varphi^2$.  Hence this family  of Higgs
bundles $(E,\varphi)$, which we will call $\tN^*$,  gives a
section of the Hitchin fibration. This section  is manifestly
holomorphic in complex structure $I$. To show that $\tN^*$  is
Lagrangian for the holomorphic two-form $\Omega_I$, we use the
explicit formula
\begin{equation}\label{duzobox}\Omega_I=\delta\left(\frac{1}{\pi}\int_C\Tr\phi_z\delta
A_{\bar z}\right).\end{equation} Since $E$, which is characterized
by $A_{\bar z}$, is fixed in the family $\tN^*$, we can take
$\delta A_{\bar z}=0$ when restricted to $\tN^*$.  So a brane
$\B_{\tN^*}$ supported on $\tN^*$ with trivial Chan-Paton bundle
is a brane of type $(B,A,A)$, as expected.

Continuing with the case $G=SU(2)$, $\tN^*$ can be naturally
identified with the Teichmuller space $C$ of the Riemann surface
$C$. (This is proved \cite{Hitchin}
 by showing
that $\tN^*$ can be interpreted as a component of the moduli space
of flat $SL(2,\R)$ bundles on $C$.)  In fact, the symplectic form
$\omega_I$ of $\M_H$, restricted to $\tN^*$, and with the standard
normalization eqn. (\ref{obox}), is the natural Weil-Petersson
symplectic form of Teichmuller space.  Since the coupling
parameter of our $A$-model is
$\hat\tau=i\varepsilon_1/\varepsilon_2$, the symplectic form of
the $A$-model is $\varepsilon_1/\varepsilon_2$ times the
Weil-Petersson form.

The relation of $\omega_I$ to the Weil-Petersson form makes
possible the following construction, described in section 4 of
\cite{GW}. Let $\B'$ be the brane with support all of $\M_H$ and
Chan-Paton curvature
$\omega_I^*=(\varepsilon_1/\varepsilon_2)\omega_I$. So $\B'$ is a
coisotropic brane of type $(B,A,A)$.

Viewing $\B'$ and $\B_{\tN^*}$ as $A$-branes of type $\omega_K$,
let $\H'$ be the space of $(\B',\B_{\tN^*})$ strings.  As reviewed
in section \ref{coiso},  $\H'$ can be interpreted as quantization
of $\tN^*$ with symplectic structure $\omega_I^*$ -- in other
words, quantization of Teichmuller space with symplectic form
$\varepsilon_1/\varepsilon_2$ times the Weil-Petersson form.

Now let us think back to the problem of four-dimensional gauge
theory that has motivated our analysis in this section: how to
relate the space $\hdef$ of supersymmetric ground states of a rank
1 generalized quiver theory on $\R\times\sdef$ to the space of
Virasoro (or Liouville) conformal blocks on $C$.  Teichmuller
space has been quantized \cite{CF} using a real polarization
(which depends on the choice of a set of $A$-cycles on $C$) in  a
way that certainly appears to give a good candidate for the space
of Virasoro conformal blocks. (There is also an analog for groups
of higher rank \cite{FG}.) If therefore we are aiming to get the
space of Virasoro conformal blocks  as the space of ground states
of open strings stretched between two branes, then $\B'$ and
$\B_{\tN^*}$ would appear to be good candidates for those branes.

These are, however, not the candidates that have emerged from our
analysis.  We have found instead that $\hdef$ is the space of
$(\B_\alpha,\B_{\tN'})$ strings (or the space of
$(\B_{\tN},\B_\beta)$ strings). Here $\B_\alpha$ is related to the
more naive candidate $\B'$  by a sort of hyper-Kahler rotation;
one is a rank one coisotropic brane of type $(A,B,A)$, and the
other is a similar object of type $(B,A,A)$. Similarly,
$\B_{\tN'}$ is related to $\B_{\tN^*}$ by a sort of dual
hyper-Kahler rotation. We do not have an intuitive understanding
of why our construction has led to the space of
$(\B_\alpha,\B_{\tN'})$ strings rather than the more obvious
space of $(\B',\B_{\tN^*})$ strings.  Perhaps these spaces are
actually naturally isomorphic.

The space of $(\B',\B_{\tN^*})$ strings, since it describes
quantization of a component of the moduli space of flat $SL(2,\R)$
connections on the Riemann surface $C$, certainly appears to be
related to $SL(2,\R)$ Chern-Simons gauge theory in three
dimensions. The construction also has an analog \cite{GW} that is
similarly related to $SU(2)$ Chern-Simons gauge theory in $2+1$
dimensions. In this analog, $\tN^*$ is replaced by the locus of
flat $SU(2)$ bundles. This locus is characterized by the condition
$\varphi=0$ and is, like $\tN^*$, the support of a Lagrangian
brane of type $(B,A,A)$.

\subsection{The Brane Of Opers}\label{operbrane}

Now we want to discuss the dual of $\B_\alpha$ (or equivalently,
modulo a $U(1)_R$ rotation, the dual of $\B_\beta$).  The
Chan-Paton curvature of $\B_\alpha$ is not zero; rather it equals
$\omega_J^*=(\varepsilon_1/\varepsilon_2)\mathrm{Re}\,\Omega_I$.
However, as the fibers of the Hitchin fibration are Lagrangian for
$\Omega_I$, the Chan-Paton bundle of $\B_\alpha$ is flat when
restricted to a fiber of that fibration.

This means that when restricted to a fiber, the $T$-dual of
$\B_\alpha$ is a point.  Therefore, the $T$-dual of $\B_\alpha$ is
supported on a section of the Hitchin fibration.  We call this
section $ \tN$ and we write $\B_\tN$ for the $T$-dual of
$\B_\alpha$.

$\B_\tN$ is a brane of type $(A,B,A)$, and as its support is
middle-dimensional, it is a Lagrangian brane.  Hence the
Chan-Paton bundle of $\B_\tN$ is flat.  It therefore is actually
trivial, since $\tN$, being a section of the Hitchin fibration, is
equivalent topologically to the base $\cmmib B$ and so is
contractible.

\def\F{{\mathcal F}}
$\tN$ is different from the section $\tN^*$ of the Hitchin
fibration that was described in eqn. (\ref{telf}), since they are
holomorphic in different complex structures.  $\tN^*$ is
holomorphic in complex structure $I$ but $\tN$ is holomorphic in
complex structure $J$.  The Hitchin fibration is holomorphic only
in complex structure $I$, and the fact that this is the complex
structure in which $\tN^*$ is holomorphic makes $\tN^*$ much
simpler to study than $\tN$.

Nevertheless, an explicit description of $\tN$ is known, in
essence, from work on geometric Langlands \cite{BD}.  (In
addition, the dual of $\B_\alpha$ has been described in
four-dimensional gauge theory language in section 4 of \cite{GW2}.
It should be possible to reduce this description to two dimensions
and recover the result of \cite{BD}, though this has not yet been
done.)  Since we do not know a quick path to this description and
will not use the details in the rest of this paper, we will here
simply state the result.

We describe the result only for gauge group $G=SU(2)$.  Also we
assume that $C$ has a negative Euler class (if its genus is 0 or
1, we assume it has at least 3 or 1 marked points, respectively).
If $E\to C$ is a flat bundle of rank 2, it can in particular be
regarded as a holomorphic bundle. A flat bundle is called an
``oper'' if, viewed as a holomorphic bundle, it is a non-trivial
extension
\begin{equation}\label{polyg}0\to K_C^{1/2}\to E\to K_C^{-1/2}\to 0,\end{equation}
where $K_C^{1/2}$ is a square root of the canonical bundle of $C$.
For a given choice of $K_C^{1/2}$, a non-trivial extension of this
kind is unique up to isomorphism. If $E$ is a rank 2 bundle with
flat connection $\A$ that is an oper, then the $(0,1)$ part of
$\A$ has the form
\begin{equation}\label{poldo}\A_{\bar z}=\begin{pmatrix}-a_{\bar z} & 0 \cr u & a_{\bar z}\end{pmatrix},\end{equation}
where $a_{\bar z}$ defines the complex structure of the line
bundle $K_C^{1/2}$, and $u$ is a $K$-valued $(0,1)$-form; the
choice of $u$ does not matter, up to gauge transformation, as long
as its cohomology class in $H^1(C,K_C))\cong\C$ is non-zero. It is
true, though not trivial, that for this  $\A_{\bar z}$, it is
possible to pick $\A_z$ so that the curvature $\F_{z\bar
z}=\partial_z\A_{\bar z}-\partial_{\bar z}\A_z+[\A_z,\A_{\bar z}]$
vanishes. One simple fact is that if we write
\begin{equation}\label{mayo}\A_z=\begin{pmatrix} f& e\cr g &
-f\end{pmatrix},\end{equation} then the upper right matrix element
$e$ is a holomorphic function on $C$, which globally must be
constant. Looking at the diagonal part of the equation $F_{z\bar
z}=0$, we learn that $e$ must be nonzero.

Another simple fact is that given any choice of $\A_z$ that makes
the curvature vanish, any other choice can be obtained by the
shift
\begin{equation}\label{nolyp}\A_z\to \A_z+\begin{pmatrix}0 & 0 \\ w & 0 \end{pmatrix},\end{equation}
where $w$ is a quadratic differential.
 So any two
opers differ by a quadratic differential.  There is actually a
canonical way to map the space of opers to the space of quadratic
differentials, since the flat $SL(2,\R)$ bundle that comes from
uniformization (that is, from the existence on $C$ of an Einstein
metric of constant negative curvature) can be interpreted as an
oper and gives a natural base point in the space of opers.

So for $G=SU(2)$, the space $\tN$ of opers is, as a complex
manifold, naturally isomorphic to the space $H^0(C,K_C^2)$ of
quadratic differentials on $C$. (If there are marked points on
$C$, a similar derivation leads to quadratic differentials that
may have a pole of a specified type at the marked point.) For
generalized quiver theories  \cite{DG} associated to $SU(2)$,
$H^0(C,K_C^2)$ is the same as the base  of the Hitchin fibration,
which as usual we call $\cmmib B$. In general, for quiver gauge
theories based on any $G$, $\tN$ is naturally isomorphic to
$\cmmib B$. (We should point out that there is something strange
about this assertion. The natural complex structure on $\tN$ is
obtained by restricting to $\tN$ the complex structure $J$ on
$\M_H$, while the natural complex structure on $\cmmib B$ is
similarly related to $I$. Nevertheless, $\tN$ with its natural
complex structure is naturally isomorphic to $\cmmib B$ with its
natural complex structure.)

Though we will not really use this information in the present
paper, the reader may find it helpful if we describe how opers are
related to conformal field theory. Again, we consider only the
case of $SU(2)$. First of all, locally it is possible to find a
gauge transformation of lower triangular form $\begin{pmatrix}1 &
0 \\ * & 1
\end{pmatrix}$ setting $u$ to zero and otherwise leaving $\A_{\bar
z}$ in the form (\ref{poldo}). In such a gauge $\A_z$ still has
the form (\ref{mayo}) and $e$ is still a nonzero constant; $f$ and
$g$ are now holomorphic sections of $K_C$ and $K_C^2$,
respectively. Without changing the form of $\A_{\bar z}$, we can
make a gauge transformation by a $2\times 2$ unimodular and
holomorphic matrix to set $e=1$ and $f=0$, whence
\begin{equation}\label{exoc} \A_z=\begin{pmatrix}0 & 1 \\ T &
0\end{pmatrix}\end{equation} where $ T$ is still holomorphic.
 One might think that $T$ would be a
quadratic differential, but actually it is more naturally
understood as a stress tensor or projective connection.  To see
why, consider an infinitesimal gauge transformation generated by
\begin{equation}\label{xoc}\begin{pmatrix} {\partial_z v}/2
& v\\ vT-\partial_z^2 v/2 & -\partial_z
v/2\end{pmatrix},\end{equation} with $v$ a holomorphic vector
field. A short calculation shows that this leaves fixed the form
of $\A_z$, and that $T$ transforms by
\begin{equation}\label{doc} T\to T+v\partial_z T +2(\partial_z v) T
-\frac{1}{2}\partial_z^3 v,\end{equation} in other words as a stress tensor.

\subsection{Physical States From The Brane Of
Opers}\label{physstate}

What we have described so far is the brane $\B_{\tN}$ that is dual
to $\B_\alpha$.  To find the dual to $\B_\beta$, which we call
$\B_{\tN'}$, we simply make a $U(1)_R$ rotation.   Since
$\B_{\tN}$ is a Lagrangian brane of type $(A,B,A)$, the $U(1)_R$
rotation maps $\B_{\tN}$ to  a Lagrangian brane $\B_{\tN'}$ of
type $(A,A,B)$, whose support $\tN'$ is obtained from $\tN$ by the
$U(1)_R$ rotation.   Thus $\tN'$ is a rotated version of the brane
of opers.

Our construction gives two related ways to describe the space
$\hdef$ of ground states of the $\N=2$ gauge theory on $\R\times
\sdef$.  It is the space of $(\B_\alpha,\B_{\tN'})$ strings (in a
description with gauge group $G$, $T$-duality on scalars related
to the $\alpha$ circle, and coupling parameter
$\varepsilon_1/\varepsilon_2$) or the space of $(\B_{\tilde
L},\B_\beta)$ strings (in a description with gauge group $G^\vee$,
$T$-duality on scalars related to the $\beta$ circle, and coupling
parameter $\varepsilon_2/\varepsilon_1$).

Either way, this sounds much like what we studied in section
\ref{compom}: one brane is a canonical coisotropic $A$-brane, and
the other is a Lagrangian $A$-brane.  But there is a crucial
difference. Let us consider first the first description. Here
$\B_\alpha$ is a coisotropic brane of type $(A,B,A)$ with
Chan-Paton curvature  $F=\omega_J$, and $\B_{\tN'}$ is  a
Lagrangian brane of type $(A,A,B)$.  Viewing these as $A$-branes
of type $\omega_I$, we want to describe the space $\hdef$ of
$(\B_\alpha,\B_{\tN'})$ strings.

If the Chan-Paton curvature $F$ were nondegenerate when restricted
to $\tN'$, then $\hdef$ would arise by quantization of $\tN'$ in
symplectic structure $F$; this was reviewed in section
\ref{coiso}.  Here we are in the opposite situation. Since
$F=\omega_J$ and  $\tN'$ is Lagrangian for $\omega_J$, $F$
actually vanishes when restricted to $\tN'$.  This situation
sounds very special, but actually it is the usual situation
considered in geometric Langlands.

\def\I{{\mathcal I}}
In such a case, rather than by quantization, $\hdef$ can be
described as follows. Let $\I=\omega^{-1}F$ be the complex
structure determined by the coisotropic brane $\B_\alpha$.  In the
present case, $\omega=\omega_I$, $F=\omega_J$, and $\I=K$.  Note
that $\tN'$ is a complex submanifold in complex structure $K$.
Then, roughly speaking, the space of physical states is the space
of holomorphic functions on $\tN'$ (actually holomorphic sections
of a certain  line bundle, as we explain momentarily), in complex
structure $K$. Since $\tN'$ is as a complex manifold the same as
the base $\cmmib B$ of the Hitchin fibration, we can think of the
space  $\hdef$ of physical states as the space of holomorphic
sections of a certain line bundle over $\cmmib B$.

To describe the relevant line bundle, observe that because of the
relation of $D$-branes to $K$-theory, by comparing the Chan-Paton
bundles of the two branes in question, one can extract a square
root $K^{1/2}_{\tN'}$ of the canonical bundle of $\tN'$.
$K^{1/2}_{\tN'}$ is a holomorphic line bundle over $\tN'$ that is
trivial, but not canonically so. $\hdef$ is the space of
holomorphic sections of $K^{1/2}_{\tN'}$.

To justify this answer, we simply quantize open strings that
stretch between the two branes $\B_\alpha$ and $\B_{\tN'}$.  Let
the string worldsheet be $\R\times I$ where $\R$ is parametrized
by the time $t$, and $I$ is an interval, with boundary conditions
at the two ends set by the two branes.  To find zero energy
states, we quantize the motion in time of the modes that have zero
kinetic energy along $I$.  We can repeat the derivation in section
2.3 of \cite{GW}, but the result is now different because $F$
vanishes when restricted to $\tN'$, rather than being
nondegenerate.  The bosonic zero modes describe maps
$x:\R\to\tN'$. There is no term first order in $\d x/\d t$,
because $F|_{\tN'}=0$.  So the low energy action  for $x$ is the
usual sort of kinetic energy $\frac{1}{2}\int \d t g_{IJ}\frac{\d
x^I}{\d t}\frac{\d x^J}{\d t}$, where $g_{IJ}$ is the induced
metric on $\tN'$.  There also are fermionic zero modes $\psi^I$,
forming a section of the pullback by $x$ of the tangent bundle of
$\tN'$. To find this, one can follow eqns. (2.8)-(2.11) of
\cite{GW}, with the difference that now $\tN'$ is real with
respect to $J$, rather than holomorphic as assumed in \cite{GW}.
While a key point in \cite{GW} was that there were no fermionic
zero modes, now zero modes $\psi^I$ do survive.  The effective
action for $x^I,\psi^J$ must have two supercharges, descending
from the unbroken supercharges $Q$, $\bar Q$ of the gauge theory.
The minimal supersymmetric action for these fields with this amount of
supersymmetry is familiar:
\begin{equation}\label{zam}I=\int\d t\left(\frac{1}{2}g_{IJ}\frac{\d
x^I}{\d t}\frac{\d x^J}{\d t}+ig_{IJ}\psi^I\frac{D \psi^J}{D
t}\right).\end{equation} This is the basic sigma-model action in
one dimension, with two supercharges when (as here) the target
space is a Kahler manifold.  In quantization, $Q$ becomes the
$\bar\partial$ operator acting on $(0,q)$-forms with values in
$K_{\tN'}^{1/2}$, and $\bar Q$ is its adjoint.  (The sum $\bar Q+
Q$ is the Dirac operator.) The cohomology of $Q$ is thus the
$\bar\partial$ cohomology of $\tN'$ with values in
$K_{\tN'}^{1/2}$.  The cohomology of degree 0 consists of
holomorphic sections of $K_{\tN'}^{1/2}$, and the higher
cohomology vanishes.

So finally, upon identifying $\tN'$ as a complex manifold with
$\cmmib B$, the base of the Hitchin fibration,  the space $\hdef$
of physical states can be identified as the space of holomorphic
sections of $K_{\cmmib B}^{1/2}$ over $\cmmib B$:
\begin{equation}\label{otung}\hdef=H^0({\cmmib B},K^{1/2}_{\cmmib
B}).\end{equation} There is a slight surprise here: the right hand
side does not depend on the parameters
$\varepsilon_1,\varepsilon_2$.   However, the action of
observables on $\hdef$ does depend on these parameters, as we will
see shortly.

We obtained this result starting from the interpretation of
$\hdef$ as the space of $(\B_\alpha,\B_{\tN'})$ strings.  The
alternative realization of $\hdef$ as the space of
$(\B_{\tN},\B_\beta)$ strings would lead after analogous steps to
a different identification of $\hdef$ with $H^0({\cmmib
B},K^{1/2}_{\cmmib B})$.

As in somewhat similar situations considered in section 2.3 of
\cite{GW}, the gauge theory on $\R\times S^3_{\varone,\vartwo}$,
or the two-dimensional sigma-model on $\R\times I$ to which it
reduces, will generate a hermitian inner product on
$\hdef=H^0({\cmmib B},K^{1/2}_{\cmmib B})$ making this space a
Hilbert space.  But, except in the limit of large
$\varepsilon_1/\varepsilon_2$, when the Kahler modulus of $\M_H$
becomes large and the two-dimensional $\sigma$-model can be
treated semiclassically, this hermitian inner product is not
necessarily given by any elementary classical formula.

\subsection{Observables}\label{oddobservables}

Now we want to describe the algebra of observables that acts on
the space $\hdef$ of ground states of the gauge theory on $\sdef$.
In the brane description, $\hdef$ is the space of
$(\B_\alpha,\B_{\tN})$ strings or alternatively the space of
$(\B_{\tN'},\B_\beta)$ strings. The first description makes it
clear that the algebra of $(\B_\alpha,\B_\alpha)$ strings acts on
$\hdef$ by attaching to the left end of a string. The second
description makes it equally clear that the algebra of
$(\B_\beta,\B_\beta)$ strings acts on $\hdef$ by attaching to the
right end of a string.  The strings in question are the physical
states in the $A$-model of type $\omega_I$.

The two algebras of $(\B_\alpha,\B_\alpha)$ strings and
$(\B_\beta,\B_\beta)$ strings are equivalent, up to the usual
steps (a  $U(1)_R$ rotation that may act on mass parameters, an
exchange $\varepsilon_1\leftrightarrow\varepsilon_2$, and duality
$G\leftrightarrow G^\vee$).  So let us just describe the algebra
of $(\B_\beta,\B_\beta)$ strings.

The brane $\B_\beta$ has Chan-Paton curvature
$F=(\varepsilon_1/\varepsilon_2)\omega_K$, and we study it in the
$A$-model of symplectic structure
$\omega=(\varepsilon_1/\varepsilon_2)\omega_I$. The
$(\B_\beta,\B_\beta)$ strings are obtained by deformation
quantization of the ring $\RR$ of holomorphic functions in complex
structure $\I=\omega^{-1}F=J$. To proceed farther, we focus on the
case of a  generalized quiver theory associated to a Riemann
surface $C$. The two-dimensional description involves a
sigma-model with target $\M_H$, and in complex structure $J$,
$\M_H$ is the moduli space of $\frak g_\C$-valued flat connections
$\A$ on $C$.  The ring of holomorphic functions on $\M_H$ is
generated by traces of holonomies.\footnote{This fact is not
obvious but is proved for classical groups in \cite{AMR}, with the
understanding that by ``holomorphic functions'' on $\M_H$, we
really mean algebraic functions on $\M_H$, viewed as a moduli
space of representations of the fundamental group of the Riemann
surface $C$.  Despite the fact that traces of holonomies suffice
to generate the ring, it may be more natural to consider also the
functions  associated to labeled graphs, as for example in
\cite{Witten3}.} In other words, for $\gamma\subset C$ a simple
closed curve and $R$ a representation of $G$, we consider the
function
\begin{equation}\label{lovely}
W_R(\gamma)=\Tr_R\,P\exp\left(-\oint_\gamma\A\right);\end{equation}
 such
functions generate $\RR$.

The deformation of the commutative ring $\RR$ to a noncommutative
but still associative ring depends on one complex parameter,
usually called $q$. This deformation, which is constructed in
\cite{AMR2} (of course there are also many related constructions
from different points of view), appears in two familiar quantum
field theory problems.  One is the problem\cite{Witten2} of
quantizing Chern-Simons gauge theory with compact gauge group $G$
on $\R\times C$, where $C$ is a Riemann surface.  The phase space
is $\M$, the moduli space of flat unitary (that is $\frak
g$-valued) connections $A$ on $C$.  The basic operators on the
physical Hilbert space are Wilson loop operators.  These are the
same traces of holonomies considered in the last paragraph, except
that they are functions on $\M$, the space of flat $\frak
g$-valued connections, rather than its complexification $\M_H$.
However, traces of holonomies regarded as functions on $\M$ have
natural analytic continuations to holomorphic functions on $\M_H$.
(The analytic continuations are given by the same traces, now
evaluated for flat connections that may be complex-valued.) So the
ring of classical observables of Chern-Simons theory with compact
gauge group $G$ is actually the same as the ring $\RR$ of complex
holonomies considered in the last paragraph. In quantizing
Chern-Simons gauge theory with gauge group $G$, $\RR$ is deformed
to a noncommutative, associative algebra of quantum Wilson loop
operators.  The deformation parameter is the Chern-Simons level
$k$, in terms of which one defines $q=\exp(2\pi i/(k+h))$, where
$h$ is the dual Coxeter number of $G$.  A physical Hilbert space
on which the quantum algebra can act exists only for positive
integer $k$. But the deformed algebra $\RR_q$ of Wilson loop
operators can be constructed as a function of a complex variable
$q$.  The procedure involved can be understood as deformation
quantization (for example, see \cite{Andersen} for this
interpretation). But the situation is much better than a typical
example of deformation quantization: this is a favorable case in
which deformation quantization gives a deformation parametrized by
a complex variable $q$, not just a deformation over a formal power
series ring.

The second related problem in which one encounters the deformed
ring $\RR_q$ of holonomy functions is two-dimensional conformal
field theory. The most basic case is current algebra of a compact
group $G$. Here a monodromy operation on the space of conformal
blocks was introduced in \cite{EVerlinde} (and this influenced
subsequent work on Chern-Simons theory). The monodromy operation
is defined by transporting a primary field in a representation $R$
of $G$ around a loop $\gamma\subset C$. In the correspondence
between two-dimensional conformal field theory and
three-dimensional Chern-Simons theory, the space of conformal
blocks  maps to the physical Hilbert space of Chern-Simons theory,
and the conformal field theory monodromies map to the action of
Wilson loop operators of gauge theory.

The conformal field theory operation just mentioned has been
adapted to Liouville theory in \cite{AGGTV,DGOT}.  The deformation
parameter is the Liouville coupling $b$, which corresponds to the
 Chern-Simons level $k$ if Liouville theory is related to
$SL(2,\R)$ current algebra.   So the same associative algebra
$\RR_q$ of quantized holonomies that can be extracted from $SU(2)$
Chern-Simons theory acts on the conformal blocks of Liouville
theory. This algebra is also seen in quantization of Teichmuller
space.  For gauge groups of higher rank, there is a similar
relation between the quantum-deformed algebra of holonomies in
Chern-Simons gauge theory with compact gauge group, and the
deformed algebra of holonomies that acts in the quantization of
the higher rank analogs of Teichmuller space  \cite{CF,FG}.

There is, however, a fundamental difference between the case of a
compact symmetry group such as $SU(2)$ and the case of a
noncompact group such as $SL(2,\R)$. In the case of a compact
symmetry group or gauge group $G$, the deformed algebra $\RR_q$ of
holonomies acts irreducibly on the space of conformal blocks of
two-dimensional conformal field theory, or equivalently the
Hilbert space of Chern-Simons theory. For Liouville theory or
$SL(2,\R)$ Chern-Simons theory, this is far from being true.
Instead, in its action on the space of Virasoro conformal blocks,
the algebra $\RR_q$ commutes with a dual algebra $\RR_{q'}$.
$\RR_{q'}$ is a second deformed algebra of holonomy operators,
with its parameter differing by the Liouville duality
$b\leftrightarrow b^{-1}$.

In our presentation here, the action of the second commuting
algebra is manifest, since the $(\B_\alpha,\B_\alpha)$ strings
acting at one end commute with the $(\B_\beta,\B_\beta)$ strings
acting at the other end.

The space of Virasoro conformal blocks can be characterized as an
irreducible module for two algebras  $\RR_q$ and $\RR_{q'}$ of
$SL(2,\R)$ holonomies at dual values of the parameters.  The two
algebras are noncommutative but associative, and commute with each
other.  The fact that we have found this structure strongly supports the idea that
the Hilbert space $\hdef$ of the generalized quiver theory on
$\sdef$ is indeed the space of Virasoro (or Liouville) conformal
blocks, as first argued in \cite{AGT}.

\subsubsection{Wilson And 't Hooft Operators}\label{wth}

Concretely, the algebras $\RR_q$ and $\RR_{q'}$ are generated by
$Q$-invariant Wilson and 't Hooft operators that act at one end of
$S^3_{\varone,\vartwo}$ or the other.  Let us return to the
explicit description (\ref{pokk}) of $S^3_{\varone,\vartwo}$. When
we toroidally reduce $S^3_{\varone,\vartwo}$ to an interval $I$
parametrized by $w$ with $0\leq w\leq\ell$, the end at $w=0$ is
really a circle $S^1_\beta$ parametrized by $\beta$, and similarly
the end at $w=\ell$ is a circle $S^1_\alpha$ parametrized by
$\alpha$.   The rings $\RR_q$ and $\RR_{q'}$ are generated by
supersymmetric loop operators wrapped on $S^1_\beta$ or
$S^1_\alpha$ at a fixed time.

In generalized quiver gauge theories associated to a Riemann
surface $C$, half-BPS Wilson and 't Hooft operators are in
one-to-one correspondence \cite{DMO} with data of the following
kind: a choice of a homotopy class of simple closed loop
$\gamma\subset C$ together with a choice of a representation $R$
of $G$.  Of course, this construction is usually made in
undeformed super Yang-Mills theory.  However, near either end of
$S^3_{\varone,\vartwo}$, the vector field that we have used to
make the $\Omega$-deformation reduces to either
$\epsilon_2\partial/\partial \beta$ (at $w=0$) or
$\epsilon_1\partial/\partial\alpha$ (at $w=\ell$).  So we can
rotate from the ordinary theory to the $\Omega$-deformed theory as
in section \ref{rethink}, and hence the half-BPS operators studied
in \cite{DMO} have analogs in our situation. The holonomy
functions (\ref{lovely}) that generate $\RR_q$ and $\RR_q'$ simply
come from the corresponding half-BPS loop operators.

It was already observed in \cite{AGT} that half-BPS loop operators
can be wrapped on the circles $S^1_\alpha$ and $S^1_\beta$ and
that in the $\Omega$-deformed theory, these operators do not
commute.  What we have contributed is to relate this fact to the
noncommutativity that arises in two-dimensional $\sigma$-models
with a $B$-field, and to formulate the problem in a way that is
closer to other occurrences of the noncommutative ring $\RR_q$ in
mathematical physics.

\subsubsection{Winding States Of The $A$-Model}\label{winda}

We now make a slight digression, aiming to spare the reader some
puzzlement by briefly answering the following question. Given the
$T$-duality between $\B_\beta$ and $\B_\tN$, the space of
$(\B_\beta,\B_\beta)$ strings must be equivalent to the space of
$(\B_\tN,\B_\tN)$ strings.  But $\B_\tN$ is an ordinary Lagrangian
$A$-brane, so the space of $(\B_\tN,\B_\tN)$ strings in the
$A$-model, as usually understood, is simply the cohomology of
$\tN$.  How can we possibly identify the space of
$(\B_\tN,\B_\tN)$ strings with a ring of holonomy functions?

The answer to this question is that it is necessary to take into
account something that is usually  not relevant -- winding states
in the $A$-model.  As a prototype of the problem, we consider a
sigma-model with target $W=\R\times S^1$, with the obvious product
metric and with coordinates $t$, $\theta$ on $\R$ and $S^1$. First
we consider the $B$-model on $W$, in the obvious complex structure
in which $t+i\theta$ is holomorphic.  Let $\B$ be a $B$-brane
whose support is all of $W$, with trivial Chan-Paton bundle. The
$(\B,\B)$ strings of zero ghost number are associated to the
holomorphic functions
\begin{equation}f_n=\exp(n(t+i\theta)),~~n\in\Z.\end{equation}
We note that apart from $f_0$, which is the constant function 1,
all other $f_n$ have non-zero momentum around $S^1$, and
exponential growth along $\R$ in one direction or the other.

Now we perform $T$-duality along the $S^1$ direction.  This maps
the $B$-model of $W$ to an $A$-model of $\tilde W=\R\times \tilde
S^1$, where $\tilde S^1$ is the dual circle.  The brane $\B$ is
mapped to a Lagrangian $A$-brane $\B'$ that is supported on
$\R\times p$, with $p$ a point in $\tilde S^1$.  The $(\B,\B)$
string corresponding to the identity function $f_0$ maps to the
$(\B',\B')$ string that corresponds to the zero-dimensional
cohomology of $\R\times p$.  What about the $f_n$ with $n\not=0$?
As they carry momentum along $S^1$, they correspond to $A$-model
states that have winding around $\tilde S^1$. Of course, these
states also have exponential growth along $\R$, since this
property is unaffected by $T$-duality.  The reason that such
$A$-model winding states are unfamiliar is that we do not usually
study $A$-model states with exponential growth.

To see that this example really is a prototype for the original
question, note that the product of two copies of $W$ is
$X=\R^2\times T^2$, which we can think of as the moduli space of
vacua for the familiar example of a free vector multiplet. The
product of two copies of the brane $\B$ considered above is the
space-filling brane $\B_*$ of type $(B,B,B)$ on $X$.  Turning on a
Chan-Paton curvature proportional to $\omega_J$, this can be
deformed to a coisotropic $A$-brane of type $(A,B,A)$, without
changing the essentials of the above discussion.  Thus, even for a
free vector multiplet, we have to go beyond the usual class of
$A$-model states to see the duality between the spaces of
$(\B_\beta,\B_\beta)$ strings and of $(\B_\tN,\B_\tN)$ strings.

\subsection{Partition Functions}\label{partfns}

The formulation in  \cite{AGT} was actually slightly different
from what we have given here and focused on partition functions.
The main claims there were that the partition function of a
generalized quiver  theory on $\rdef$ gives a chiral conformal
block in Liouville theory, and that the partition function of such
a theory on $\ssdef$  (at least for $\epsilon_1=\epsilon_2$) gives
the modular-invariant partition function of Liouville theory with
left- and right-movers included.

It makes sense to compare the various spaces that are involved
here, because $\R\times S^3$ is equivalent topologically to $\R^4$
with a point at the origin omitted, while $S^4$ can be viewed as
$\R^4$ with a point at infinity added. Comparing the three spaces
in this way, the $U(1)\times U(1)$ action that we have used to
make an $\Omega$-deformation on $\R\times S^3$ can be extended
over $\R^4$ or $S^4$.  Then we make the $\Omega$-deformation on
all three spaces with the same parameters
$\varepsilon_1,\varepsilon_2$, and we refer to the result as a
generalized quiver theory on $\R\times \sdef$, $\rdef$, or
$\ssdef$.

In general, in any four-dimensional quantum field theory, let  $B$ be  a four-manifold with boundary $S^3$.
Then the path
integral on $B$ gives a physical state in the Hilbert space associated to $S^3$.
In the context of topologically twisted four-dimensional gauge theory, we have to make
a choice, as was remarked in section \ref{gl}: on $\R\times S^3$, topological twisting conserves
 two supercharges,
$Q$ and its adjoint $\bar Q$, but only one can be conserved on a more general four-manifold
$B$.  Let us make a choice and conserve $Q$.
If in addition the $U(1)\times U(1)$ action of $S^3$ extends over $B$, we can make the
  $\Omega$-deformation on $B$ and in that case,
the path integral on $B$ will give a vector in the $Q$-cohomology
of the $\Omega$-deformed theory on $S^3$, in other words, a
Virasoro (or Liouville) conformal block.

The most simple choice of $B$ that has the right properties is a
four-dimensional ball $B^4$, as explained at the end of section
\ref{gl}.  Writing $\bdef$ for the $\Omega$-deformed version of
$B^4$, the path integral over $\bdef$ will give a Virasoro
conformal block $\mathcal W$. Observing that $\bdef$ is the same
topologically as $\rdef$, and that this equivalence is compatible
with the $U(1)\times U(1)$ action, what we have just said is
equivalent to the ``chiral'' version of the claim in \cite{AGT}.

For the non-chiral version of their claim, we want to glue
together two copies of $\bdef$, with opposite orientation, along
their common boundary $\sdef$.  If we preserve the same
topological supercharge in both copies of $\bdef$, we will get
Donaldson theory on $\ssdef$.  This is not what we want.  Instead,
to get the claim of \cite{AGT}, we must preserve one supercharge
$Q$ on one copy of $\bdef$ and its conjugate $\bar Q$ on the other
copy.  Then, the partition function on $\ssdef$ gives the norm
squared of the Virasoro conformal block $\mathcal W$, in the
Hilbert space $\hdef$.

In general, it is not clear how to make opposite topological
twists in the two hemispheres of $\ssdef$.  For  $\epsilon_1=\epsilon_2=1/R$, it has
been shown \cite{Pestun1} that the $\Omega$-deformed theory with opposite twists
on the two sides is equivalent to
physical Yang-Mills theory on a four-sphere of radius $R$,
with  no $\Omega$-deformation at all.   This fact was exploited in \cite{AGT}. We hope
that this fact can be adapted for our derivation and  has a useful generalization to other values of the
deformation parameters.   Until this is found, we may fall back on
the approach of \cite{CV} in two dimensions.  In that approach,
between the two copies of $\bdef$, one places a very long cylinder
$I\times \sdef$.  In the limit that the length of $I$ becomes much
greater than the radius of $\sdef$, the path integral on
$I\times\sdef$ projects onto quantum ground states, and the path
integral on $\ssdef$ can be evaluated in the space $\hdef$ of
quantum ground states even though we make opposite topological
twists at the two ends.

\subsection{Flatness}\label{flatness}

One basic fact about conformal field theory
on a Riemann surface $C$ is that the space of conformal blocks is locally independent
of the complex structure of $C$.  There is a projectively flat connection that can be used
to transport the space of conformal blocks as the complex structure of $C$ varies.
This holds for  theories with finite-dimensional spaces of conformal
blocks as well as for more sophisticated theories such as Liouville theory.  See \cite{FS} for more
discussion.

In our present context, this implies that although the definition
of the variety $\tN$ of opers depends on the complex structure of
$C$, the space $H^0({\tN},K^{1/2}_{\tN})$ does not.  More
precisely, the bundle of Hilbert spaces over the moduli space of
complex structures on $C$ whose fiber is $H^0({\tN},K^{1/2}_{\tN})$ should admit a projectively flat connection.   In principle, the $\sigma$-model must generate this connection
but we do not know what sort of semiclassical formula can be given for it.

One may ask whether the underlying $(0,2)$ model in six dimensions predicts the existence of
this flat connection.  The $R$-symmetry group of this theory is $SO(5)_R$.  There is no non-trivial
homomorphism from $SO(6)$ to $SO(5)$, so the $(0,2)$ model does not have a twisted version
that would lead to a topological field theory in six dimensions.  However, suppose that
we specialize to six-manifolds of the form $\R\times W$, with a product
metric; here $W$ is a five-manifold.  The structure group of the tangent bundle of $W$ is
$SO(5)$, which of course does admit an isomorphism with $SO(5)_R$.
So a twisted theory can be constructed for six-manifolds of the from $\R\times W$.  This theory
has a supercharge $Q$ obeying $Q^2=0$, and it is plausible that the cohomology of $Q$
is locally independent of the metric of $W$.

This is close to what we want.  What we have been studying is an
$\Omega$-deformed version of the cohomology of the twisted theory
just described for the case $W=S^3\times C$. Of course, we did not
introduce the subject in precisely this way.  Rather we started
with a four-dimensional generalized quiver theory, which can be
obtained from the $(0,2)$ theory in six dimensions by
compactifying on $C$ with a topological twist that preserves
supersymmetry. Then we compactified on $S^3$, again with a
topological twist that preserves supersymmetry. The net effect was
to compactify from six dimensions to one dimension on $W=S^3\times
C$ with a topological twist that preserves supersymmetry.  There
is essentially only one way to do this -- the twist mentioned in
the last paragraph that preserves five-dimensional symmetry.

This seems promising.  The only catch is that it is not clear how
to include the $\Omega$-deformation in this analysis, since one
does not have a good understanding of the $\Omega$-deformation in
six-dimensional terms.   To explain the existence of a flat
connection on the bundle of Hilbert spaces $\hdef$ over the
Teichmuller space of $C$, we obviously need an argument that takes
the $\Omega$-deformation into account.

\subsection{Including Surface Operators}\label{surfop}

The correspondence between gauge theory and two-dimensional
conformal field theory becomes richer \cite{SurfOp,SurfOp2} if
surface operators are included.  Here we consider a half-BPS
surface operator inserted at one of the two ends of
$S^3_{\varone,\vartwo}$.  Near the support of the surface
operator, the spacetime looks like $\Sigma\times D$ where $\Sigma$
is a two-manifold and $D$ is the usual cigar geometry.  The
surface operator is inserted at $\Sigma\times p$, where $p$ is the
tip of the cigar.  In our applications, $\Sigma=\R\times S^1$.
 The $\Omega$
deformation is made using a vector field that generates the
rotation of $D$ around $p$ (section \ref{compom}) or one that also
acts at the same time by rotation of $S^1$ (the present section).
Far from the tip, $D$ looks like a cylinder $\R\times \tilde S^1$.

A half-BPS surface operator inserted on $\Sigma\times p$ preserves
all supersymmetry that is compatible with the $\Omega$-deformation
(or with the curvature of $D$), so it can be included naturally in
our analysis.  First we consider the abelian case $G=U(1)$.

We consider the simplest half-BPS surface operator.  It is defined
by introducing a Lie algebra valued parameter $\alpha$ and
requiring that the curvature have a delta-function singularity
$F/2\pi \sim \alpha\delta_{\Sigma\times p}$. Equivalently, the
holonomy around the singularity is $\exp(2\pi i\alpha)$.

Now let us consider the two-dimensional sigma-model, with target
$\M_H$, that we get by compactification on $S^1\times \tilde S^1$.
The tip of $D$ together with the surface operator will give a
brane in this model.  In the absence of the surface operator, we
observed that supersymmetry requires that the curvature along $D$
vanishes.  Hence the holonomy around $\tilde S^1$ is trivial, and
the circle-valued field that we called $A_3$ in section
\ref{compom} vanishes at the boundary. To get a more useful
description, in section \ref{otto}, we $T$-dualized this field to
another angle-valued field $\acalpha$. Vanishing of $A_3$ meant
that $\acalpha$ obeys Neumann boundary conditions. All other
bosonic fields in the sigma-model obey Neumann boundary conditions
for more obvious reasons.  So altogether, the brane $\B$ that
comes from the tip of the cigar has all of $\M_H$ for its support.
Of course, this is familiar from section \ref{otto}.

\def\N{{\mathcal N}}
 Including the surface operator changes this analysis in only
one way.  Vanishing of curvature along $D$ now requires that the
holonomy around $\tilde S^1$ should equal $\exp(2\pi i\alpha)$
(rather than 1) or equivalently that the boundary value of $A_3$
is equal to $\alpha$.  After $T$-duality, the dual field
$\acalpha$ still obeys Neumann boundary conditions, but the circle
that it parametrizes is equipped with a flat Chan-Paton line
bundle $\mathcal S$.  The sole effect of the surface operator on
the brane $\B$ that comes from the tip of the cigar is that the
Chan-Paton bundle of that brane is tensored by $\mathcal S$. There
is no problem in describing $\S$ explicitly as a line bundle over
$\M_H$.  For $G=U(1)$, $\M_H=\R^2\times T^2$ and $\S$ is simply a
flat line bundle with holonomy around one of the directions in
$T^2$.  As $\S$ is flat, tensoring with $\S$ preserves all
supersymmetry.

Now let us consider the nonabelian case.  We will consider the
case of a generalized quiver theory \cite{DG} based on $SU(2)$,
but we start with the case of a single $SU(2)$ gauge group. The
parameter $\alpha$ now takes values in the Lie algebra of a
maximal torus of $SU(2)$. There is really only one major change in
the above analysis.  To describe it, let us work over a generic
part of the base $\cmmib B$ of the Hitchin fibration where $SU(2)$
is broken to an abelian subgroup, isomorphic to $U(1)$. Then the
monodromy around $\tilde S^1$ is again $U(1)$-valued, computed
from the zero mode of an effective scalar field that we will again
call $A_3$. And $A_3$ is again $T$-dualized to another scalar
$\acalpha$ to go to a more useful description.  There is only one
major difference from the case of $G=U(1)$.  Vanishing of the
curvature along $D$, instead of requiring $A_3$ and $\alpha$ to be
equal (as circle-valued objects), now requires only that they
should be equal up to a Weyl transformation.  Thus (for
$G=SU(2)$), there are now two allowed boundary values of $A_3$,
namely $A_3=\pm\alpha$.

\def\U{{\mathcal U}}
After $T$-duality, each of the allowed boundary values of $A_3$
maps to a flat line bundle over the circle parametrized by
$\varrho$.  As $A_3$ has two allowed boundary values, this gives a
rank two flat bundle over the circle.  As the other fields vary,
this will give a rank two bundle $\U\to\M_H$ (at least in the
region where our approximations apply). The effect of the surface
operator is to tensor the Chan-Paton bundle of the brane $\B$ that
comes from the tip of the cigar with $\U$.

Our description so far has been based on an approximation valid
far from the discriminant locus in $\cmmib B$.  In this
approximation, $\U$ is a flat vector bundle.  It is not possible
for this to be the whole story, since the total space of the
Seiberg-Witten or Hitchin fibration for any gauge theory with a
semi-simple gauge group has a finite fundamental group. The most
optimistic hypothesis is that the full description of the surface
operator involves a rank two bundle $\U\to\M_H$ that is flat far
from the discriminant locus and in general preserves all
supersymmetry.  We will assume this to be the case, though it is
also imaginable that something more complicated happens near the
discriminant locus in $\cmmib B$.

Before making a proposal for a more concrete description of $\U$,
we will consider $SU(2)$ generalized quiver theories that arise
from six dimensions by compactification on a Riemann surface $C$.
The gauge group is a product of $SU(2)$ factors.  According to
\cite{SurfOp}, in such a theory, there is a natural family of
surface operators parametrized by the choice of a point $q\in C$.
In suitable local descriptions, an operator in this family can be
described precisely as above with the relevant part of the gauge
group being simply $SU(2)$.  Hence the above analysis applies, and
the insertion at the tip of the cigar of a surface operator
associated to a point $q\in C$ will generate a rank two bundle
$\U_q\to \M_H$.

As $\U_q$ varies in $C$, the $\U_q$ will fit together as fibers of
a rank two holomorphic vector bundle $\U\to\M_H\times C$.  There
is an obvious candidate for what this bundle may be: the universal
bundle, or more exactly, the bundle part of the universal Higgs
bundle. In other words, every point $r\in \M_H$ parametrizes a
solution of Hitchin's equations -- a holomorphic bundle $E\to C$
together with a Higgs field $\varphi\in
H^0(C,{\mathrm{ad}}(E)\otimes K_C)$ and a hermitian metric such
that Hitchin's equations are satisfied. (Here we emphasized a
holomorphic point of view, but of course there are many equivalent
descriptions of this data.) The universal Higgs bundle is a triple
consisting of a holomorphic bundle ${\mathcal E}\to\M_H\times C$,
 a Higgs field $\hat\varphi:{\mathcal E}\to {\mathcal E}\otimes K_C$,
and a hermitian metric on ${\mathcal E}$, such that the
restriction of this data to $r\times C$, for any $r\in \M_H$, is
the solution of Hitchin's equations corresponding to $r$. (For an
elementary description of universal bundles, including subtleties
involving the center of the gauge group, see section 7.1 of
\cite{KW}.)  By restricting $\mathcal E$ to $\M_H\times q$, we get
a very plausible candidate for $\U_q$.

In the absence of the $\Omega$ deformation, the condition that
tensoring with $\U_q$  preserves all supersymmetry means that
$\U_q$ must be  holomorphic in every complex structure on $\M_H$.
This follows from standard properties of the universal Higgs
bundle. Upon making the $\Omega$ deformation, further properties
are needed, and they are not well-understood, since the theory of
coisotropic $A$-branes of rank greater than 1 has not been
developed. Possibly the necessary properties follow from the
existence and properties of the universal Higgs field
$\hat\varphi$.

\vskip 1 cm

\noindent{\it Acknowledgments}
Part of this work has been done while NN visited the  IAS and while
EW visited the IHES and the SCGP. We thank these institutes for their hospitality.
Research of NN was partly supported by  {\it l'Agence Nationale de la
Recherche} under  grants
ANR-06-BLAN-3$\_$137168 and ANR-05-BLAN-0029-01, and by the Russian
Foundation for Basic Research through the grants
RFBR N 09-02-00393 and NSh-3036.2008.2.
Research of EW was partly supported by
NSF Grant Phy-0503584.  We thank R. Donagi, E. Frenkel, D.
Gaiotto, S. Gukov, G. Moore, S. Shatashvili,  J. Teschner, and Y.
Tachikawa for comments and discussions.

\appendix

\section{The General $\Omega$-Deformation In Toroidal Compactification }\label{gencompgeom}
\def\IT{T_{\circ}}
\def\d{{\mathrm d}}
\def\t{{\theta}}
\def\a{{\alpha}}
\def\om{\omega}
\def\im{{\mathrm {Im}}}
\def\vt{{\vartheta}}
\def\re{{\mathrm{Re}}}
\def\ve{{\epsilon}}
\def\p{\partial}

\subsection{The Setup}
The purpose of this appendix is to generalize various computations
involving compactification on a two-torus which in the text were
done in special cases.  Here we will allow an arbitrary flat
metric on the two-torus, not necessarily rectangular, and we allow
arbitrary $\epsilon$ parameters for the $\Omega$-deformation.
Since we focus on the two-torus, we will simplify somewhat the
notation used in the text.

We denote the torus as $\IT$ and endow it with a constant metric
$G_{IJ}$, $I,J=1,2$:
 \begin{equation}\label{dstwo}\d s_{{\IT}}^{2}
= G_{ij} \d{\t}^{i} \d{\t}^{j}, \qquad {\t}^{i} \sim {\t}^{i} +
2\pi .\end{equation} The gauge theory probes the dual torus
${\IT}^{\vee}$, the moduli space of flat $U(1)$-connections on
${\IT}$.  We write such a connection as
\begin{equation}\label{flt}A = i{\a}_{1} \d{\t}^{1} + i{\a}_{2}
\d{\t}^{2},\end{equation} with constant hermitian matrices $\a_1$,
$\a_2$ (our connection form $A$ is anti-hermitian).  The gauge
transformations generated by the $U(1)$-valued functions
\begin{equation}\label{uone}u_{n_{1}, n_{2}} = \exp\left({i n_1 {\t}^1 + i
n_2 {\t}^2}\right)\end{equation} shift the components ${\a}_{1,2}$ by
$n_{1,2}$, respectively. The natural metric on ${\IT}^{\vee}$ is
given by:
\begin{equation}\label{natmet}\d s^2_{{\IT}^{\vee}} = \frac{1}{
(2{\pi}i )^2} \int_{{\IT}} {\d} A \wedge \star {\d} A =
\sqrt{\det(G)} G^{ij} \d{\a}_{i}\d{\a}_{j}. \end{equation}  It
depends only on the complex structure of $\IT$.

It is convenient to parametrize $G^{ij}$ by two complex numbers
${\om}_{1,2}$, \begin{equation}\label{invme} G^{ij}\d{\a}_{i}
\d{\a}_{j} = \vert {\om}_{1} \d{\a}_{1} + {\om}_{2} \d{\a}_{2}
\vert^2\end{equation} defined up to a simultaneous phase rotation,
so that the invariants are:
\begin{align}\label{mtrx}\notag \vert {\om}_{1} \vert^2& =
G^{11}, \\ \notag \vert {\om}_{2} \vert^2& = G^{22}, \\  {\mathrm{Re}}(
{\om}_{1}{\bar\om}_{2} )& = G^{12}.\end{align} Let us assume
$$
{\im}\left( \frac{{\om}_{2} }{ {\om}_{1}} \right) > 0
$$
We then have: \begin{equation}\label{durf}\sqrt{\det (G)} =
\frac{1}{\im \,({\om}_{2}{\bar\om}_{1})}.\end{equation} For a
rectangular torus,
$$
{\om}_{1} = \frac{1}{ {\rho}_{1}}, \qquad
{\om}_{2} = \frac{i }{ {\rho}_{2}}
$$
Let us first consider the case of gauge group $U(1)$.  We take the
Maxwell action to be
\begin{equation}\label{lmax}I = \frac{1}{ 8\pi} \int_{X^{4}} \d^{4}x
\sqrt{g} \,  \left( \frac{4\pi }{ {\rm g}_{4}^{2}} F_{mn}F^{mn} +
\frac{i \vt }{ 4\pi} {\ve}_{mnpq} F^{mn}F^{pq}
\right).\end{equation} If we take the four-manifold to be
${\Sigma} \times {\IT}$, with the product metric  $h \times G$,
with $h$ being the metric on $\Sigma$, and denote the Riemannian
measure of $\Sigma$ as $\d\mu$, then, in the low-energy
approximation, (\ref{lmax}) reads as:
\begin{align}\label{lmaxi} I \ = \notag& \frac{(2\pi )^2 }{ 8\pi}
\sqrt{\det(G)} \int_{\Sigma} \d\mu\, \left(  \frac{8\pi }{ {\rm
g}_{4}^{2}} h^{ab}G^{ij} {} \left( {\p}_{a}A_{i} {\p}_{b}A_{j}
\right) - \frac{i{\vt} }{ \pi } {\ve}^{ab}{\ve}^{ij} {} \left(
{\p}_{a}A_{i} {\p}_{b}A_{j} \right) \right)\\  =& - {i{\vt}}
\int_{\Sigma} \d{\a}_1 \wedge \d{\a}_2 + \frac{4\pi^2 \sqrt{\det(G)} }{
{\rm g}_{4}^{2}} \int_{\Sigma} \d ( {\om}_{1} {\a}_{1} + {\om}_{2}
{\a}_{2} ) \wedge \star \d ( {\bar\om}_{1} {\a}_{1} + {\bar\om}_{2}
{\a}_{2} ).\end{align}
\subsection{The Familiar Story}\label{famstory}

\def\re{{\mathrm{Re}}}
\def\bT{{\mathbf T}}
\def\Lie{{\rm Lie}}
\def\lieg{{{\underline{\bf g}}}}
\def\liet{{{{\frak t}}}}
\def\rm{\mathrm}
\def\bR{{\Bbb R}}
\def\bZ{{\Bbb Z}}
\def\bC{{\Bbb C}}
\def\pb{\bar{\partial}}
\def\CN{{\mathcal N}}
\def\tr{\,{\mathrm {tr}}\, }
\def\Tr{{\rm {Tr}}\,}
 We study the pure ${\N}=2$ gauge theory with the simple gauge
group $G$, its maximal torus ${\bT}$ and Cartan subalgebra $\frak
t = {\rm {Lie}}{\bT}$, and with a gauge coupling ${\rm g}_{4}$ and
theta angle $\vt$. The complex scalar in the vector multiplet is
denoted by $\phi$.

In compactifying this theory on $\IT$, in the small radius or low
energy approximation, one can impose the constraints
\begin{equation}\label{lwencn}[ {\phi}, {\bar\phi} ] = 0, \qquad [
A_{1}, A_{2} ] = 0,\end{equation} and assume everything is
independent of the angular coordinates $\theta^1,\theta^2$.

\def\cwt{{\mathrm{cwt}}}
Thus
\begin{equation}\label{thepaper}A_{1} = i {\a}_{1}, \quad A_{2} = i
{\a}_{2},\end{equation} where from now on ${\a}_{1,2} \in ({\frak
t} \otimes {\bR})/({\Lambda}_{\cwt} \otimes {\bZ})$, with
$\Lambda_{\cwt}$ the coweight lattice of $G$.  One must divide by
the Weyl group  acting diagonally on $\a_{1,2}$ as well as $\phi$.

\subsubsection{The Bosonic Lagrangian}

The bosonic part of the pure ${\CN}=2$ gauge theory Lagrangian
reduced on the torus ${\IT}$ is given at low energies by
\begin{align}\label{boslag}L =\notag & \frac{8{\pi}^2 }{ 2{\rm g}_{4}^{2}}
\sqrt{\det (G)} {\tr} \biggl\lbrace \left( {\om}_{1} \d{\a}_{1} +
{\om}_{2} \d{\a}_{2} \right) \wedge \star \left( {\bar\om}_{1}
\d{\a}_{1} + {\bar\om}_{2} \d{\a}_{2} \right) + \d{\phi} \wedge
\star \d{\bar\phi} \biggr\rbrace \\ & \qquad\qquad\qquad\qquad -
i{\vt} {\tr} \d{\a}_{1} \wedge \d{\a}_{2},\end{align} where ``tr''
denotes the induced metric on $\frak t$.  The gauge theory part of
this Lagrangian can be borrowed from  (\ref{lmaxi}).

We view here ${\a}_{1,2} \in {\liet} \otimes {\bR}/(
{\Lambda}_{\cwt} \otimes {\bZ})$ as real, and ${\phi} \in {\liet}
\otimes {\bC}$ as complex, with ``tr'' defining a positive
definite inner product on $\liet$. The Euclidean path integral
measure is given by
\begin{equation}\label{emes} e^{-\int L }.\end{equation}  The
condition for a field configuration to be invariant under the
supercharge $Q$ that is relevant to Donaldson theory and the
$\Omega$-deformation is
\begin{equation}\label{dw}\d \phi = 0, \qquad
{\pb} ( {\om}_{1}{\a}_{1} + {\om}_{2} {\a}_{2} ) =
0,\end{equation}  where the second equation is anti-selfduality of
the gauge field in our low energy approximation. For such fields,
(\ref{emes}) evaluates to:
\begin{equation}\label{Linst}\exp\left({-\int L}\right) = \exp\left({2\pi i \tau_{0} \int {\tr} \d{\a}_{1}
\wedge \d{\a}_{2}}\right), \end{equation} where the complexified gauge coupling is equal
to
\begin{equation}\label{taub} {\tau}_0 =  \frac{\vt}{ 2\pi} + \frac {4\pi i}{ {\rm
g}_{4}^{2}}.\end{equation}

\def\x{{\mathbf x}}
\def\CM{{\mathcal M}}
\def\m{{\mu}}
\def\CZ{{\mathcal Z}}
\def\CW{{\mathcal W}}
\def\CT{{\cal T}}
\subsubsection{$T$-Duality to ${\CM}_{H}$}

In what follows, we use the notation
\begin{equation}\label{mun}\boxed{{\m}_{0} = \frac{8{\pi}^{2}}{ {\rm
g}_{4}^{2}} \sqrt{\det (G)} = 2\pi \frac{{\im}({\tau}_{0})}{ {\im}({\om}_{2}{\bar\om}_{1}) }}\end{equation}
  The Lagrangian (\ref{boslag}) describes a sigma-model with target
   the product of a torus and $\frak t\otimes \Bbb C$, all divided
   by the Weyl group.
   Upon $T$-duality along the ${\a}_1$ direction,
    we map it to a sigma-model on ${\CM}_{H}$, after taking into account the nonlinear corrections.

  The $T$-duality is performed in the standard fashion. The first step is to replace $\d{\a}_1$ in (\ref{boslag})
   by an independent $\liet$-valued one-form $p_1$ and  add the term $ -2\pi i {\tr}
   ( p_1 \wedge \d{\tilde \a}_{1}) $ to $L$, with the understanding that ${\tilde\a}_{1}$ takes values in a
    circle of circumference $1$:
 \begin{align}\label{boslagt}
 \notag L^{\prime} = & \frac{{\m}_{0} }{2} {\tr} \biggl\lbrace \left( {\om}_{1} p_1
 + {\om}_{2} \d{\a}_{2} \right) \wedge \star \left( {\bar\om}_{1} p_1 + {\bar\om}_{2} \d{\a}_{2} \right)
 + \d{\phi} \wedge \star \d{\bar\phi} \biggr\rbrace \\
& \qquad\qquad\qquad\qquad +\, 2\pi i\, {\tr}\biggl\lbrace \left(
\d{\tilde \a}_1 + \frac{\vt}{ 2\pi}  \d{\a}_{2} \right) \wedge p_{1}
\biggr\rbrace \end{align} Integrating over $\tilde\alpha_1$ would
lead us back to (\ref{boslag}).  Instead we integrate over $p_1$.
 The path integral over $p_{1}$ is Gaussian, with the saddle point
for $p_1$ at:
\begin{equation}\label{poonsh}p_{1} = - {\rm {Re}}\left(
\frac{\om_{2}} {{\om}_{1}} \right) \d{\a}_{2} + i \star \,
\frac{2\pi}{ {\m}_{0} | {\om}_{1} |^{2}} \left( \d{\tilde \a}_{1} +
\frac{\vt }{ 2\pi} \d{\a}_{2} \right).\end{equation} In terms of
the left- and right-moving components of ${\a}_1$, (\ref{poonsh})
reads as follows:
\begin{align}\label{lraone}\notag & {\a}_1^{L} = \frac{2\pi}{{\m}_{0} |
{\om}_{1} |^{2} } \left( {\tilde {\a}}_{1} + \frac{\vt}{  2\pi}
{\a}_2 \right) - {\re }\left(\frac{{\om}_{2}}{ {\om}_{1}} \right)
{\a}_2 \\ & {\a}_{1}^{R} = -\frac{2\pi}{ {\m}_{0} | {\om}_{1}
|^{2} } \left( {\tilde \a}_{1} + \frac{\vt}{ 2\pi} {\a}_2 \right) -
{\re }\left(\frac{{\om}_{2}}{ {\om}_{1}} \right)
{\a}_2.\end{align} The $T$-dual Lagrangian is given by:
\begin{align}\label{boslagti}\notag L^{T} =
 &\frac{\mu_0}{2}\biggl\lbrace{\tr\,\frac{|\im(\omega_2\bar\omega_1)|^2}{|\omega_1|^2}\d \alpha_2\wedge
\star\d\alpha_2+\d\phi\wedge\star\d\phi+(2\pi)^2\frac{(\d\tilde\alpha_1+\frac{\vartheta}{2\pi}\d\alpha_2)
\wedge \star
(\d\tilde\alpha_1+\frac{\vartheta}{2\pi}\d\alpha_2)}{\mu_0^2|\omega_1|^2}\biggr\rbrace}\\ &
-2\pi i\,{\rm{Re}}\left(\frac{\omega_2}{\omega_2}\right)\tr\left(\d\tilde\alpha_2+\frac{\vartheta}{2\pi}\d\alpha_2
\right)\wedge\d\alpha_2.
\end{align}
 Introduce the $\liet \otimes {\bC}$-valued
dimensionless coordinates ${\CZ}, {\CW}$:
\begin{align}\label{zvf} \notag {\CZ} &= {\tilde \a}_{1} + \frac{\vt }{
2\pi} {\a}_{2} + \frac{4\pi i }{ {\rm g}_{4}^{2}} {\a}_{2}  \\
{\CW}& = \frac{1}{ 2\pi} {\m}_{0} {\bar\om}_{1} {\phi}.\end{align}
The choice of phase in the definition of $\CW$ is a matter of
convenience. Most of the time we assume ${\om}_{1} = | {\om}_{1} |
\in {\bR}_{+}$ anyway. In terms of ${\CW}$ and ${\CZ}$, eqn.
(\ref{boslagti}) takes the form:
\begin{align}\label{boslagtii}L^{T} = & \frac{(2{\pi})^2 }{ 2 {\m}_{0}
|{\om}_{1}|^2}  {\tr} \biggl\lbrace   \d{\CZ} \wedge \star \d{\bar
\CZ} + \d{\CW} \wedge \star \d{\bar\CW}  \biggr\rbrace \\ &
\qquad\qquad\qquad\qquad - 2\pi i\, {\re } \left(
\frac{{\om}_{2}}{ {\om}_{1}}  \right) {\tr} \left( \d{\tilde \a}_1
+ \frac{\vt }{ 2\pi}  \d{\a}_{2} \right) \wedge
\d{\a}_{2}.\end{align} Note that
\begin{equation}\label{mzome}{\boxed{ \frac{2\pi^2 }{ {\m}_0 |
{\om}_{1}|^{2}} = {\pi} \frac{{\im}\left( \frac{ \om_2}{ \om_1}
\right)}{{\rm {Im}} ({\tau}_{0} )} }}\end{equation} We deduce from
(\ref{boslagtii}) the target space metric
\begin{equation}\label{hkmt}\d s^{2}_{{\CM}_{H}} = {2 \pi} \frac{\rm
{Im}\left( {\om}_{2}/ {\om}_{1} \right) }{ {\rm {Im}} ({\tau}_{0}
)} \left( \d{\CZ} \d{\bar\CZ} + \d{\CW} \d{\bar\CW}
\right).\end{equation} In our approximation, the target space
metric is flat; in the exact theory, it is a complete hyper-Kahler
metric on what we usually call $\M_H$. We also deduce from
(\ref{boslagtii}) a $B$-field, which, up to exact terms, is given
by:
\begin{equation}\label{bfld}B = \frac{2\pi}{  2i} \frac{{\mathrm{Re}}(\omega_2/\omega_1)}{\im(\tau_0)}
\left( \d{\CZ}
\wedge \d{\bar\CZ} + \d{\CW} \wedge \d{\bar\CW} \right) = \left({\rm
{Re}}\left( {{\om}_{2}/ {\om}_{1}} \right)\right) {\om}_{I}.\end{equation}
Here ${\om}_{I}$ is the topologically normalized symplectic form
on ${\CM}_{H}$, which is K\"ahler in the complex structure $I$.
The functions of ${\CZ}, {\CW}$ are holomorphic in complex
structure $I$.

As we explained in section \ref{twod}, the supersymmetry of
Donaldson theory or the $\Omega$-deformation is that of the
$B$-model in complex structure $I$. Indeed, the equations
(\ref{dw}) read:
\begin{align}\label{dwi}& {\pb} ( {\om}_{1}{\a}_{1}^{L} +
{\om}_{2}{\a}_{2} ) = 0 \ \leftrightarrow \cr & \qquad\qquad
{\om}_{1} {\pb}   \left( \frac{2\pi }{ {\m}_{0}|{\om}_{1}|^{2}}
\left( {\tilde \a}_{1} +\frac{\vt }{ 2\pi} {\a}_{2} \right) +  i \,
{\rm {Im}} ( {\om}_{2}/{\om}_{1} ) {\a}_{2}  \right) = 0
\leftrightarrow {\pb}{\CZ} = 0 \cr & {\p} ( {\bar\om}_{1}
{\a}_{1}^{R} + {\bar\om}_{2} {\a}_{2} ) = 0 \leftrightarrow  \cr &
\qquad\qquad {\bar\om}_{1} {\p}  \left( - \frac{2\pi }{
{\m}_{0}|{\om}_{1}|^{2}} \left( {\tilde \a}_{1} +\frac{\vt }{ 2\pi}
{\a}_{2} \right)   -   i \, {\rm {Im}} ( {\om}_{2}/{\om}_{1} )
{\a}_{2} \right)=0. \leftrightarrow {\p}{\CZ} = 0  \end{align} Here
we used (\ref{mzome}) and:
$$
{\om}_{2} - {\re }( {\om}_{2}/{\om}_{1}) {\om}_{1} = i {\om}_{1} {\rm {Im}} ( {\om}_{2}/{\om}_{1} ).
$$
This, together with the obvious $\d{\phi} =0\leftrightarrow
\d{\CW} =0$ implies \begin{equation}\label{dwiii}\boxed{\d{\CW} =
0, \qquad \d{\CZ} = 0}.\end{equation}

To summarize, the sigma-model that comes from compactification on
a two-torus has ${\CM}_{H}$ as its target space and  the usual
supercharge is the one associated to the $B$-model of type $I$.
Let us write $\omega_I^*$ for the Kahler form of $\M_H$ in complex
structure $I$. Then we have
\begin{equation}\label{shelter}
{\om}_{I}^{*} = {\rm {Im}}({\om}_{2}/{\om}_{1}) {\om}_{I}, B =
{\rm {Re}} ({\om}_{2}/{\om}_{1}) {\om}_{I}.\end{equation} The
complexified gauge coupling ${\tau}_0$ maps to the complex
structure of the (asymptotic) fiber of Hitchin's fibration. This
is to be compared to the alternative approach \cite{KW} to the
same class of models by compactification of $\N=4$ super
Yang-Mills theory from four to two dimensions on a Riemann
surface. Not surprisingly, the r\^ole of the complexified gauge
coupling of ${\CN}=4$ super Yang-Mills is played in the present
approach by ${\om}_{2}/{\om}_{1}$:
\begin{equation}\label{nef}\boxed{{\tau}_{{\CN}=4} = \frac{{\om}_{2} }{
{\om}_{1}}}.\end{equation}

\subsection{The $\Omega$-Deformation}
\def\bi{{\mathbf i}}
\def\bk{{\mathbf k}}
\def\bj{{\mathbf j}}
\def\bH{{\Bbb H}}
\def\re{{\mathrm{Re}}}
\def\ba{{\mathbf a}}
\def\bb{{\mathbf b}}
\def\bX{{\mathbf X}}
\def\bt{{\mathfrak t}}
\def\bg{{\mathfrak g}}
Now let us turn on the $\Omega$-deformation, with the parameters
${\ve}_{1}, {\ve}_{2}$, corresponding to the shifts along the
$\theta^{1}$, $\theta^{2}$ directions. The Lagrangian becomes
\begin{align}\label{boslagoi}\notag & \frac{{\m}_{0}}{ 2}  {\tr} \biggl\lbrace \left( {\om}_{1}
\,\d{\a}_{1} + {\om}_{2} \,\d{\a}_{2} \right) \wedge \star \left( {\bar\om}_{1}\, \d{\a}_{1} + {\bar\om}_{2} \,
\d{\a}_{2} \right) \biggr\rbrace   \\ \notag
& \qquad\qquad +\frac{{\m}_{0}}{ 2} {\tr} \biggl\lbrace  \left( \d{\phi} + {\ve}_{1} \,\d{\a}_{1} + {\ve}_{2}\,
 \d{\a}_{2} \right) \wedge \star
\left( \d{\bar\phi} + {\bar\ve}_{1} \,\d{\a}_{1} + {\bar\ve}_{2}\, \d{\a}_{2} \right) \biggr\rbrace \\
& \qquad\qquad\qquad\qquad - i  {\vt} {\tr} \d{\a}_{1} \wedge \d{\a}_{2}. \end{align}
It is convenient to rewrite this formula using quaternions. Let ${\bi}, {\bj}, {\bk}$ denote the usual quaternion imaginary units,
\begin{equation}\label{ijk}{\bi}{\bj} = {\bk}, {\bi}^{2} = {\bj}^{2} = {\bk}^{2} = - 1, {\bi}{\bj} = - {\bj}{\bi}, \ldots .\end{equation}
The elements of
${\bH} = {\bR} \oplus {\bR}{\bi} \oplus {\bR}{\bj} \oplus {\bR} {\bk}$
can be conveniently represented as $z + w {\bj}$,
$z, w \in {\bC} = {\bR} \oplus {\bR} {\bi}$.
We have the trace map ${\re}: {\bH} \to {\bR}$,
$$
{\re } ( x_0 + x_1 {\bi} + x_2 {\bj} +  x_3 {\bk} ) = x_{0}, \ {\re } ( {\ba}{\bb} ) = {\re } ( {\bb}{\ba} ),
$$
and the conjugation map:
$$
( x_0 + x_1 {\bi} + x_2 {\bj} +  x_3 {\bk} )^{\dagger} = ( x_0 - x_1 {\bi} - x_2 {\bj} -  x_3 {\bk} )
$$
Introduce the following ${\liet} \otimes {\bH}$
valued field:
\begin{equation}\label{qkf}{\bX} = {\om}_{1} {\a}_{1} +
{\om}_{2} {\a}_{2} + \left( {\phi} + {\ve}_{1} {\a}_{1} +
{\ve}_{2} {\a}_{2} \right) {\bj}.\end{equation} Recall that we
view $\alpha_1,\alpha_2$ as real variables, i.e. they commute with
the quaternions. On the other hand, ${\phi}$
 obeys:
$$
{\phi} {\bi} = {\bi} {\phi}, ~~~
{\phi} {\bj} = {\bj} {\bar\phi}, ~~~{\phi} {\bk} = {\bk} {\bar\phi}
$$
The conjugated field is
\begin{equation}\label{qkfc}{\bX}^{\dagger} = {\bar\om}_{1} {\a}_{1} +
{\bar\om}_{2} {\a}_{2} - {\bj} \left( {\bar\phi} + {\bar\ve}_{1} {\a}_{1} + {\bar\ve}_{2} {\a}_{2} \right) .
\end{equation}
The action (\ref{boslagoi})
 reads, simply:
\begin{equation}\label{boslagoii}L = \frac{{\m}_{0}}{  2} {\re } \left( {\tr} \left(\ \d{\bX} \wedge \star
\d{\bX}^{\dagger} \right) \right) - i {\vt}{\tr}  \d{\a}_{1}
\wedge\ d{\a}_{2}.\end{equation} The advantage of our definition
of $\bX$ was to remove explicit deformation parameters from the
Lagrangian.  We recognize in (\ref{boslagoii})  the Lagrangian of
a sigma-model on $({\bt} \otimes {\bH})/({\Lambda}_{\cwt} \otimes
{\Gamma}) $ where now ${\Gamma} \cong {\bZ} \oplus {\bZ}$ is the
two-dimensional lattice embedded in ${\bH} \approx {\bC} \oplus
{\bC}{\bj}$ as follows:
\begin{equation}\label{imb}{\G} =   {\bZ}{\varpi}_{1}  +  {\bZ}{\varpi}_{2},\end{equation}
where
\begin{equation}\label{embi}{\varpi}_{1} = {\om}_{1} + {\ve}_{1} {\bj}, \qquad\qquad {\varpi}_{2} = {\om}_{2} + {\ve}_{2} {\bj}.\end{equation}

{\it A Remark On Relative Phases~} In deciding to combine ${\phi}$
and ${\a}_{1,2}$ to ${\bX}$ as in (\ref{qkf}) and ${\om}_{1,2}$
and ${\ve}_{1,2}$ as in (\ref{embi}), we made a choice of a
relative phase of ${\ve}_{1,2}$ and ${\om}_{1,2}$. This choice has
no intrinsic meaning, as long as we consider a theory with
$U(1)_R$ symmetry.   In fact, there are two arbitrary phases; one
can be fixed by requiring ${\om}_{1} = | {\om}_{1} |$, while the
other involves the phases of the deformation parameters. The
appearance of these phases is related to the impossibility to
decide canonically, within an $\N=2$ gauge theory that has
$U(1)_R$ symmetry, which complex structure on $\M_H$ is $J$ and
which is $K$ (of course, once $J$ is agreed upon, $K = IJ$
follows).  In general,  the distinction between $J$ and $K$
becomes meaningful upon incorporation of mass parameters that
break $U(1)_R$ symmetry. However, we will not incorporate such
parameters here.

\subsubsection{The Supersymmetry of the $\Omega$-Deformed Theory}

\def\CI{{\mathcal I}}
\def\CM{{\mathcal M}}
The supercharge which is favored by the $\Omega$-background enforces the following field equations:
\begin{align}\label{fld}\notag  \d ( {\phi} + {\ve}_{1} {\a}_{1} + {\ve}_{2} {\a}_{2} )& = 0
\\ \notag
 {\pb} ( {\om}_{1} {\a}_{1} + {\om}_{2} {\a}_{2} ) &= 0
\\
{\p} ( {\bar\om}_{1} {\a}_{1} + {\bar\om}_{2} {\a}_{2} )& = 0
.\end{align}
Upon performing  $T$-duality along the ${\a}_{1}$-direction, the equations (\ref{fld}) become:
\begin{align}\label{fldii}\notag {\pb} W^{+} &= {\pb} Z^{+} = 0, \\
 {\p} W^{-} &= {\p} Z^{-} = 0 ,\end{align}
with
\begin{align}\label{fldix}
\notag  W^{+} &= \left({\phi} + {\ve}_{1} {\a}_{1} + {\ve}_{2} {\a}_{2}\right)^L \\ \notag
 Z^{+}& = \left({\om}_{1} {\a}_{1} + {\om}_{2} {\a}_{2}\right)^L \\ \notag W^{-}&
 =  \left({\phi}  + {\ve}_{1} {\a}_{1} + {\ve}_{2} {\a}_{2}\right)^R \\
 Z^{-}& = \left(-{\bar\om}_{1} {\a}_{1} - {\bar\om}_{2} {\a}_{2}\right)^R
.\end{align}
The minus sign in the definition of $Z^{-}$ is chosen for convenience.
We need to express $W^{\pm}, Z^{\pm}$ in terms of ${\tilde \a}_{1}, {\a}_{2}, {\phi}, {\bar\phi}$, and relate them to the geometry of ${\CM}_{H}$.
We shall find that $W^{\pm}, Z^{\pm}$
are the holomorphic coordinates in two distinct complex
structures ${\CI}^{\pm}$ on ${\CM}_{H}$.

\subsubsection{Undoing the $\Omega$-Deformation}

We can find two unit quaternions $g_{\pm} \in SU(2)$, $g_{\pm} = {\ba}_{\pm} + {\bb}_{\pm} {\bj}$, $|{\ba}_{\pm}|^{2} + |{\bb}_{\pm}|^{2} = 1$, $g_{+}g_{+}^{\dagger} = g_{-}g_{-}^{\dagger} = 1$, such
that
\begin{equation}\label{omvarp} \boxed{\begin{matrix}\qquad {\Omega}_{1} = g_{+}^{\dagger} ( {\om}_{1} + {\ve}_{1} {\bj} ) g_{-} \\
\qquad {\Omega}_{2} = g_{+}^{\dagger} ( {\om}_{2} + {\ve}_{2}
{\bj} ) g_{-}  \end{matrix}}\end{equation} for two complex numbers
$\Omega_1,\Omega_2$.  (To accomplish this, pick $g_+^\dagger$ to
diagonalize the quaternion $(\omega_1+\epsilon_1\bj)(\bar\omega_2-
\epsilon_2\bj)$, and similarly choose $g_-^\dagger$ to diagonalize
$(\bar\omega_2-\epsilon_2\bj)(\omega_1+\epsilon_1\bj)$. To
``diagonalize'' a quaternion means to conjugate it into $\Bbb C$.)

Some easy consequences of (\ref{omvarp}) are
\begin{align}\label{zustard}\notag
& | {\Omega}_{1} |^{2} = | {\om}_{1}|^{2} + | {\ve}_{1} |^{2} \, , \quad | {\Omega}_{2} |^{2} = | {\om}_{2}|^{2} + | {\ve}_{2} |^{2} \\ \notag
& \qquad\qquad {\re} ( {\Omega}_{1} {\bar\Omega}_{2} ) = {\re} ( {\om}_{1} {\bar\om}_{2} + {\ve}_{1} {\bar\ve}_{2} ) \\ \notag
& {\bar\Omega}_{1} = g_{-}^{\dagger} ( {\bar\om}_{1} - {\ve}_{1} {\bj} ) g_{+} \, , \qquad
 {\bar\Omega}_{2} = g_{-}^{\dagger} ( {\bar\om}_{2} - {\ve}_{2} {\bj} ) g_{+} \\
& g_{+} {\Omega}_{1} g_{-}^{\dagger} = {\om}_{1} + {\ve}_{1} {\bj}  \, , \qquad
 g_{+} {\Omega}_{2} g_{-}^{\dagger} = {\om}_{2} + {\ve}_{2} {\bj}  .\end{align}
Again, we have some phase ambiguities. The periods ${\Omega}_{1,2}$ can be simultaneously
rotated, i.e. multiplied by a phase. This is equivalent to the right multiplication of $g_{-}$ and $g_{+}$
by  arbitrary phases:
\begin{equation}\label{multr}g_{\pm} \mapsto g_{\pm} e^{i {\a}_{\pm}}.\end{equation}
We can, as usual, normalize ${\Omega}_{1,2}$
in such a way that:
\begin{equation}\label{nrmlz}{\Omega}_{1} = | {\Omega}_{1} |, \qquad {\im}\,{\Omega}_{2} > 0.
\end{equation}
This reduces (\ref{multr}) down to a single phase ambiguity:
\begin{equation}\label{multrr}( g_{+}, g_{-} ) \mapsto (g_{+} e^{i {\a}}, g_{-} e^{i \a} ).\end{equation}

\subsubsection{Rotating the Fields}
 The Lagrangian (\ref{boslagoii}) is unchanged in form if ${\bX}$
replaced by ${\tilde\bX}$,
\begin{equation}\label{tldx}
{\tilde\bX} = g_{+}^{\dagger} {\bX} g_{-} =
{\Omega}_{1} {\a}_{1} + {\Omega}_{2} {\a}_{2} +
{\phi}_{\Vert} + {\phi}_{\perp}{\bj},\end{equation}
where
\begin{equation}\label{tldph}{\phi}_{\Vert} + {\phi}_{\perp}{\bj} = g_{+}^{\dagger} {\phi} {\bj} g_{-}.
\end{equation}
One can express ${\phi}_{\Vert}$ as
\begin{equation}\label{phvrt}{\phi}_{\Vert} = {\varphi}_{1} {\Omega}_{1} + {\varphi}_{2} {\Omega}_{2}, \end{equation}
where ${\varphi}_{1,2} \in {\liet}$ are real components that can
be deduced from
\begin{align}\label{sclpr}\notag  {\re} ( {\phi}_{\Vert} {\bar\Omega}_{2} )& = {\re} ({\phi}{\bar\ve}_{2} ) \\
\notag {\re} ( {\phi}_{\Vert} {\bar\Omega}_{1} ) &= {\re} ({\phi}{\bar\ve}_{1} ) \\
 {\bar\phi}_{\Vert} - {\phi}_{\perp} {\bj}& = - g_{-}^{\dagger} {\phi} {\bj} g_{+} .\end{align}
We will not need to compute these components explicitly.

The Lagrangian (\ref{boslagoii}) becomes
 \begin{align}\label{boslagoiii} \notag &{{\m}_{0} \over 2}  {\tr} \left\{ \left(
 \d{\phi}_{\Vert} + {\Omega}_{1} \d{\a}_{1} + {\Omega}_{2} \d{\a}_{2} \right) \wedge \star
 \left( \d{\bar\phi}_{\Vert} + {\bar{\Omega}}_{1} \d{\a}_{1} + {\bar{\Omega}}_{2} \d{\a}_{2} \right) +
 \d{\phi}_{\perp}  \wedge \star
\d{\bar\phi}_{\perp} \right\} \\
& \qquad\qquad\qquad\qquad\qquad\qquad - i  {\vt} {\tr} \d{\a}_{1}
\wedge \d{\a}_{2}.\end{align} By shifting ${\a}_{1,2}$ by
$-{\varphi}_{1,2}$, we could eliminate ${\phi}_{\Vert}$ at the
expense of adding exact terms to the $B$-field.

\subsubsection{$T$-Duality, Again}
We now $T$-dualize ${\a}_{1}$, as in section \ref{famstory}.  The
steps are the same; after a replacement $\d{\a}_{1} \mapsto
p_{1}$, we shift $L \to L^{\prime} = L_{\d{\a}_{1} \mapsto p_{1}}
- 2\pi i\, {\tr} p_{1} \wedge \d{\tilde \a}_{1}$, and  then
integrate out $p_{1}$.

Since (up to a trivial $B$-field shift),  (\ref{boslagoiii}) looks
like (\ref{boslag}) with ${\om}_{1,2}$ replaced by
${\Omega}_{1,2}$ (and no change in ${\m}_{0}$), we must end up
with the sigma-model on ${\CM}_{H}$ with the complexified Kahler
class
\begin{equation}\label{newnf}
{\tau}_{{\CN}=4} = {{\Omega}_{2} \over {\Omega}_{1}}\end{equation}
and the asymptotic complex structure of the fibers
given by:
\begin{equation}\label{tauff}{\tau} = {\vt \over 2\pi} + {4\pi i \over {\rm g}_{4}^{2}} {{\im}({\Omega}_{2}{\bar\Omega}_{1}) \over {\im}({\om}_{2}{\bar\om}_{1})}.\end{equation}
One can compute the right hand side of (\ref{tauff}), using (\ref{zustard}), but we shall not need these formulas.

Let us now perform the steps of the $T$-duality more explicitly.
The on-shell value of $p_1$ is given by:
\begin{equation}\label{ponesh}p_{1} = - \d {\re}\left( \frac{{\phi}_{\Vert} }{ {\Omega}_{1} } \right) - {\re}\left( \frac{{\Omega}_{2} }{ {\Omega}_{1} } \right) \d{\a}_{2} + 2\pi i\, \star \, {{\d{\tilde \a}_{1} +\frac{\vt }{ 2\pi} \d{\a}_{2}} \over {\m}_{0} | {\Omega}_{1} |^{2}}.\end{equation}
Note that here and below ${\m}_{0}$ is given by (\ref{mun}), with ${\om}_{1,2}$, not ${\Omega}_{1,2}$. In terms of the left- and right-moving components of ${\a}_{1}^{L,R}$, (\ref{ponesh}) reads:
\begin{align}\label{aonelr}\notag{\a}_{1}^{L} &= {{2\pi} \over {\m}_{0} | {\Omega}_{1} |^{2}} \left( {\tilde \a}_{1} + {\vt \over 2\pi} {\a}_{2} \right)  -  {\re}\left( {{\phi}_{\Vert} \over {\Omega}_{1} } \right) - {\re}\left( {{\Omega}_{2} \over {\Omega}_{1} } \right) {\a}_{2} \\
 {\a}_{1}^{R} &= -{{2\pi} \over {\m}_{0} | {\Omega}_{1} |^{2}} \left( {\tilde \a}_{1} + {\vt \over 2\pi} {\a}_{2} \right)  -  {\re}\left( {{\phi}_{\Vert} \over {\Omega}_{1} } \right) - {\re}\left( {{\Omega}_{2} \over {\Omega}_{1} } \right) {\a}_{2}.\end{align}
Having eliminated $p_1$ from $L^{\prime}$ we arrive at (we write $| \d x |^2$ instead of $\d x \wedge \star \d{\bar x}$ to avoid clutter):
\begin{align}\label{boslagoiv}L^{T}\  =  \  &
{{\m}_{0} \over 2} {\tr} \left\lbrace\ | \d{\phi}_{\perp} |^{2} +
|{\Omega}_{1} |^{2} \left\vert \d \left( {\im} \left(
{{\phi}_{\Vert} \over {\Omega}_{1}} \right) + {\im}\left(
{\Omega}_{2} \over {\Omega}_{1} \right) {\a}_{2} \right)
\right\vert^{2} \right\rbrace  + {(2\pi)^2 \over 2{\m}_{0}
|{\Omega}_{1}|^{2}} \left\vert \d \left( {\tilde \a}_{1} + {\vt
\over 2\pi}  {\a}_{2} \right) \right\vert^{2} \cr & \qquad\qquad +
2\pi i\, \d \left(  {\re}\left( {{\phi}_{\Vert} \over {\Omega}_{1}
} \right) + {\re}\left( {{\Omega}_{2} \over {\Omega}_{1} } \right)
{\a}_{2} \right) \wedge \d \left( {\tilde \a}_{1} + {\vt \over
2\pi} {\a}_{2} \right) .\end{align} We now introduce the complex
coordinates ${\CZ}, {\CW}$ on the target space of the effective
sigma-model:
\begin{align}\label{czcwo}\notag  {\CZ}& = {\tilde \a}_{1} + {\vt \over 2\pi} {\a}_{2} + {i {\m}_{0}\over 2\pi}\,  \left( {\im} ( {\Omega}_{2} {\bar\Omega}_{1})  {\a}_{2} +
{\im} \left( {\phi}_{\Vert} {\bar\Omega}_{1} \right) \right) \\
{\CW} &= {1\over 2\pi} {\m}_{0} {\bar\Omega}_{1} {\phi}_{\perp} .\end{align}
In terms of ${\CZ}, {\CW}$ the Lagrangian (\ref{boslagoiv}) reads simply:
\begin{equation}\label{boslagov}L^{T} = {(2\pi)^2 \over 2{\m}_{0} |{\Omega}_{1} |^{2}} {\tr} \left( \left\vert \d{\CZ} \right\vert^{2} + \left\vert \d{\CW} \right\vert^{2} \right) - i\, {\re}\, ( {\Omega}_{2}/{\Omega}_{1} ) {\om}_{I},\end{equation}
up to exact terms in the $B$-field. Note that ${\CZ} = {\tilde
\a}_{1} + {\tau} {\a}_{2}$ up to a shift by ${\im}({\phi}_{\Vert}
{\bar\Omega}_{1})$, with  $\tau$  given in (\ref{tauff}). This
shows that the fibers of Hitchin's fibration indeed have the
complex structure determined by $\tau$, not ${\tau}_{0}$.

\subsection{Rotated Complex Structures}

Now comes the hard work --  the determination of  the left- and
right-moving complex structures ${\CI}^{\pm}$ relevant to the
$T$-dualized theory, and of the Kahler class of the effective
$A$-model.

\begin{figure}
 \begin{center}
   \includegraphics[width=3in]{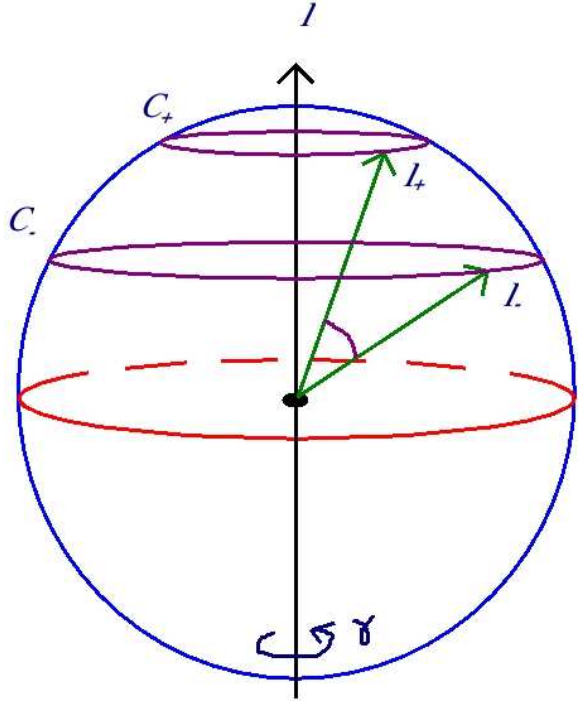}
 \end{center}
\caption{\small The complex structures $\mathcal I^{\pm},\,I$.}
 \label{complexstructures}
\end{figure}

With $W^{\pm}, Z^{\pm}$ given in (\ref{fldii}) and
${\a}_{1}^{L,R}$ given in (\ref{aonelr}), we compute:
\begin{align}\label{wzpm}\notag Z^{+} + W^{+} {\bj}& =
{2\pi \over {\m}_{0} |{\Omega}_{1}| } \, g_{+} \, \left( {\CZ} + {\CW} {\bj} \right) g_{-}^{\dagger}, \\
Z^{-} + W^{-} {\bj} &=
{2\pi \over {\m}_{0} |{\Omega}_{1}| } \, g_{-} \, \left( {\CZ} + {\CW} {\bj} \right) g_{+}^{\dagger}, \end{align}
where we used the normalization ${\Omega}_{1} = | {\Omega}_{1} |$. Otherwise, the formulas (\ref{wzpm}) are corrected by the multiplication of ${\CZ}, {\CW}$ by the phase of ${\Omega}_{1}^{\pm 1}$.

From this, it is straightforward to determine the effective
complex structures $\mathcal I^\pm$.  The complex structure that
multiplies $Z^+$ and $W^+$ by $i$ amounts to left multiplication
of the quaternion $Z^++W^+\bj$ by $\bi$.  This is equivalent to
left multiplication of the quaternion $\CZ+\CW\bj$ by
$g_+^\dagger\bi g_+$.  So $\mathcal I^+=g_+^\dagger\bi g_+$, and
by the same reasoning, $\mathcal I^-=g_-^\dagger \bi g_-$:
\begin{equation}\label{pmcmplx}\boxed{\begin{matrix} &
\CI^{+} = g_{+}^{\dagger} {\bi} g_{+}
 \\
&  {\CI}^{-} = g_{-}^{\dagger} {\bi} g_{-} \end{matrix}}\end{equation}

In the $U(1)_R$-symmetric case, there is an undetermined phase in
$g_\pm$, and hence the $\CI^\pm$ are only naturally determined up
to a rotation around the $I$ axis.  This is good enough for
determining the effective $A$-model, since in $U(1)_R$-invariant
theories (and more generally in any $\N=2$ model without
hypermultiplet bare masses, even if $U(1)_R$ is anomalous), only
the part of the complexified Kahler class proportional to
$\omega_I$ is relevant. This was explained in section
\ref{distinguished}, and reflects the fact that, in such examples,
the other hyper-Kahler forms $\omega_J$ and $\omega_K$ are
cohomologically trivial.  In a model that is not
$U(1)_R$-symmetric, $\omega_J$ and $\omega_K$ may be
cohomologically non-trivial, in which case more detailed  input
about the model is needed to fully determine the effective
$A$-model.

\subsubsection{Mapping to the $A$-Model}

\def\Cl{{\mathcal I}}
\def\z{{\zeta}}
\def\b{\beta}
Suppose that we fix a general pair
$({\CI}^{+}, {\CI}^{-})$ of  left- and right- moving complex structures in a model of
this type with $(4,4)$ supersymmetry.   The general formulas for the effective parameters
of the resulting $A$-model were given in eqn. (\ref{golf}).

It is convenient to make these formulas more explicit as follows.  First, write
\begin{equation}\label{cipm}{\CI}^{\pm} = {\z}^{\pm}_{1} {\bi} + {\z}^{\pm}_{2} {\bj} + {\z}^{\pm}_{13} {\bk},\end{equation}
where ${\vec\z}^{\pm} = ( {\z}^{\pm}_{1}, {\z}^{\pm}_{2}, {\z}^{\pm}_{3})$ both have unit norm, ${\vec\z}^{\pm} \cdot {\vec\z}^{\pm} = 1$.
Then expand the symplectic form $\omega$ and the two-form field $B$ of the effective $A$-model
\begin{align}\label{sbf}\notag {\om} &= {\xi}_{1} {\om}^*_{I} + {\xi}_{2} {\om}^*_{J} +
{\xi}_{3} {\om}^*_{K} \\
 B &=  {\b}_{1} {\om}^*_{I} + {\b}_{2} {\om}^*_{J} +
{\b}_{3} {\om}^*_{K} \end{align} in a basis of hyper-Kahler forms
$\om_{I,J,K}^*$ normalized with respect to the hyper-Kahler metric
$g$ which one reads off from the Lagrangian (thus,
$\omega_I^*=Ig$, etc.). As in eqn. (\ref{retso}), these forms
differ by a factor $\im(\Omega_2/\Omega_1)$ from the topologically
normalized symplectic forms $\om_{I,J,K}$ of eqns. (\ref{zobox})
and (\ref{obox}). In particular,
\begin{equation}\label{ompht}{\om}_{I}^{*} = {\im}( {\Omega}_{2} / {\Omega}_{1} ) {\om}_{I}.
\end{equation}
The vectors ${\vec\xi}, {\vec\b}$ of eqn. (\ref{sbf}) are given by the simple formulas:
\begin{align}\label{bexi}\notag {\vec\xi}& = {{\vec\z}^{+} - {\vec\z}^{-} \over
1 - {\vec\z}^{+} \cdot {\vec\z}^{-}} \\
 {\vec\b}   &=  {{\vec\z}^{+} \times {\vec\z}^{-} \over 1 - {\vec\z}^{+} \cdot {\vec\z}^{-}},\end{align}
which follow from eqn. (\ref{golf}). To be more exact, the
$B$-field described in (\ref{bexi}) is generated by the
transformation to an effective $A$-model.  We should not forget to
add to it the $B$-field already present in (\ref{boslagov}).

Now let us compute what we can, in eqn. (\ref{bexi}):
\begin{align}\label{scprd}\notag{\vec\z}^{+} \cdot {\vec\z}^{-} &=
- {\re}(  {{\CI}^{+} {\CI}^{-}} ) = - {\re} ( g_{+}^{\dagger} {\bi} g_{+} g_{-}^{\dagger} {\bi} g_{-} ) \\\ \notag
 {\z}^{+}_{1} - {\z}^{-}_{1} &= - {\re} \left( {\bi} ( {\CI}^{+} - {\CI}^{-} ) \right) = - {\re} ( g_{+}{\bi} g_{+}^{\dagger} {\bi} - g_{-} {\bi} g_{-}^{\dagger} {\bi} )\\
 ({\vec\z}^{+} \times {\vec\z}^{-})_{1}& = - {1\over 2} {\re} ( {\bi} [ {\CI}^{+}, {\CI}^{-} ] ) = {\re} \left( {\bi} g_{+}^{\dagger} {\bi} g_{+} g_{-}^{\dagger} {\bi} g_{-} \right).\end{align}
Here we used the cyclic symmetry of the quaternion trace $\x\to \re\,\x$.
 From the last two equations in (\ref{zustard}), using the normalization ${\Omega}_{1} = | {\Omega}_{1} |$, we derive:
 \begin{align}\label{nrmg}\notag g_{+}g_{-}^{\dagger} & = {1\over |{\Omega}_{1} | } \left( {\om}_{1} + {\ve}_{1} {\bj} \right) \\
  g_{+} {\bi} g_{-}^{\dagger} &= {1\over {\im} \,{\Omega}_{2}  } \left( {\om}_{2} + {\ve}_{2} {\bj} \right) - {{\re}{\Omega}_{2}  \over {\im} {\Omega}_{2}  } {1\over |{\Omega}_{1} | } \left( {\om}_{1} + {\ve}_{1} {\bj} \right) ,\end{align}
 while from the first equations in (\ref{zustard}), we compute:
 \begin{align}\label{gmgp}\notag g_{+} {\bi} g_{+}^{\dagger} = \frac{1}{ {\im}\, ( {\Omega}_{2} {\bar\Omega}_{1}) } &  \times \begin{pmatrix} & {\im}( {\om}_{2}{\bar\om}_{1} + {\ve}_{2} {\bar\ve}_{1} ) {\bi} \ +  \  \cr
 & \quad  +\  {\re} ( {\ve}_{2} {\om}_{1} - {\ve}_{1} {\om}_{2} ) {\bj}\   \cr
 &  \quad\qquad + \ {\im} ( {\ve}_{2} {\om}_{1} - {\ve}_{1} {\om}_{2} ) {\bk}\quad \end{pmatrix} \\
 g_{-} {\bi} g_{-}^{\dagger} = \frac{1}{ {\im} ( {\Omega}_{2} {\bar\Omega}_{1}) } &  \times\begin{pmatrix}&  {\im}( {\om}_{2}{\bar\om}_{1} + {\ve}_{1} {\bar\ve}_{2} ) {\bi} \ + \ \cr
 &  \quad + \ {\re} ( {\ve}_{2} {\bar\om}_{1} - {\ve}_{1} {\bar\om}_{2} ) {\bj} \quad  \cr
 & \quad\qquad + \ {\im} ( {\ve}_{2} {\bar\om}_{1} - {\ve}_{1} {\bar\om}_{2} ) {\bk}  \quad \end{pmatrix}.\end{align}

 Armed with eqns. (\ref{nrmg}) and (\ref{gmgp}), we finish the computation in (\ref{scprd}).  We compute
 \begin{align}\label{scprdii}\notag {\vec\z}^{+}\cdot {\vec\z}^{-}& =
\frac {| {\om}_{1} |^{2} - |{\ve}_{1}|^{2} }{ | {\om}_{1} |^{2} + |{\ve}_{1}|^{2}} \Longrightarrow 1 - {\vec\z}^{+}\cdot {\vec\z}^{-} =\frac {2 |{\ve}_{1}|^{2} }{ \vert {\Omega}_{1} \vert^{2}} \\
 & \cr
  {\z}^+_{1} - {\z}^-_{1} &= 2 \frac{{\im} ( {\ve}_{2}{\bar\ve}_{1} ) }
  { {\im}({\Omega}_{2}{\bar\Omega}_{1})} \notag \\
 & \cr
  ({\vec\z}^{+}\times {\vec\z}^{-})_{1} & = \frac{1}{ {\im}({\Omega}_{2}{\bar\Omega}_{1}) } \left( - {\re}({\om}_{2}{\bar\om}_{1}) + {\re}({\ve}_{2}{\bar\ve}_{1})
 + \left( | {\om}_{1}|^{2} - | {\ve}_{1} |^{2} \right) {\re} \left( {\Omega}_{2}/{\Omega}_{1} \right) \right) \notag \\
 &  = \frac{2| {\ve}_{1} |^{2} }{{\im}({\Omega}_{2}{\bar\Omega}_{1}) }  \left(  {\re}({\ve}_{2}/{\ve}_{1})
 - {\re} \biggl( {\Omega}_{2}/{\Omega}_{1} \right) \biggr).  \end{align}
 These combine to give us:
 \begin{equation}\label{bexao}\boxed{\begin{matrix}& {\xi}_{1} =\frac {\im({\ve}_{2}/{\ve}_{1})}{ {\im}({\Omega}_{2}/{\Omega}_{1})} \\
 & \\
 &  {\b}_{1} = \frac{{\re}({\ve}_{2}/{\ve}_{1}) -
 {\re}({\Omega}_{2}/{\Omega}_{1})}{ {\im}({\Omega}_{2}/{\Omega}_{1})} \end{matrix}}
 \end{equation}
 and
 \begin{equation}\label{final}\boxed{\begin{matrix}& {\om} = {\im}({\ve}_{2}/{\ve}_{1} ) {\om}_{I} + \ldots
 \cr
 & B = {\re}({\ve}_{2}/{\ve}_{1}) {\om}_{I} + \ldots \end{matrix}}\end{equation}
 In arriving at this $B$-field, we have included the bare $B$-field from (\ref{boslagov}),
 and of course we have also taken (\ref{ompht}) into account.
 Finally, we get the complexified Kahler class of the effective
 $A$-model:
 \begin{equation}\label{class}B + i {\om} = \frac{{\ve}_{2} }{ {\ve}_{1}} {\om}_{I} + \ldots .\end{equation}
This formula was obtained much more simply for a special choice of the $\Omega$-deformation
parameters in eqn. (\ref{roperly}).  Obtaining it in general has been the main goal of the present
appendix.

 \subsection{Dehn Twists}

 In our setup, the torus $T_{\circ}$ had a preferred $1$-cycle. The second 1-cycle is not defined uniquely.
 We could have made a Dehn twist of the torus:
 \begin{equation}\label{chnas}({\a}_{1}, {\a}_{2}) \mapsto ( {\a}_{1} + m {\a}_{2}, {\a}_{2}),\end{equation}
 for any integer $m \in {\bZ}$.   This is a symmetry of (\ref{boslagoi}) provided the parameters $({\om}_{1}, {\om}_{2})$, $({\ve}_{1}, {\ve}_{2})$ are transformed accordingly:
 \begin{equation}\label{omvetr}({\om}_{1}, {\om}_{2}) \mapsto ({\om}_{1}, {\om}_{2} - m {\om}_{1}), \qquad
 ( {\ve}_{1}, {\ve}_{2}) \mapsto ( {\ve}_{1}, {\ve}_{2} - m {\ve}_{1}).\end{equation}
 The effect of this transformation on our sigma-model is clear from (\ref{class}): the $B$-field is shifted by
 an integer multiple of ${\om}_{I}$,
 which corresponds to the presence of a Chan-Paton curvature. Note also that
 ${\tau}_{{\CN}=4}$ is shifted by $m$.

\bibliographystyle{unsrt}

\begin{thebibliography}{99}



\bibitem{MNS}
  G.~W.~Moore, N.~Nekrasov and S.~Shatashvili,
  ``Integrating over Higgs Branches,''
  Commun.\ Math.\ Phys.\  {\bf 209} (2000) 97, arXiv:hep-th/9712241.

\bibitem{GS}
  A.~A.~Gerasimov and S.~L.~Shatashvili,
  ``Higgs Bundles, Gauge Theories and Quantum Groups,''
  Commun.\ Math.\ Phys.\  {\bf 277} (2008) 323, arXiv:hep-th/0609024.

\bibitem{GS2}
  A.~A.~Gerasimov and S.~L.~Shatashvili,
  ``Two-Dimensional Gauge Theories and Quantum Integrable Systems,''
  arXiv:0711.1472 [hep-th].

\bibitem{NS}   N.~A.~Nekrasov and S.~L.~Shatashvili,
  ``Supersymmetric Vacua and Bethe Ansatz,''
  arXiv:0901.4744 [hep-th].

 \bibitem{NSmore} N.~A.~Nekrasov and S.~L.~Shatashvili,
  ``Quantum Integrability and Supersymmetric Vacua,''
  Prog.\ Theor.\ Phys.\ Suppl.\  {\bf 177} (2009) 105, arXiv:0901.4748 [hep-th].


\bibitem{NS2}   N.~A.~Nekrasov and S.~L.~Shatashvili,
  ``Quantization of Integrable Systems and Four Dimensional Gauge Theories,''
  arXiv:0908.4052 [hep-th].

\bibitem{Nek} N.~A.~Nekrasov,
  ``Seiberg-Witten Prepotential From Instanton Counting,''
  Adv.\ Theor.\ Math.\ Phys.\  {\bf 7} (2004) 831,
  arXiv:hep-th/0206161.


\bibitem{LNS1}
  A.~Losev, N.~Nekrasov and S.~L.~Shatashvili,
  ``Issues in Topological Gauge Theory,''
  Nucl.\ Phys.\  B {\bf 534} (1998) 549, arXiv:hep-th/9711108.


\bibitem{LNS2}
A.~Losev, N.~Nekrasov and S.~L.~Shatashvili,
  ``Testing Seiberg-Witten Solution,''
{\it NATO Advanced Study Institute on Strings, Branes and
Dualities} (Cargese, 1997), hep-th/9801061.





\bibitem{KO} A. Kapustin and D. Orlov, ``Remarks on $A$-Branes, Mirror Symmetry, and the Fukaya Category,''
  J.\ Geom.\ Phys.\  {\bf 48} (2003) 84,
  arXiv:hep-th/0109098.

\bibitem{Kap}
  A.~Kapustin,
  ``$A$-Branes and Noncommutative Geometry,''
  arXiv:hep-th/0502212.



\bibitem{AZ} M.~Aldi and E.~Zaslow,
  ``Coisotropic Branes, Noncommutativity, and the Mirror Correspondence,''
  JHEP {\bf 0506} (2005) 019,
  arXiv:hep-th/0501247.


\bibitem{KW} A. Kapustin and E. Witten,   ``Electric-Magnetic Duality and the Geometric Langlands Program,''
Commun. Numb. Th. Phys. {\bf 1} (2007) 1-236
  arXiv:hep-th/0604151.

\bibitem{GW}   S.~Gukov and E.~Witten,
  ``Branes and Quantization,''
  arXiv:0809.0305 [hep-th].

\bibitem{AGT}
  L.~F.~Alday, D.~Gaiotto and Y.~Tachikawa,
  ``Liouville Correlation Functions From Four-Dimensional Gauge Theories,''
  arXiv:0906.3219 [hep-th].

\bibitem{DG}
D. Gaiotto, ``$\N=2$ Dualities,'' arXiv:0904.2715 [hep-th].



  \bibitem{Hitchin}
  N. Hitchin, ``The Self-Duality Equations On A Riemann Surface,'' Proc. London Math.
  Soc. (3) {\bf{33}} (1987) 59-126.

\bibitem{CN}
  N.~C.~Leung and C.~Vafa,
  ``Branes and Toric Geometry,''
  Adv.\ Theor.\ Math.\ Phys.\  {\bf 2} (1998) 91, arXiv:hep-th/9711013.

\bibitem{F1}
E. Frenkel, ``Affine Algebras, Langlands Duality, And Bethe
Ansatz,'' {\it XIth International Congress on Mathematical
Physics,} Paris, 1994 (Int. Press, Cambridge, MA, 1995),
 pp.
606--642, arXiv:q-alg/9506003.

\bibitem{FFR}
B. Feigin, E. Frenkel, and N. Reshetikhin, ``Gaudin Model, Bethe
Ansatz, and Critical Level,'' Comm. Math. Phys. {\bf {166}} (1994)
27-62.

\bibitem{F2}
E. Frenkel, ``Gaudin Model And Opers,'' Progr. Math. {\bf 237}
(2005) 1-58, arXiv:math/0407524.


\bibitem{F3}
B. Feigin, E. Frenkel, and V. Toledano-Laredo, ``Gaudin Models
with Irregular Singularities,'' Adv.  Math. {\bf  223} (2010)
873-948, arXiv:math/0612798.





\bibitem{FF}
B.Feigin and E.Frenkel, ``Affine Kac-Moody Algebras at the
Critical Level and Gelfand-Dikii Algebras,'' Int. J. Mod. Phys. A
Suppl. 1A (1992) 197-215.

\bibitem{F}
E. Frenkel, ``Lectures on the Langlands Program and Conformal
Field Theory,'' in {\it Frontiers in Number Theory, Physics and
Geometry II,} eds. P. Cartier, et. al., pp. 387-533
(Springer-Verlag, 2007), hep-th/0512172.


\bibitem{Teschner} J. Teschner, ``Quantum Geometric Langlands vs. Non-Perturbative
Dualities In Sigma Models,''  \url{http://online.kitp.ucsb.edu/online/duality09/teschner1/}.

\bibitem{Teschner2} J. Teschner, ``Notes On Liouville Theory, Quantization Of The Hitchin
System, And The Geometric Langlands Correspondence,'' to appear.

\bibitem{LNS}
  A.~Losev, N.~Nekrasov and S.~L.~Shatashvili,
  ``The Freckled Instantons,''
  arXiv:hep-th/9908204.


\bibitem{Pestun}
  V.~Pestun,
  ``Topological Strings in Generalized Complex Space,''
  Adv.\ Theor.\ Math.\ Phys.\  {\bf 11} (2007) 399, arXiv:hep-th/0603145.

  \bibitem{CDS}
  A.~Connes, M.~R.~Douglas and A.~S.~Schwarz,
  ``Noncommutative Geometry and Matrix Theory: Compactification on Tori,''
  JHEP {\bf 9802} (1998) 003,
  arXiv:hep-th/9711162.

  \bibitem{SW}
  N.~Seiberg and E.~Witten,
  ``String Theory and Noncommutative Geometry,''
  JHEP {\bf 9909} (1999) 032,
  arXiv:hep-th/9908142.

\bibitem{DonW}
  R.~Donagi and E.~Witten,
  ``Supersymmetric Yang-Mills Theory And Integrable Systems,''
  Nucl.\ Phys.\  B {\bf 460} (1996) 299, arXiv:hep-th/9510101.




\bibitem{HitchA}
N. Hitchin, ``Flat Connections And Geometric Quantization,''
Commun. Math. Phys. {\bf 191} (1990) 347-380.

\bibitem{BD}
A. Beilinson and V. Drinfeld, ``Quantization Of Hitchin's
Integrable System and Hecke Eigensheaves,'' preprint (ca. 1995),
available at \url{http://www.math.uchicago.edu/~mitya/langlands.html}.


\bibitem{SW2}
  N.~Seiberg and E.~Witten,
  ``Gauge Dynamics and Compactification to Three Dimensions,''
  arXiv:hep-th/9607163.

\bibitem{WD}
  E.~Witten,
  ``Topological Quantum Field Theory,''
  Commun.\ Math.\ Phys.\  {\bf 117} (1988) 353.

\bibitem{HM}
  J.~A.~Harvey, G.~W.~Moore and A.~Strominger,
  ``Reducing $S$-Duality To $T$-Duality,''
  Phys.\ Rev.\  D {\bf 52} (1995) 7161
  arXiv:hep-th/9501022.



\bibitem{BJV}
  M.~Bershadsky, A.~Johansen, V.~Sadov and C.~Vafa,
  ``Topological Reduction of 4-d SYM to 2-d Sigma Models,''
  Nucl.\ Phys.\  B {\bf 448} (1995) 166
  arXiv:hep-th/9501096.

\bibitem{GMN}
  D.~Gaiotto, G.~W.~Moore and A.~Neitzke,
  ``Four-Dimensional Wall-Crossing Via Three-Dimensional Field Theory,''
  arXiv:0807.4723 [hep-th].

\bibitem{Rocek} S. J. Gates, C. M. Hull, and M. Rocek, ``Twisted
Multiplets And New Supersymmetric Nonlinear Sigma Models,'' Nucl.
Phys. {\bf B248} (1984) 157-186.

\bibitem{Hitchin2}
N. Hitchin, ``Generalized Calabi-Yau Manifolds,'' Quart. J. Math.
{\bf 54} (2003) 281-308, math.DG/0209099.

\bibitem{Gualtieri} M. Gualtieri, ``Generalized Complex
Geometry,'' math.DG/0401221.

\bibitem{GaiWit}
  D.~Gaiotto and E.~Witten,
  ``Supersymmetric Boundary Conditions in $\N=4$ Super Yang-Mills Theory,''
  arXiv:0804.2902 [hep-th].

\bibitem{Wyllard}
  N.~Wyllard,
  ``$A_{N-1}$ Conformal Toda Field Theory Correlation Functions From Conformal
  $\N=2$ $SU(N)$ Quiver Gauge Theories,''
  JHEP {\bf 0911} (2009) 002
  arXiv:0907.2189 [hep-th]

  \bibitem{MMMM}
  A.~Mironov, S.~Mironov, A.~Morozov and A.~Morozov,
  ``CFT Exercises for the Needs of AGT,''
  arXiv:0908.2064 [hep-th].

\bibitem{HT}
 T. Hausel and M. Thaddeus, ``Mirror Symmetry, Langlands Duality,
 And The Hitchin System,'' Invent. Math. {\bf 153} (2003) 197-229.

 \bibitem{CF}
 L. Chekhov and V. V. Fock, ``Quantum Teichmuller Spaces,'' Teor. Mat. Fiz. {\bf 120}
 (1999) 511-528, arXiv:math/9908165.

 \bibitem{FG}
 V.  V. Fock and A. B. Goncharov, ``The Quantum Dilogarithm And Representations Of
 Quantum Cluster Varieties,''  arXiv:math/05103.

\bibitem{GW2}
D. Gaiotto and E. Witten,
``Supersymmetric Boundary Conditions in $\N=4$ Super Yang-Mills Theory,''
arXiv:0804.2902 [hep-th].

\bibitem{AMR}
J. E. Andersen, J. Mattes, and N. Reshetikhin, ``The Poisson Structure on
the Moduli Space of Flat Connections and Chord Diagrams,'' Topology
{\bf 35} (1996) 1069--1083.

\bibitem{Witten3}
E. Witten, ``Gauge Theories, Vertex Models, And Quantum Groups,''
 Nuclear Phys. {\bf B330}  (1990)  285-346.

\bibitem{AMR2}
J.E. Andersen,  J. Mattes, and N. Reshetikhin, ``Quantization of the
Algebra of Chord Diagrams,'' Math. Proc. Camb. Phil. Soc. {\bf 124}
(1998) 451-467.



 \bibitem{Witten2}
E. Witten, ``Quantum Field Theory And The Jones Polynomial,''
  Commun.\ Math.\ Phys.\  {\bf 121} (1989) 351.

\bibitem{Andersen}
J. E. Andersen, ``Hitchin's Connection, Toeplitz Operators, And
Symmetry-Invariant Deformation Quantization,'' arXiv:math/0611126.

\bibitem{EVerlinde}
E. Verlinde, ``Fusion Rules And Modular Transformations In 2D
Conformal Field Theory,'' Nucl. Phys. {\bf B300} (1988) 360.

\bibitem{AGGTV}
  L.~F.~Alday, D.~Gaiotto, S.~Gukov, Y.~Tachikawa and H.~Verlinde,
  ``Loop and Surface Operators in ${\mathcal N}=2$ Gauge Theory and Liouville Modular
  Geometry,''
  arXiv:0909.0945 [hep-th].

  \bibitem{DGOT}
    N.~Drukker, J.~Gomis, T.~Okuda and J.~Teschner,
  ``Gauge Theory Loop Operators and Liouville Theory,''
  arXiv:0909.1105 [hep-th].


\bibitem{DMO}
  N.~Drukker, D.~R.~Morrison and T.~Okuda,
  ``Loop Operators and $S$-Duality from Curves on Riemann Surfaces,''
  JHEP {\bf 0909} (2009) 031,
  arXiv:0907.2593 [hep-th].



\bibitem{Pestun1}
  V.~Pestun,
  ``Localization of Gauge Theory on a Four-Sphere and Supersymmetric Wilson
  Loops,''
  arXiv:0712.2824 [hep-th].

  \bibitem{CV}
    S.~Cecotti and C.~Vafa,
  ``Topological Antitopological Fusion,''
in {\it   Trieste 1991, Proceedings, High Energy Physics and Cosmology, vol. 2}, pp. 682-784.

\bibitem{FS}
  D.~Friedan and S.~H.~Shenker,
  ``The Analytic Geometry of Two-Dimensional Conformal Field Theory,''
  Nucl.\ Phys.\   {\bf B281} (1987) 509.

\bibitem{SurfOp}
  L.~F.~Alday, D.~Gaiotto, S.~Gukov, Y.~Tachikawa and H.~Verlinde,
  ``Loop and Surface Operators in $\N=2$ Gauge Theory and Liouville Modular
  Geometry,''
  arXiv:0909.0945 [hep-th].

\bibitem{SurfOp2}
  D.~Gaiotto,
  ``Surface Operators in $\N=2$ $4d$ Gauge Theories,''
  arXiv:0911.1316 [hep-th].



\end{thebibliography}

\end{document}